\documentclass{mscs-n}

\usepackage{natbib}
\let\cite=\citep

\usepackage{amsmath}
\usepackage{amssymb}
\usepackage{meq}
\usepackage{xspace}
\usepackage{url}
\usepackage[inline]{enumitem}
\setlist{itemsep=.5em,topsep=.5em}
\usepackage{wrapfloat}
\usepackage{fancyvrb}
\usepackage{graphicx}
\usepackage{framed}

\usepackage{header}
\usepackage[all]{xy}
\usepackage[UglyObsolete]{diagrams}
\renewcommand{\hole}{\square}

\newcommand{\eps}{\ensuremath{\epsilon}}

\defmath{\Lab}{L}
\newcommand{\amarker}[1]{\ensuremath{#1}}
\newcommand{\defaultM}{\dmark}

\newenvironment{tarray}[2][c]{%
\settowidth{\dimen1}{${} = {}$}%
\setlength{\arraycolsep}{0.125\dimen1}
\hspace{-0.125\dimen1}\array[#1]{#2}\relax
}
{%
\endarray\hspace{-0.125\dimen1}
}
\newcommand{\bb}{\tarray{lllll}}
\newcommand{\bbt}{\tarray[t]{lllll}}

\newcommand{\ee}{\endtarray}

\newcommand{\sem}[1]{\den{#1}}

\newcommand{\gc}{\cat{Gr}}
\newcommand{\gs}[2]{\underline{\gc}(#1,#2)}
\newcommand{\gsb}[2]{\gc(#1,#2)}
\newcommand{\com}{\circ}
\newcommand{\pf}[2]{\left\langle #1 , #2 \right\rangle}

\begin{document}
\pagestyle{plain}

\title{The Algebra of Recursive Graph Transformation Language UnCAL:
Complete Axiomatisation and Iteration Categorical Semantics}
\author[Hamana, Matsuda and Asada]
   {Makoto Hamana$^1$, Kazutaka Matsuda$^2$ and Kazuyuki Asada$^3$\\
$^1$ Department of Computer Science, Gunma University, Japan 
\addressbreak
$^2$ Graduate School of Information Sciences, Tohoku University, Japan
\addressbreak
$^3$ Department of Computer Science, University of Tokyo, Japan
}

\pubyear{2015}
\pagerange{\pageref{firstpage}--\pageref{lastpage}}
\volume{00}
\doi{}

\maketitle

\y{-3em}
\begin{abstract} 

The aim of this paper is to provide mathematical foundations of a graph
transformation language, called UnCAL, using categorical semantics of
type theory and fixed points.  About twenty years ago, Buneman et
al. developed a graph database query language UnQL on the top of a
functional meta-language UnCAL for describing and manipulating
graphs. Recently, the functional programming community has shown renewed
interest in UnCAL, because it provides an efficient graph transformation
language which is useful for various applications, such as bidirectional
computation.

In order to make UnCAL more flexible and fruitful for further extensions
and applications, in this paper, we give a more conceptual understanding
of UnCAL using categorical semantics.  Our general interest of this paper
is to clarify what is the {algebra of UnCAL}. Thus, we give an equational
axiomatisation and categorical semantics of UnCAL, both of which are new.
We show that the axiomatisation is complete for the original bisimulation
semantics of UnCAL.  Moreover, we provide a clean characterisation of the
computation mechanism of UnCAL called ``structural recursion on graphs''
using our categorical semantics.  We show a concrete model of UnCAL given
by the \lmdG-calculus, which shows an interesting connection to lazy
functional programming.

\end{abstract}


\tableofcontents
\ifprodtf \newpage \else \vspace*{-1\baselineskip}\fi

\section{Introduction}\label{sec:review}

Graph databases, which store graphs rather than relations in 
ordinary relational database,
are used as back-ends of various web and net services.
Therefore it is one of the important software systems in the Internet
society. About twenty years ago, \citet{SIGMOD1996,Buneman2,Buneman} developed a graph
database query language UnQL (Unstructured data Query Language)
on top of a functional meta-language
\textbf{UnCAL} (Unstructured Calculus)
for describing and manipulating graph data. 
The term ``unstructured'' refers to
unstructured or semi-structured data, i.e.,
data having no assumed format 
in a database
(in contrast to relational database).
Recently, 
the functional programming community has expressed 
renewed interest in UnCAL.
A research group in Tokyo
discovered a new application area of UnCAL
in so-called bidirectional transformations on graph data
\cite{ICFP10,LOPSTR,ICFP13,PPDP13},
because it provides an efficient graph transformation
language.
It is also suitable 
for various applications, including software engineering \cite{ICSE12,Hu-tram}.
UnCAL has been extended and refined in various directions.
Those developments have enhanced the importance of UnCAL.

In order to make UnCAL more 
flexible and fruitful for further extensions and applications,
in this paper, 
we present a more conceptual understanding of UnCAL using 
semantics of type theory and fixed points.
Our general interest in this study is clarification of the \W{algebra of UnCAL}. 
Therefore, we give
an equational axiomatisation  and categorical semantics
of UnCAL, both of which are new.
We show that the axiomatisation exactly matches with the original formulation.
That is, it is complete
for the original bisimulation of UnCAL graphs.
Moreover, we provide a clean characterisation of 
the computation mechanism of UnCAL called
``structural recursion on graphs'' using our categorical semantics.

\subsection{Why the algebra of UnCAL? --- models of graphs, bisimulation and structural recursion}

Besides the practical importance of UnCAL, 
this study is also conceptually and theoretically motivated.
UnCAL deals with graphs \Hi{modulo bisimulation}, rather than
isomorphisms or strict equality.
Numerous studies have examined
\Hi{algebraic} or \Hi{categorical characterisation} of graphs
(or DAGs)
modulo \Hi{isomorphism} in the field of semantics of programming languages, such as 
\cite{ETGRS,cyclic-lmd,HasseiTLCA,Hassei,GS-monoidal,
ActionCalc,ax-bigraph,PureBigraph,LMCS,FLOPS,Gibbons,FiorePROP}.
Algebraic characterisation is associated directly 
to the constructive nature of graphs. Thereby these can be applied nicely to 
functional programming concepts
such as the datatype of graphs and its ``fold'' or structural recursion
\cite{Gibbons,LMCS} and
\lmd-calculus \cite{cyclic-lmd,HasseiTLCA,Hassei}.
In this respect, algebraic characterisation has both
theoretical benefits (e.g. new semantical structures)
and practical benefits (e.g. programming concepts).

Graphs modulo \Hi{bisimulation} have mainly been studied in the context of
concurrency \cite{MilnerRegular,CCS}, 
rather than functional programming and program semantics.
In concurrency, graphs are used to represent traces (or behavior) of processes,
and bisimulation are used to compare them.
Semantic tools to model them are often coalgebraic because
a graph can be regarded as a coalgebraic structure
that outputs node information along its edges, 
e.g.\ \cite{AczelAdamek,NeilCoalg}.
Then bisimulation is modelled as a relation on coalgebras,
e.g.\ \cite{RelatingBisim}.

The design of UnCAL is unique from the viewpoint of
existing works related to semantics of graphs, bisimulation, and 
programming described above.
In fact, UnCAL is similar to functional programming 
(because of computation by ``structural recursion on graphs''), 
but the basic data structure is non-standard,
i.e. graphs modulo bisimulation.
One reason for this design choice of UnCAL is due to efficiency.
Therefore, the original UnCAL's formulation is mainly graph algorithmic.
Basic graph theory, algorithmic evaluation,
and ordinary relational treatment of bisimulation are the main tools.

Clarifying more mathematical semantics of UnCAL is challenging.
Three key concepts of UnCAL 
--- graphs, bisimulation, and functional programming ---
have been modelled in different settings as described above, but they
have not been examined in a single framework. 
Consequently, our central question is
\begin{center}
\Hi{What is the mathematical structure for modelling UnCAL?}
\end{center}
Finding the structure has additional importance because
it tightly connects to the notion of structural recursion in UnCAL.
In principle, a structural recursive function
is a mapping that preserves ``structures''.
However, such structures of UnCAL have not been pursued seriously and 
have remained vague in the original and subsequent developments of UnCAL.

Our approach to tackle this problem is 
algebraic and categorical, rather than a coalgebraic approach.
The reason is described below.
Our central idea is to follow the methodology of modelling 
structural recursion or the ``fold'' of algebraic datatype
by the universality of datatypes in functional programming, 
{e.g.} \cite{Hagino}.
It is the established methodology to characterise datatypes and recursion in a single setting,
where the general structure is functor-algebra,
and datatype is characterised as an initial functor-algebra.
Structural recursion follows automatically from the universality.
This methodology can be achieved only using an algebraic approach.
Therefore, the coalgebraic approach is unsuitable for our purposes.

In the case of UnCAL, our methodology is
\begin{enumerate}[label=(\arabic*)]
\item to identify a suitable category for UnCAL,
which has all required constructs and properties, and
\item to characterise the ``datatype'' of UnCAL graphs as 
a universal one in it.
\end{enumerate}
Of course, these 
are not merely an algebraic datatype, functor-algebras 
and an initial algebra, unlike \cite{LMCS}, which modelled
rooted graphs without any quotient by an initial functor-algebra
in a presheaf category.
It is necessary to seek more involved and suitable mathematical structure.

As described herein, we identify the mathematical structure of UnCAL
to be \Hi{iteration categories} \cite{IterCat,BE} and 
our notion of \Hi{$L$-monoids} (Def. \ref{def:L-monoid}).
We characterise UnCAL's graphs (regarded as the ``datatype'')
by the universal property of certain $L$-monoid and iteration category.
Structural recursion follows from the universality.

This universal characterisation of UnCAL
shows that UnCAL has a standard status, i.e.,
it is not an ad-hoc language of graphs transformation.
The simply-typed \lmd-calculus is a standard functional
language because it is universal in
cartesian closed categories.
The universal characterisation of UnCAL also provides syntax-independent formulation. 
Therefore, the current particular syntax of UnCAL is unimportant.
What is most important we think is that 
the necessary mathematical structures for 
a language of graph transformation
with bisimulation and structural recursion are
iteration categories and
$L$-monoids.
In this respect, 
UnCAL is one such language, but
it is also \Hi{the most standard language} because
it has exactly these required structures for graph transformation.

\subsection{Bridge among database, categorical semantics, and
  functional programming}

This universal characterisation provides another benefit.
We mentioned that the current syntax of UnCAL is unimportant.
We will actually give another syntax of UnCAL using a \lmd-calculus,
called the \lmdG-calculus (\Sec \ref{sec:lambdaG}).
It is because the syntax and equational theory of the \lmdG-calculus 
forms a categorical model of UnCAL,
Therefore, by universality, there exists a structure preserving translation from
the universal one, i.e. the original UnCAL syntax, to the \lmdG-calculus.
The translation is also invertible (Prop. \ref{th:corrspond-ELu}).
Therefore, one can use the \lmdG-calculus as an alternative syntax,
computation and reasoning method of UnCAL.
Moreover, an efficient abstract machine for \lmdG-calculus has been implemented
as a functional language FUnCAL \cite{MA}.

\bigskip

UnCAL originated in the field of
database, but its language design also gained from
ideas in
concurrency and functional programming, as mentioned recently
in Buneman's retrospect on the relationship between 
database and programming languages \cite{BunemanPOPL}.
Our categorical approach \Hi{sharpens} 
this line of good relationship between database and programming languages,
not only at the level of language design and implementations, but also 
at the level of rigorous mathematical semantics.
It bridges different areas of computer science, i.e.
graph database (UnCAL), categorical semantics (iteration category), 
and functional programming (the \lmdG-calculus)
by categorical interpretation.


\subsection{UnCAL overview}
\subsubsection{}\label{sec:ireview}
We begin by introducing UnCAL.
UnCAL deals with graphs. 
Hence, it is better to start
with viewing how concrete semi-structured data
is processed in UnCAL. Consider the
semi-structured data \code{sd} below
which is 
taken from \cite{Buneman}.
It contains information about country, e.g. geography, people, 
government,
etc. 
It is depicted as a tree
\begin{wrapfigure}[8]{r}{.7\linewidth}\y{-1em}
\begin{Verbatim}[fontsize=\small,frame=single,baselinestretch=.9,
commandchars=\\\[\],codes=\mathcom]
sd $\;\triangleleft\;\,$ country:{name:"Luxembourg",
 geography:{coordinates:{long:"49 45N", lat:"6 10E"},
            area:{total:2586, land:2586}},
 people:{population:425017,
         ethnicGroup:"Celtic",
         ethnicGroup:"Portuguese",
         ethnicGroup:"Italian"},
 government:{executive:{chiefOfState:{name:"Jean",..}}}}
\end{Verbatim}
\end{wrapfigure}
\noindent
above, in which  edges and leaves are labelled.
Using UnCAL's term language for describing graphs (and trees),
this is 
defined by \code{sd} shown at the right.
Then we can define functions in UnCAL to 
process data. For example, a function that retrieves all ethnic groups in the graph can be 
defined simply by 
\[
\arraycolsep = 1mm
\begin{array}[h]{llcll}
  \code{sfun} &\code{f1}(\code{ethnicGroup}\co t) &=& \code{result}\co t\\
              &\code{f1}(\ell\co t) &=& \code{f1}(t) &\qquad\text{for
                }\ell\not= \code{ethnicGroup} 
\end{array}
\]

\noindent
The keyword \code{sfun} denotes a function definition by 
\W{structural recursion on graphs}.
Executing it, we certainly extract:  
\begin{Verbatim}[commandchars=\\\[\],codes=\mathcom]
  f1(sd) \narone {result:"Celtic", result:"Portuguese", result:"Italian"}
\end{Verbatim}
The computation by structural recursion on graphs is the 
\Hi{only} computational mechanism of UnCAL. 
There is no other evaluation mechanism.
That situation is analogous to
a functional 
programming language in which
the only computation mechanism is ``fold''.
Although it is restrictive,
a benefit of this design choice is that the result of 
computation by structural recursion
always terminates even when an input graph involves cycles.
The execution of structural recursion on graphs
is different from ordinary evaluation of functional programming.
It was realized by two graph algorithms,
called the \Hi{bulk semantics} and the \Hi{recursive semantics}. Although they are
called semantics, these are actually algorithms.

\begin{rulefigt}
  \y{-2em}
    \includegraphics[scale=.9]{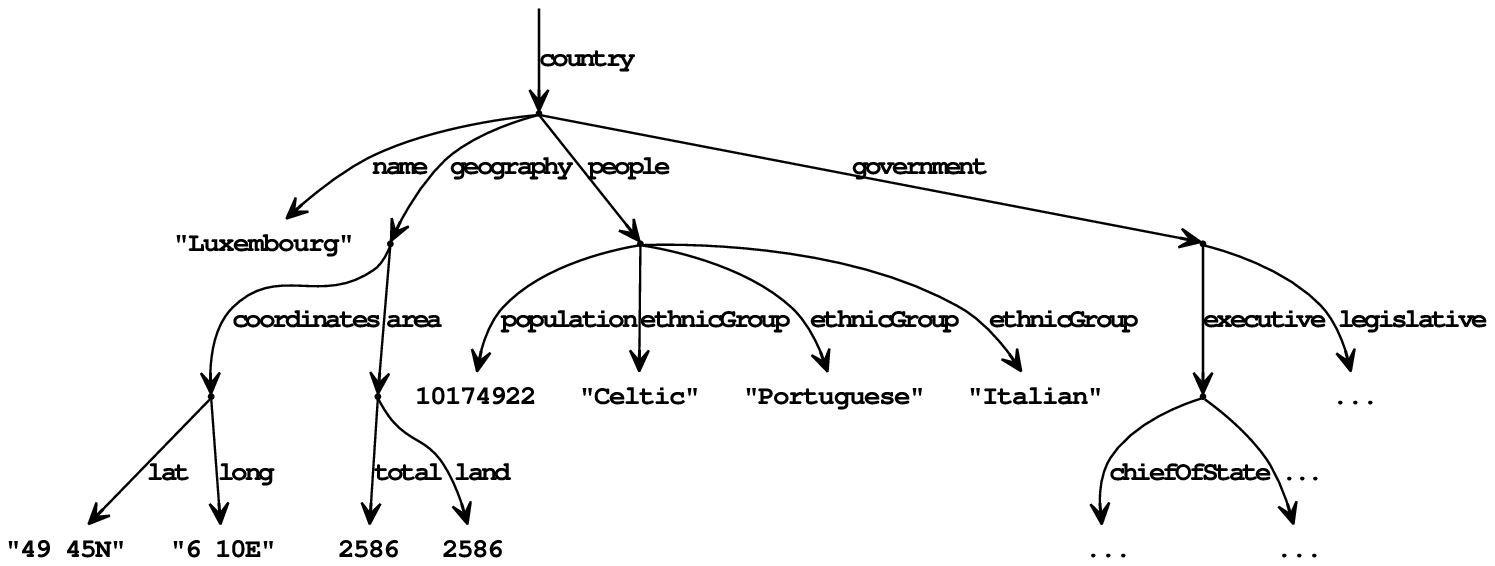}

  \caption{An example of semi-structured data}
\end{rulefigt}

\begin{rulefigw}
  \begin{center}
\includegraphics[scale=.44]{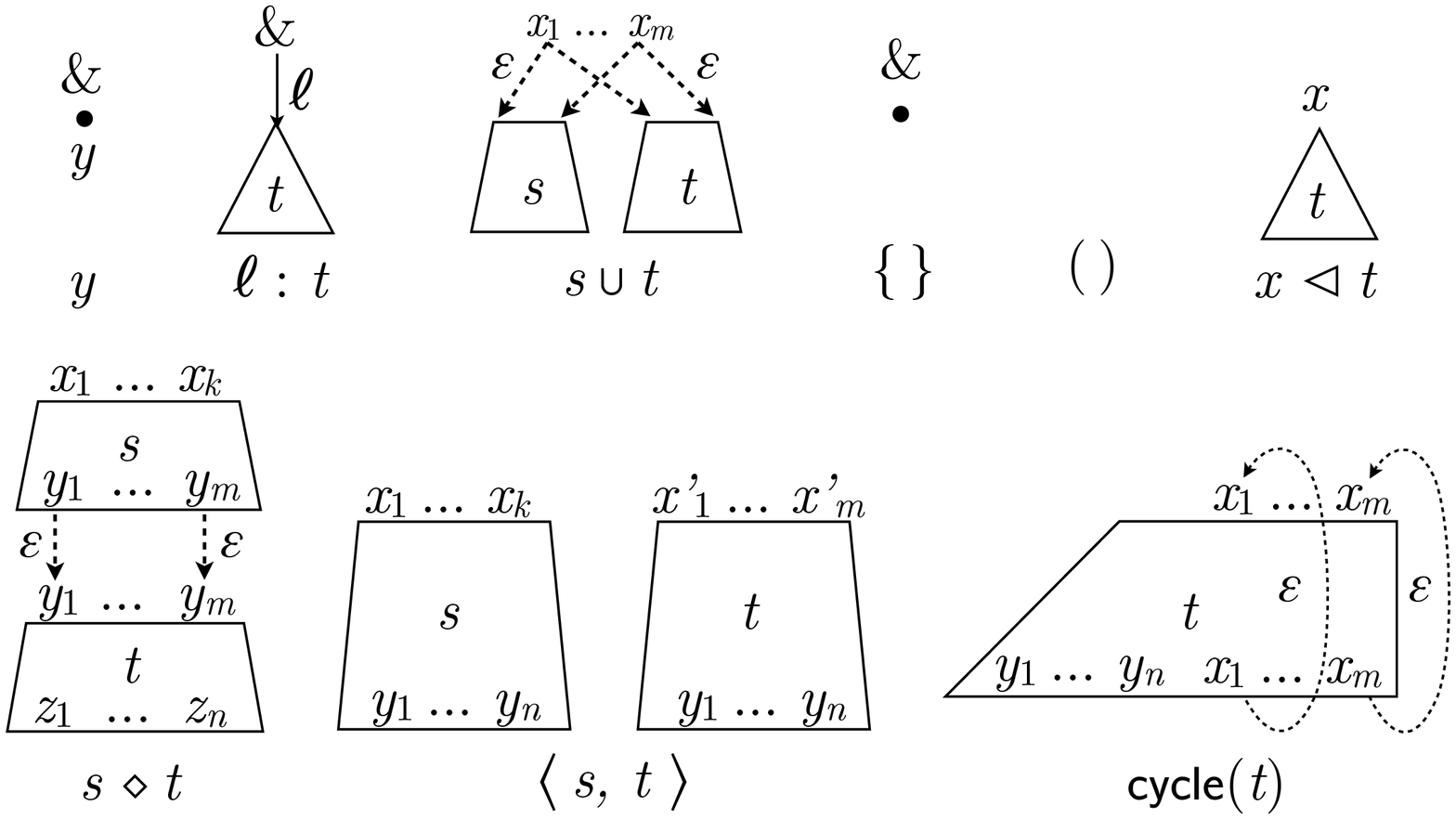}      
  \end{center}
\caption{\small Graph theoretic definitions of constructors \cite{Buneman}}
The notation is slightly changed from the original. The correspondence
between the original and this paper's is:\\
$\&y = y,\quad @ = \at,\quad \oplus=<-,->,\quad (-:=-) = \xsub{-}{-}$.
\label{fig:graph-constr}
\end{rulefigw}

The notation 
$\set{t_1,\ooo,t_n}$ is used as an abbreviation of
$\set{t_1} \;\uniSym\;\ccc \;\uniSym\; \set{t_n}$.
UnCAL's term language consists of markers $\var x$, 
labelled edges ${\,\ell\co{t}\,}$,
vertical compositions  $s \at t$,
horizontal compositions   $\pa{s\opl t}$,   
other horizontal compositions 
${s} {\;\RM{\scriptsize $\union$}\;} {t}$ merging roots,
forming cycles $\cy{t}$,
constants $\nil$,$\EMP$, and definitions $\xsub{x}{t}$.
These term constructions have underlying graph 
theoretic meaning shown in Fig. \ref{fig:graph-constr}.
Namely, these are officially defined as operations 
on the ordinary representations of graphs:
(vertices set, edges set, roots, leaves)-tuples $(V,E,I, O)$,
but we do not choose 
this representation as a foundation in this paper.
Rather, we reformulate UnCAL in
more simpler \W{algebraic} and \W{type-theoretic} manner.

\begin{rulefigt}
  \begin{center}
\includegraphics[scale=.36]{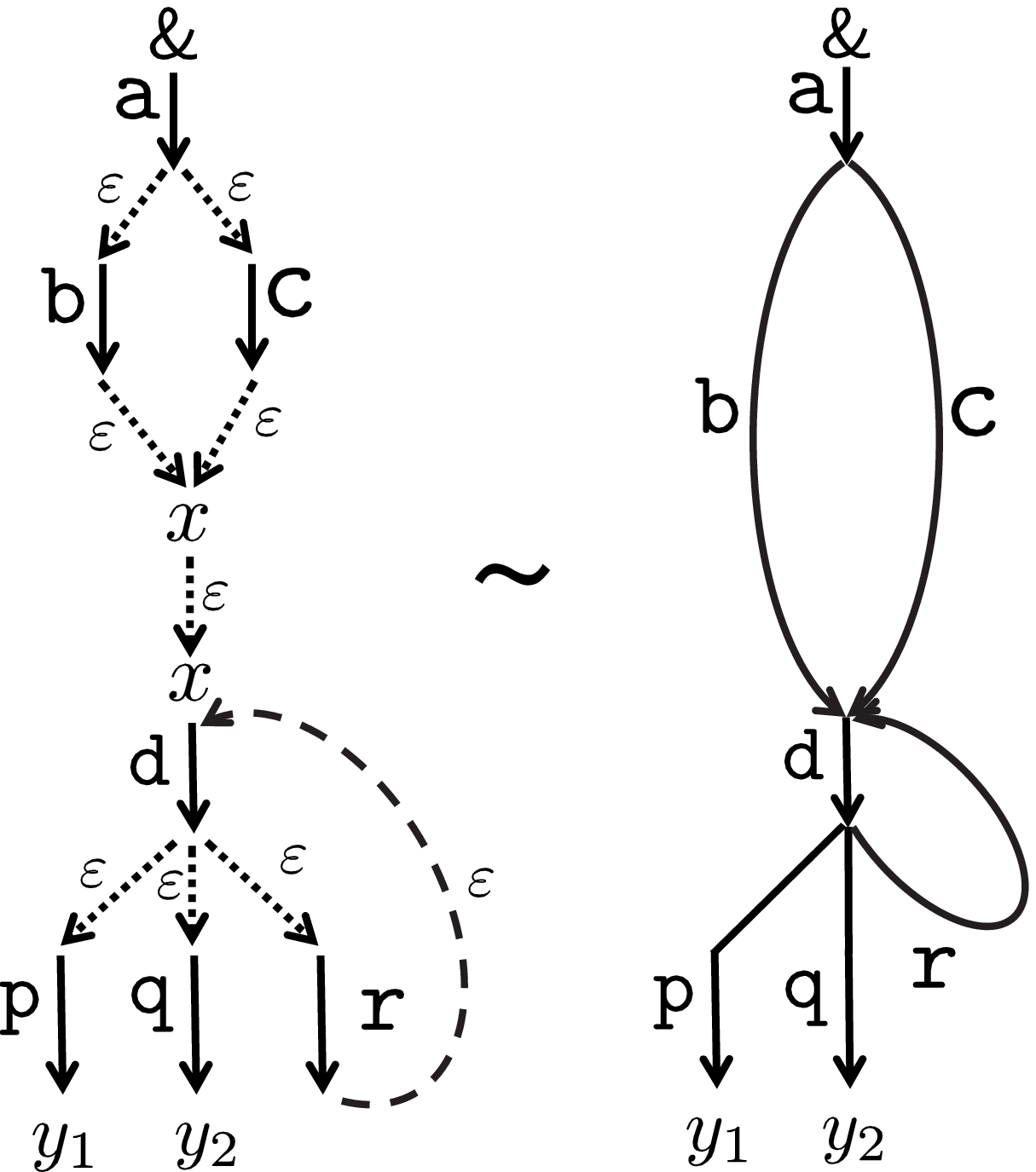}  
   \end{center}\y{-1em}
\caption{\small A graph $G$ and bisimilar one.
}
\label{fig:G}
\end{rulefigt}

UnCAL deals with graphs 
\Hi{modulo bisimulation} (i.e., not modulo graph isomorphism).
An UnCAL graph is directed and has (possibly multiple) root(s)
written $\dmark$ (or multiple $\var{x_1}\ccc\var{x_n}$) and leaves (written $\var{y_1}\ccc\var{y_m}$),
and with
the roots and leaves
drawn pictorially at the top and bottom, respectively.
The symbols $x,y_1,y_2,\dmark$ in the figures and terms
are called markers, which are
the names of nodes in a graph
and are used for references for cycles and sharings.
Here, the edges \code{b} and \code{c} share
the subgraph below the edge \code{d}.
Also, they are used as port names to connect two graphs.
A dotted line labelled \epsilon
is called an \epsilon-edge, which is a ``virtual'' edge 
connecting two nodes 
directly.
This is achieved by identifying graphs by \Hi{extended bisimulation},
which ignores \epsilon-edges suitably in UnCAL.
The UnCAL graph $G$ 
shown at the left in Fig. \ref{fig:G} 
is an example. 
This is extended bisimilar to a graph shown at the right in Fig. \ref{fig:G} 
that reduces all \epsilon-edges.
Using UnCAL's language, 
$G$ is represented as the following term $\mathbf{t}_G$ 
$$
\mathbf{t}_G \; =\; 
 \code a \cc{\Uni{\code b\co{\var x}}{\code c\co{\var x}}} \;\at\;
 {\cy{\nxsub{x}{\code  d\cc{ \UniT{\code p\co{\var{y_1}}} {\code q\co{\var{y_2}}} {\code r\co{\var{x}}}}
   \;} }}.$$
UnCAL's structural recursive function works also on cyclic data.
For example, define another function 
\begin{Verbatim}
                      sfun f2(L:T) = a:f2(T)
\end{Verbatim}

\noindent
that
replaces every edge with \code a. 
As expected,
$$\code{f2}(\,\mathbf{t}_G\,) \;\;\narone\;\; \code a \cc{\Uni{\code a\co{\var x}}{\code a\co{\var x}}} \;\at\;
   {\cy{\nxsub{x}{ \code a\cc{ \UniT{\code a\co{\var{y_1}}} {\code
   a\co{\var{y_2}}} {\code a\co{\var{x}}}}
   \;} }}$$
where all labels are changed to \code{a}.

Another characteristic role of bisimulation is 
that it identifies expansion
of cycles.
For example, 
a term $\cy{\nxsub \dmark{{\code a}\co \dmark}} $ corresponds to
the graph shown below at the leftmost.
It is bisimilar to the right ones, especially
the infinitely expanded graph shown at the rightmost, which has no cycle.
\medskip
\begin{center}
\ \qquad\includegraphics[scale=.37]{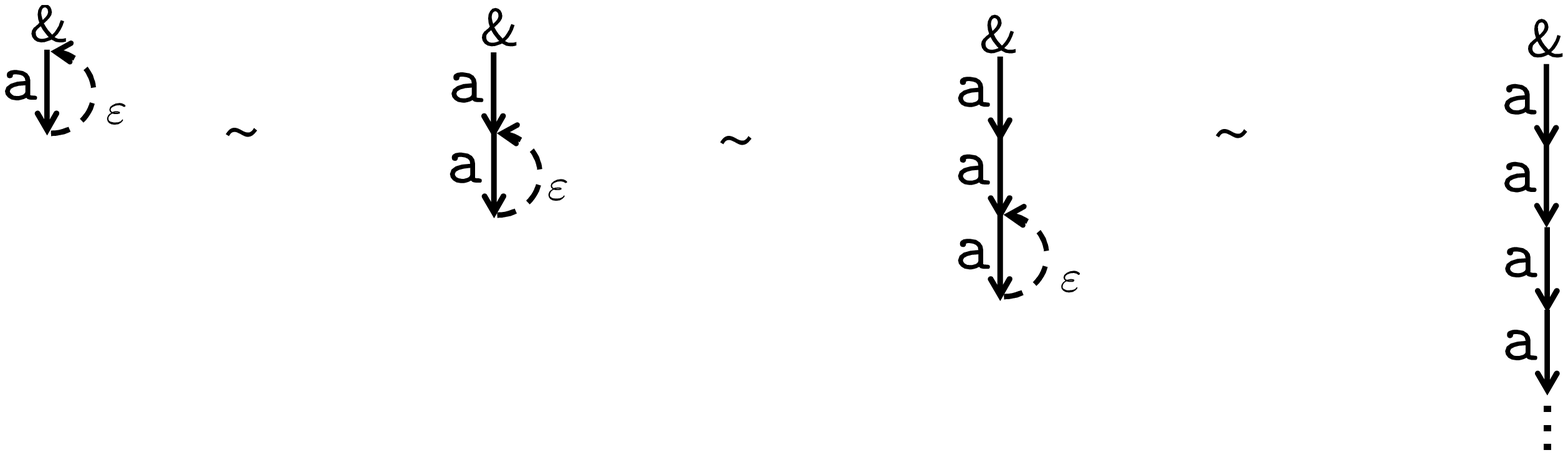}  
\end{center}
These are in term notation:
$$\cy{\nxsub \dmark{{\code a}\co \dmark}}
\;\;\bisim\;\;  \code a \co \cy{\nxsub \dmark{{\code a}\co \dmark}}
\;\bisim\;  \code a \co \code a \co \cy{\nxsub \dmark{{\code a}\co \dmark}}
\ \x{5em}
$$

\subsec{The aims of this paper}\label{sec:problems}
There have been no algebraic laws that establish the above expansion of
\cyc. Namely, these are merely bisimilar, and not a consequence of 
any algebraic law.
But obviously,
we expect that
it should be a consequence of the algebraic law of \Hi{fixed point property} of \cyc.

In the original and subsequent formulation of UnCAL 
\cite{Buneman,ICFP13,ICFP10,PPDP13},
there are complications of this kind.
The relationship between terms and graphs in UnCAL
is not a one-to-one correspondence.
There is no explicit term notation for \epsilon-edges.
Moreover, \epsilon-edges appear in several different constructions 
(vertical composition $\at$, merging roots $\union$, and \cyc).
No term notation exists for infinite graphs.
Hence the rightmost infinite graph of the above expansion of \code{a}
cannot be expressed in syntax.
But such an infinite graph is allowed as a possible graph in 
the original formulation of UnCAL.
Consequently, instead of terms,
one must use graphs and graph theoretic reasoning with
care of bisimulation to reason about UnCAL.
Therefore, a property in UnCAL could not be established 
only by using induction on terms.
That fact sometimes makes the proof of a proposition
about UnCAL quite complicated.
For instance, the proof of the fusion law~\cite[Theorem 4]{Buneman} 
is 3.5 pages long 
(see also our alternative shorter proof in \Sec \ref{sec:fusion}).

Because UnCAL graphs are identified by bisimulation,
it is necessary to
use a procedure or algorithm to check the bisimilarity
as in the cycle example above.
Listing some typical valid equations for the bisimulation 
can be a shortcut \cite{Buneman,LOPSTR},
but it was only sound and {was not complete reasoning method} for bisimulation.
In this paper, we give a complete equational axiomatisation of UnCAL graphs
as an equational logic on the term representation, which
captures the original bisimulation.

\subsection{Related work}

Other than the term representation,
various representations of graphs have been known.
One of the frequently used representations is by
a \Hi{system of equations}, also known as
an {equational term graph} \cite{ETGRS}.
It is essentially the same as the representation of graphs using
\textbf{letrec}-expressions used in \cite{Hassei}.
Semantically, the least solution of a system of equations is
regarded as a (unfolded) graph.
Syntactically, a system of (flat)
equations can be seen as adjacency lists, e.g.
an equation $x = f(y_1,\ooo,y_n)$ in a system can be regarded as an 
adjacency list of vertex $x$ pointed to 
other vertices $y_1,\ooo,y_n$ 
(cf. discussions in \citet[Sect. 8]{LMCS}).
This representation has
also been used as normal forms of process graphs in 
\cite{MilnerRegular}.
But the combination of it with
structural recursion on graphs has not usually been considered.

One notable exception is a work by \citet{NiOT96}
on
a query language for an object-oriented database.
An object in the object-oriented database 
is represented as a \Hi{system of equations}, which
may point to other objects, i.e., it forms a \Hi{graph}.
Although the proposal is done independently of UnCAL, 
they also developed the ``reduce'' operation on objects
based on a mechanism
for data-parallelism on recursive data given in \cite{NiOh99}, which is 
a structural recursion on graphs in our sense
(although bisimilar graphs are not identified in their work).
This shows that our results would not be restricted to UnCAL,
but common in transformations of graph data.

The original formulation of UnCAL focuses a single concrete model,
graphs modulo bisimulation. Having this concrete model eases
discussions on the complexity of UnCAL
transformations. \citet{Buneman} showed that UnCAL transformations are
in the class FO+TC of problems identified by the
first-order logic with transitive closure ~\cite{Immerman87}.   
The class FO+TC is the
same as $\mathrm{NL}$~\cite{Immerman87}, and thus UnCAL queries are
performed in polynomial time in nature.  They also showed that, for
tree-encoded relational data, UnCAL has the exactly same power as the
relational algebra, i.e., first-order formulas, because finite
unfolding of transitive closure suffices for such trees.

As an extension of UnCAL to handle graphs other than unordered ones,
\citet{ICFP13} proposed an UnCAL variant for ordered graphs, and then
\citet{PPDP13} extended the result to more general graphs whose
branching structure is specified by a certain sort of
monads. Accordingly, they gave new formalization of bisimulation to
such graph models.

An earlier version of this paper appeared in \cite{FICS}, which used 
iteration algebras for the completeness proof.
The present paper refines the mathematical setting
throughout and investigates it further in various directions. 
We use iteration categories 
as the foundations instead of iteration algebras,
which provide a unified framework
for graphs, bisimulation, and structural recursion.

\medskip
\subsection{Organisation}
This paper is organised as follows. We first 
give an equational theory
for UnCAL graphs
by reformulating UnCAL graph data in a type theoretic manner
in Section \ref{sec:theory}.
We then give categorical semantics of UnCAL using iteration categories
in Section \ref{sec:interp}.
We further model 
the computation mechanism ``structural recursion on graphs''
in Section \ref{sec:str-rec}.
We prove completeness of our axioms for UnCAL graphs for bisimulation
in Section \ref{sec:alg}.
We consider several instances of categorical models 
in Section \ref{sec:models}.
In particular, in Section \ref{sec:lambdaG},
we present the \lmdG-calculus as a model, and show that
it induces a sound and complete translation from
UnCAL to the \lmdG-calculus.
In Section \ref{sec:concl}, we discuss variations of UnCAL and 
related graph calculi.


\section{UnCAL and its Equational Theory}\label{sec:theory}

In this section, we establish
a new framework of equational reasoning
for UnCAL graphs without explicitly touching 
graphs.
Namely,
we \W{reformulate} UnCAL 
as an equational logic in a type theoretic manner.
We do not employ the graph-theoretic or operational concepts
(such as \epsilon-edges, bisimulation, and the graph theoretic definitions in
Fig.\ \ref{fig:graph-constr}).
Instead, we give a \W{syntactic} equational axiomatisation of  UnCAL graphs
following the tradition of categorical type theory (e.g.\ \cite{Crole}).

The idea of our formulation is as follows.
The general methodology of categorical type theory 
is to
interpret a term-in-context
$\Gamma \pr t : \tau$ in a type theory
as a morphism $\den{t}:\den{\Gamma} \to \den{\tau}$ in a category.
Drawing the morphism $\den{t}$ as a tree diagram
in which the top is the root for $\den{\tau}$,
it is a graph 
consisting directed edges that go from the bottom $\den{\Gamma}$ 
to the top $\den{\tau}$.
We aim to model UnCAL graphs in this way.

The syntax of UnCAL in this paper
is slightly modified from the original presentation
(cf.\ Fig.\ \ref{fig:graph-constr})
to reflect the categorical idea, and
to make UnCAL more readable for the reader familiar with 
categorical type theory, but this is essentially just a notational change.
This replacement is one-to-one, and one can systematically recover the original
syntax from the syntax used here.

\subsection{Syntax}\label{sec:syntax}

\subsec{Markers and contexts}\label{sec:markers}
We assume an infinite set of symbols called \Hi{markers}, denoted by typically
$x,y,z,\ooo$. One can understand markers as variables in a type theory.
The marker named $\dmark$ is called the {default marker} 
(cf. Remark \ref{sec:dmark}).
Let 
$L$ be a set of {labels}. 
A \W{label} $\ell$ is a symbol (e.g.\ $\code a,\code b,\code c,\ooo$ 
in Fig.\ \ref{fig:G}).
A \W{context}, denoted by $\seq{\v{x}_1,\v{x}_2,\ooo}$,
is a finite sequence of pairwise distinct markers .
We typically use $X,Y,Z,\ooo$ for contexts.
We use $\seq{}$ for the empty contexts, 
$X\conc Y$ or $X,Y$ for the concatenation, and $|X|$ for its length.
We may use the vector notation $\;\vec{\v x}\;$ for sequence $\v{x_1},\ooo,\v{x_n}$.
The outermost bracket $\seq{\;}$ of a context may be omitted.
We may use an alternative notation for the empty context: $0=\seq{}$. 
Note that the concatenation may need suitable renaming to 
satisfy pairwise distinctness of markers.

\subsec{Raw terms}\label{sec:rawterms}
\[
\arraycolsep = 1.4mm
\begin{array}[h]{lclllllllllllllllllll}
t &::=& {\v{y}}_{\,Y} 
  &|&   \ell\c{t} 
  &|&   s \at t 
  &|&   \pa{s\opl t  }
  &|&   \cyX{t} 
  &|&   \nil_Y
  &|&   \EMP_Y &|& \mult_Y
  &|&   \xsub{{x}}{t}
\end{array}
\]
We assume several conventions to simplify the presentation of theory.
We often omit subscripts or superscripts such as $Y$ when they are unimportant or inferable.
We identify $<t, \EMP>$ and $<\EMP, t>$ with $t$, and
$\pa{\pa{s \oppl t} \oppl u}$ with $\pa{s \oppl \pa{t \oppl u}}$;
thus we will freely omit parentheses as $\pa{t_1\opl \ooo \opl t_n}$.
When writing $\pa{t_1\opl \ooo \opl t_n}$ for any $n\in\Nat$,
we mean that it includes the cases $n=0,1$. If $n=0$, it means $()$. If $n=1$,
it means $t_1$ (not a tuple).

A constant $\mult$ is used to
express a branch in a tree,
and we call the symbol $\mult$ a \W{man}, because it is similar to the shape of a kanji 
or Chinese character meaning a man.
A definition $\xsub{{x}}{t}$ does not bind $x$.
As a graph, it is intended to name the root as $x$.
See Fig. \ref{fig:graph-constr} and the typing rule (Def) in 
(Fig. \ref{fig:typing}).

\subsection{Typed syntax}

\begin{rulefigw}
\begin{meq}
\ninfrule{Nil}{}{
\ju{ \nil_Y  }{Y}{\dmark}
}
\qquad
\ninfrule{Emp}{
}{
\ju{ \EMP_Y  }{Y}{\seq{}}
}
\qquad
\ninfrule{Man}{
}{
 \ju{ \mult_{\seq{\var{y_1},\var{y_2}}} }{{\var{y_1},\var{y_2}}}{\dmark}
}
\qquad
\\[1em]
\ninfrule{Com}{
 \ju{ s }{Y}{Z}\infspc  \ju{ t }{X}{Y}
}{
 \ju{ s \at t }{X}{Z}
}
\qquad
\ninfrule{Label}{
 \ell \in L \infspc \ju{ t }{Y}{\dmark}  
}{
 \ju{ \ell \co{t} }{Y}{\dmark}
}
\qquad
\ninfrule{Mark}{
Y = \seq{\var{y_1},\ooo,\var{y_n}}
}{
 \ju{ {\var y_i}_Y }{Y}{\dmark}
}
\\[1em]
\ninfrule{Pair}{
 \ju{ s }{Y}{X_1}\qquad \ju{ t }{Y}{X_2}
}{
 \ju{ \pa{s \opl t }}{Y}{X_1\conc X_2}
}
\qquad
\ninfrule{Cyc}{
 \ju{ t }{Y \conc X}{ X}
}{
 \ju{ \cyX{t} }{Y}{X}
}
\qquad
\ninfrule{Def}{
\quad      \ju{ t }{Y}{ \dmark}
}{
 \ju{ \xsub{x}{t} }{Y}{ {\var x}}
}
\end{meq}
\caption{Typing rules}
\label{fig:typing}
\end{rulefigw}


For contexts $X,Y$,
we inductively define 
a judgment relation $$\ju{t}{Y}{X}$$ of terms
by the typing rules
in Fig. \ref{fig:typing}.
Now $Y$ is similar to a variable context in ordinary type theories,
which we call the \W{source context}
and $X$ is the names of roots, which we call the \W{target context} or \W{type}.
We identify $t$ of type $\dmark$ with $\xsub{\dmark}{t}$.

The intuitive meaning of the term constructors (i.e. $\at,\; -\c -$, etc.)
is the corresponding operations on graphs described in Fig. \ref{fig:graph-constr}.
But we warn that in our formulation, Fig. \ref{fig:graph-constr} is not
the definitions of constructors, 
so the reader should use it just for intuition.
If the reader is familiar with category theory, 
it may be helpful to refer 
the categorical interpretation we will define in Fig. \ref{fig:interp}
(but not mandatory).
After establishing our categorical semantics,
we can give a graph theoretic model (Appendix \ref{sec:graphmodel}).

For example, the term $\mathbf{t_G}$ in \Sec \ref{sec:review} is well-typed
as
$$
\ju{\mathbf{t_G}} {\v{y_1},\v{y_2}}{\dmark},
$$
which corresponds
a graph in Fig.\ \ref{fig:G}.
When a raw term $t$ is well-typed by the typing rules, 
we call $t$ a (well-typed UnCAL) term.

\Remark\label{sec:dmark}
In UnCAL, we always need to name the root of a term by a marker.
By default,
we name the root as ``\dmark'' as the above $\mathbf{t_G}$, hence
we call \dmark the \Hi{default marker}.
This convention is reflected to the typing rules.
\oRemark

\subsec{On (Pair)}\label{sec:pair}
We have assumed in \Sec \ref{sec:markers} that if contexts $X_1$ and $X_2$
share some markers, the concatenation $X_1+X_2$ need renaming
of markers to satisfy pairwise disjointness.
We do not explicitly define how to rename markers to avoid clutter.
This also gives a simplified notation for pair terms.

For example, suppose that $\seq{\dmark}+\seq{\dmark}$ is renamed to $y_1,y_2$. 
Consider (Pair) construct
\begin{meq}
\ninfrule{Pair}{
 \ju{t_1}{ Y }{\dmark}\qquad \ju{t_2}{ Y }{\dmark}
}{
 \ju{ \pa{t_1 \opl t_2 }}{Y}{\seq{\dmark}+\seq{\dmark}}
}
\end{meq}
Hence
$\ju{\pa{t_1 \opl t_2 }}{Y}{y_1,y_2}$. 
We can derive essentially the same term
\begin{meq}
\ninfrule{Pair}{
 \ju{\xsub{y_1}{t_1}}{Y}{ y_1 }\qquad 
 \ju{\xsub{y_2}{t_2}}{Y}{ y_2 }
}{
 \ju{ \pa{\xsub{y_1}{t_1} \opl \xsub{y_2}{t_2} }}{Y}{y_1,y_2}
}
\end{meq}
without renaming markers in the target context.
We identify these two term judgements 
and we often choose the first representation
$\ju{\pa{t_1 \opl t_2 }}{Y}{y_1,y_2}$,
because it is simpler and the part of 
definitions $y_1 \triangleleft - $ and $y_2 \triangleleft -$ can be 
recovered from the target context $y_1,y_2$.
Hence hereafter, writing simply $\jud X {<t_1,\ooo,t_n>} {y_1,\ooo,y_n}$,
we mean $\jud X {<\xsub{y_1}{t_1},\ooo,\xsub{y_n}{t_n}>} {y_1,\ooo,y_n}$.
As long as well-typed terms, there is no confusion.

Even if the markers in the target context are omitted,
still the information for this renaming can be recovered from contexts
around the term. For example,
when we consider a well-typed term \(s \at \pa{t_1 \opl t_2 }\) where the source context of 
 \(s\) is \({y_1,y_2}\),
then we can rename the term correctly as \(s \at \pa{\xsub{y_1}{t_1} \opl \xsub{y_2}{t_2} }\).

\Remark
\label{rmk:rootMarker}
We could omit the marker in a target context
and also the syntax \(\xsub{x}{-}\)
and then, we could replace, e.g., (Cyc) with:
\[
\infrule{}{
 \ju{ t }{Y \conc X}{n} \qquad |X|=n
}{
 \ju{ \cyX{t} }{Y}{n}
}
\]
But in this paper we keep the UnCAL-style syntax to accommodate with the existing
work on UnCAL.
\oRemark

\subsec{Abbreviations}
We define several abbreviations for terms. We define
\[
\pi_{X,Y}^X  \deq <{x_1},\ooo,{x_n}>
\]
for $X=\seq{x_1,\ooo,x_n}$.
When $n=1$, $\pi_{x_1,Y}^{x_1} \deq {x_1}$.
When $n=0$, $\pi_{0,Y}^{0} \deq \EMP$.
Similarly for the case $\pi_{X,Y}^Y$, but
because of the convention of $<-,->$ describe in \Sec \ref{sec:pair},
it is better to explicitly mention the source and target contexts
when defining an abbreviation. So we use the following style
\begin{meq}
\infrule{proj}{X=\seq{x_1,\ooo,x_n}}{
\ju{ \pi_{X,Y}^X  \deq <{x_1},\ooo,{x_n}> }{X+Y}{X}
}
\qquad
\infrule{proj}{Y=\seq{y_1,\ooo,y_n}}{
\ju{ \pi_{X,Y}^Y  \deq <{y_1},\ooo,{y_n}> }{X+Y}{Y}
}
\end{meq}
These mean to define the abbreviations and 
to indicate also the source and target contexts of the abbreviations.
\begin{meq}
\infrule{times}{ \ju{t_1}{Y_1}{X_1}\qquad \ju{t_2}{Y_2}{X_2} }
{\ju{ t_1\X t_2 
  \deq <t_1\at\pi_{Y_1,Y_2}^{Y_1}, t_2\at\pi_{Y_1,Y_2}^{Y_2}> }
{Y_1+Y_2}{X_1+X_2}
}
\\[.5em]
\infrule{times}{ \ju{t_1}{Y_1}{\dmark}\qquad \ju{t_2}{Y_2}{\dmark} }
{\ju{ \Uni{t_1}{t_2}
  \deq 
\mult \at \pa{{t_1}\opl{t_2}} }
{Y_1+Y_2}{\dmark}
}
\\[.5em]
\infrule{id}{ }
{\ju{ \Id_{\seq{\,}} \deq \EMP_{\seq{\,}} }0 0
}
\infspc
\infrule{id}{ }
{\ju{ \Id_{\seq x} \deq \xsub x x}x x
}
\infspc
\infrule{id}{ X = {\seq{x_1,\ooo,x_n}}}
{\ju{ \Id_X\deq <{x_1},\ooo,{x_n}>}
X X}
\\[.5em]
\infrule{id}{ \ju{\Id_X}X X }
{\ju{\Delta_X \deq  <\Id_X,\Id_X>}
{X}{X+ X}}
\infspc
\infrule{c}{ \ju{\pi_{x,y}^x}{x,y}{x} \infspc \ju{\pi_{x,y}^y}{x,y}{y}  }
{\ju{ \fn{c}_{x,y} \deq <\pi_{x,y}^y,\pi_{x,y}^x >}
{x, y}{y, x}}
\end{meq}
Inheriting the convention of $\pa{-,-}$, 
we also identify 
$(s \X t) \X u$ with $s \X (t \X u)$,
thus we omit parentheses as $t_1\X \cdots \X t_n$.
We will see 
that these named terms are actually intended arrows in a cartesian category (\Sec \ref{sec:interp}).

\DefTitled[def:subst]{Substitution}
Suppose $$\ju t {y_1,\ooo,y_n} X,\qquad \ju {s_i} Z {\dmark} \qquad (1\le i \le n).$$
Then a substitution $\;\ju {t \sub{\vec{{y}}}{\vec s}} {Z} X\;$
is inductively defined as follows.
\[
\arraycolsep = 1mm
\begin{array}[h]{lllll} 
\begin{array}[h]{rclllllllllllllllllll}
\var{y_i}        \qesub &\deq& s_i \\
\nil_Y      \qesub &\deq& \nil_{Z}\\
\EMP_Y      \qesub &\deq& \EMP_{Z}\\
(\ell\co t)    \qesub &\deq& \ell\co{(\, t\qesub\,) }
\end{array}
\;
\begin{array}[h]{rclllllllllllllllllll}
(t_1\at t_2)  \qesub &\deq& t_1\at (t_2 \qesub)\\
\pa{t_1\opl t_2} \qesub &\deq& \pa{t_1\qesub \opl t_2\qesub}\\
\cyX{t}      \qesub &\deq& \cyX{t\; [\vec y\maps\vec s,\vec x\maps\vec x]} \\
\xsub{x}{t} \qesub &\deq& \xsub{x}{ t \qesub }
\end{array}
\\
\mult_{\seq{\var{y_1},\var{y_2}}} \subst{\var{y_1}\maps s_1, \var{y_2}\maps \var{s_2}} \deq
 \mult_{\seq{\var{y_1},\var{y_2}}} \at \pa{s_1\opl s_2}
\end{array}
\]
\oDef
Note that ${t \sub{\vec{{y}}}{\vec s}}$ denotes 
a meta-level substitution operation,
not an explicit substitution.
This simultaneous substitution is a generalisation of the single variable case 
given in \cite{LOPSTR}, where 
the contexts and types for terms were not explicit.
Explicating the context and type information is important 
in our development to
connect it with the categorical interpretation 
(cf.\ Lemma \ref{th:subst-compo}).

\begin{rulefigw}
\begin{meq}
\arraycolsep = 0mm
\ninfrule{Ax}
{(\ju{s=t}Y X) \in E}
{\ju{s=t}Y X}
\infspc
\ninfrule{Sub}
{\ju {t=t'} {y_1,\ooo,y_n} X\infspc \ju {s_i=s'_i} Z {\dmark} \infspc (1\le i \le n)}
{\ju {t \sub{\vec{{y}}}{\vec s} = t' \sub{\vec{{y}}}{\vec s'}} {Z} X}
\\[.7em]
\ninfrule{Com}
{\ju{ s=s' }{Y}{Z} \infspc \ju{t = t'}Y X  \quad 
}
{\ju{s\at t = s'\at t'}{X}{Z}}
\qquad
\ninfrule{Pair}
{ \ju{ s=s' }{Y}{X_1}\qquad \ju{ t=t' }{Y}{X_2}
}
{\ju{<s,t> = <s',t'>}{Y} {X_1\conc X_2}}
\\[.7em]
\ninfrule{Cyc}
{ \ju{ t=t' }{Y \conc X}{ X}
}
{\ju{\cyX{t} = \cyc^{X'}(t)} {Y} {X}}
\qquad
\ninfrule{Def}
{\ju{t = t'}Y \dmark  \quad }
{\ju{\xsub x t = \xsub x{t'}}Y{x}}
\\[.7em]
\ninfrule{Ref}
{}
{\ju{t = t}Y X}
\infspc
\ninfrule{Sym}
{\ju{s = t}Y X}
{\ju{t = s}Y X}
\infspc
\ninfrule{Tr}
{\ju{s = t}Y X \infspc \ju{t = u}Y X}
{\ju{s = u}Y X}
\end{meq}
\caption{Equational Logic \ELu}
\label{fig:ELu}
\end{rulefigw}


\subsection{Equational axiomatisation}\label{sec:el-ax}

For terms $\ju{ s }{Y}{X}$ and $\ju{ t }{Y}{X}$, an \W{(UnCAL) equation} is of the form
$$
Y \pr s = t \;\;: X.
$$
Hereafter, 
we often omit the source $X$ and target $Y$ contexts, and 
simply write $s = t$ for an equation, 
but even such an abbreviated form,
we assume that
it has implicitly suitable source and target contexts
and is of the above judgemental form.

A set of \W{UnCAL axioms} $E$ is given by the axioms \AxG in 
Fig.\ \ref{fig:axioms} 
plus a set of arbitrary equations called \Hi{additional axioms}.
The axioms \AxG characterises UnCAL graphs.
These axioms are chosen to soundly and completely represent the original 
bisimulation of graphs by the equality of this logic.
Checking soundness is straightforward: 
for every axiom $s = t$, we see that $s$ and $t$ are bisimilar.
But completeness is not clear only from the axioms.
We will show it in \Sec \ref{sec:alg}.

The equational logic \ELu for UnCAL is a logic
to deduce formally proved 
equations, called \W{(UnCAL) theorems}.
The equational logic is almost the same as ordinary one for algebraic terms.
The inference system of equational logic for UnCAL terms
is given in Fig.\ \ref{fig:ELu}.
The structural rules (weakening, contraction, and permutation of 
source context $Y$) are derivable from (Sub).
The set of all theorems deduced from axioms $E$ is
called a \W{(UnCAL) theory}.
When $E$ has no additional axioms, we call the theory \W{pure}.

\begin{rulefigh}
\[
\arraycolsep = .1mm
\renewcommand{\arraystretch}{1.5}
\normalsize \x{-2em}
\begin{array}[h]{lrlllllllllllll}
\axtit{Composition}\\[-.5em]
\urule{(sub)} &
\aju{t \at \pa{s_1\opl\ccc\opl s_n} = t\qesub}{Z}{X} &
\FFOR 
\renewcommand{\arraystretch}{.8}
\begin{array}[h]{lll}
\ju t Y X \infspc Y=\seq{y_1,\ooo,y_n} \\  
\ju{s_1}Z{\dmark} \;\ccc\; \ju{s_n}Z{\dmark}
\end{array}
\\
\urule{(SP)} &
\aju{\pa{\pi_{X,Y}^X \at t,\; \pi_{X,Y}^Y \at t} = t}Z {X+Y}
&
\FOR \ju t Z {X+Y}
\\
\urule{(\eta\!\mult)} &
\aju{\mult\at<x,y>=\mult_\seq{x,y}}{x,y}\dmark
\\
\axtit{Parameterised fixed point}
\\[-.5em]
\urule{(fix)} & 
\aju{\cy t = t \at \cpair{\Id_Y}{\cy t}}Y X
&
\FOR \ju t {Y\!+\!X}X
\\
\urule{(nat)} & 
\aju{\cy t \at s  =  \cy{t \at (s\X \Id_X)}}Z X
&
\FOR
\ju t {Y\!+\!X} X \qquad \ju s Z     Y
\\
\urule{(dinat)} & 
\aju{\cy{s \at t} =  s \at \cy{ t \at (\Id_Y \X s) }}Y X
&\FOR \ju s Z X \qquad  \ju t {Y+X} Z
\\[.5em]
\urule{(Beki\u{c})} & \aju{\cyc^{X\!+\!Y}(\ccpair{t}{s})
= 
\renewcommand{\arraystretch}{.8}
\begin{array}[h]{lll}
  \cpair{\pi_{Z,X}^X}{\cyY{s}} \,\at\, \\
  \cpair{\Id_Z} {\cyX{t \at \cpair{\Id_{Z\!+\! X}}{\cyY{s}}}}
\end{array}
}
Z
{X\!+\!Y}
&\FOR
   \ju t {Z\!+\!X\!+\!Y} X \qquad
   \ju s {Z\!+\!X\!+\!Y} Y
\\[1em]
\urule{(CI)} & 
\aju{\cyc( <
  \renewcommand{\arraystretch}{.8}
  \begin{array}[h]{clll}
    t \at (\Id_X \!\X\! \rho_1), \\
    \ccc, \\
    t \at (\Id_X \!\X\! \rho_m) 
  \end{array}
  >        
)
=
\Delta_m \at \cy{t \at (\Id_X \X \Delta_m)}}
X Y
&\FOR \ju t {X+Y} \dmark
\\
\axtit{Deleting trivial cycle}
\\[-.5em]
\urule{(c1)} & \aju{\cyc^\dmark(\dmark) = \nil}{}{\dmark} \\
\orule{(c2)} & \aju{\cyc^\dmark(\mult_{\seq{\var{y},\dmark}}) = \var{y}}{y}\dmark
\\
\axtit{Degenerated commutative bialgebra}
\\
\urule{(unitL$\mult$)} & \aju{\mult_{\seq{\dmark,x}} \at (\nil_0\X\Id_x) = x}x\dmark
\\
\urule{(unitR$\mult$)} & \aju{\mult_{\seq{x,\dmark}} \at (\Id_x\X\nil_0) = x}x\dmark
 \\
\urule{(assoc$\mult$)} & \aju{\mult \at (\Id\X \mult)=\mult \at (\mult\X \Id)}
 {x,y,z}\dmark \\
\orule{(com$\mult$)}  & \aju{\mult \at \fn{c}_{x,y} =\mult_\seq{x,y}}{x,y}\dmark   \\
\orule{(degen)}        & \aju{\mult \at<x,x> = x}
{x}\dmark
\end{array}
\]
\begin{minipage}[t]{\textwidth}\small
In (sub), $n$ is possiblely $0$ or $1$. 
If $n=0$, the left-hand side is $t\at \EMP$.
If $n=1$, the left-hand side is $t\at s_1$.
In (CI), for $Y=\seq{\var{y_1},\ooo,\var{y_m}},\;
\ju{\Delta_m \deq \pa{\Id_\dmark\opl\ccc\opl \Id_\dmark}}{\dmark}{Y},$
$\ju{\rho_i \deq \pa{y_{i_1}\opl\ooo\opl y_{i_m}}}{Y}{Y}$,
where $i_1,\ooo,i_m \in \set{1,\ooo,m}$, meaning that
some of $i_1,\ooo,i_m$ may be equal.
\end{minipage}
\caption{Axioms \AxG} 
\label{fig:axioms}
\end{rulefigh}


\begin{rulefigh}
\includegraphics[scale=.51]{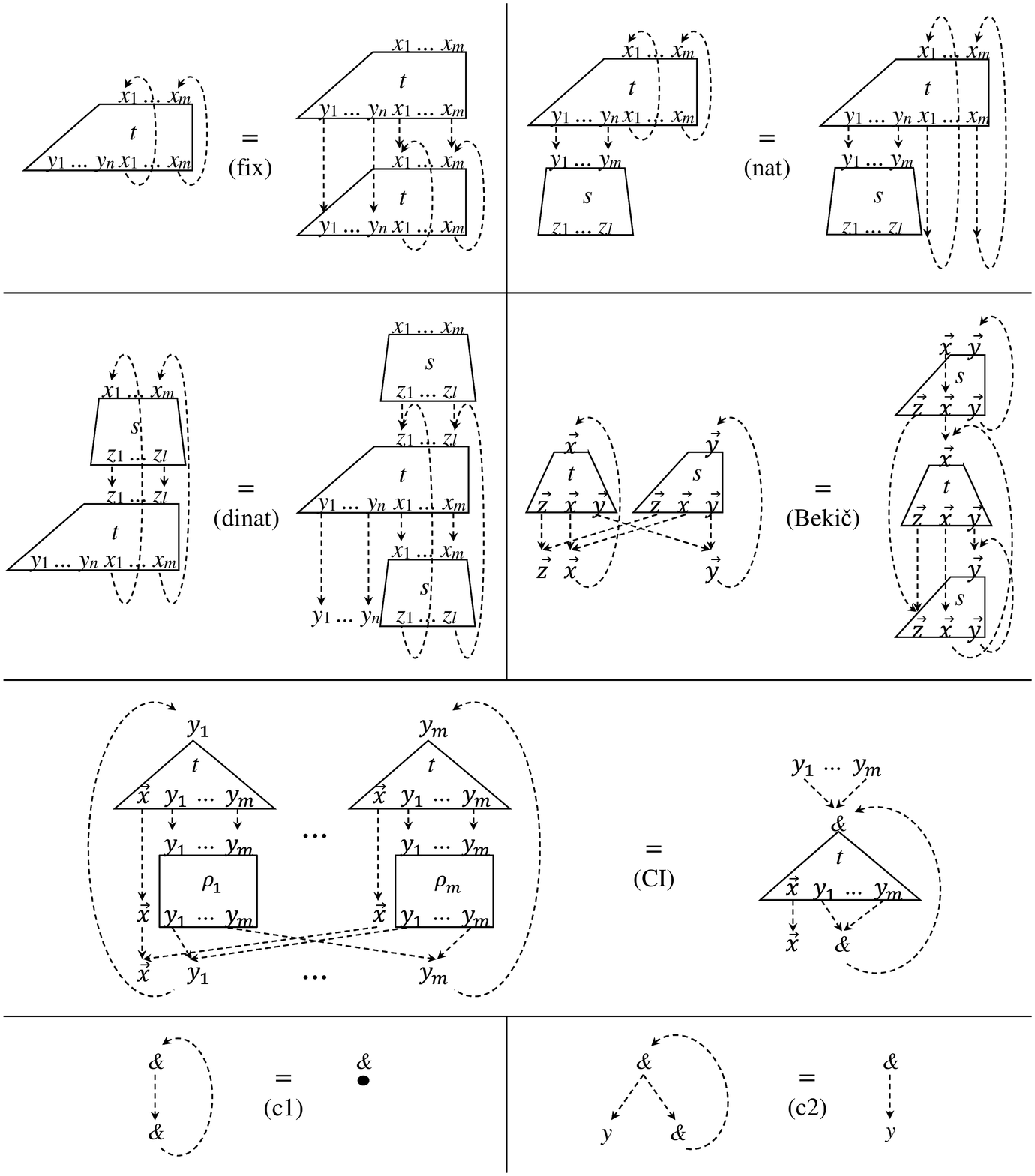}
\caption{Graphical representation of the axioms \AxG}
\label{fig:graphAx}
\end{rulefigh}
\begin{rulefigh}
\includegraphics[scale=.51]{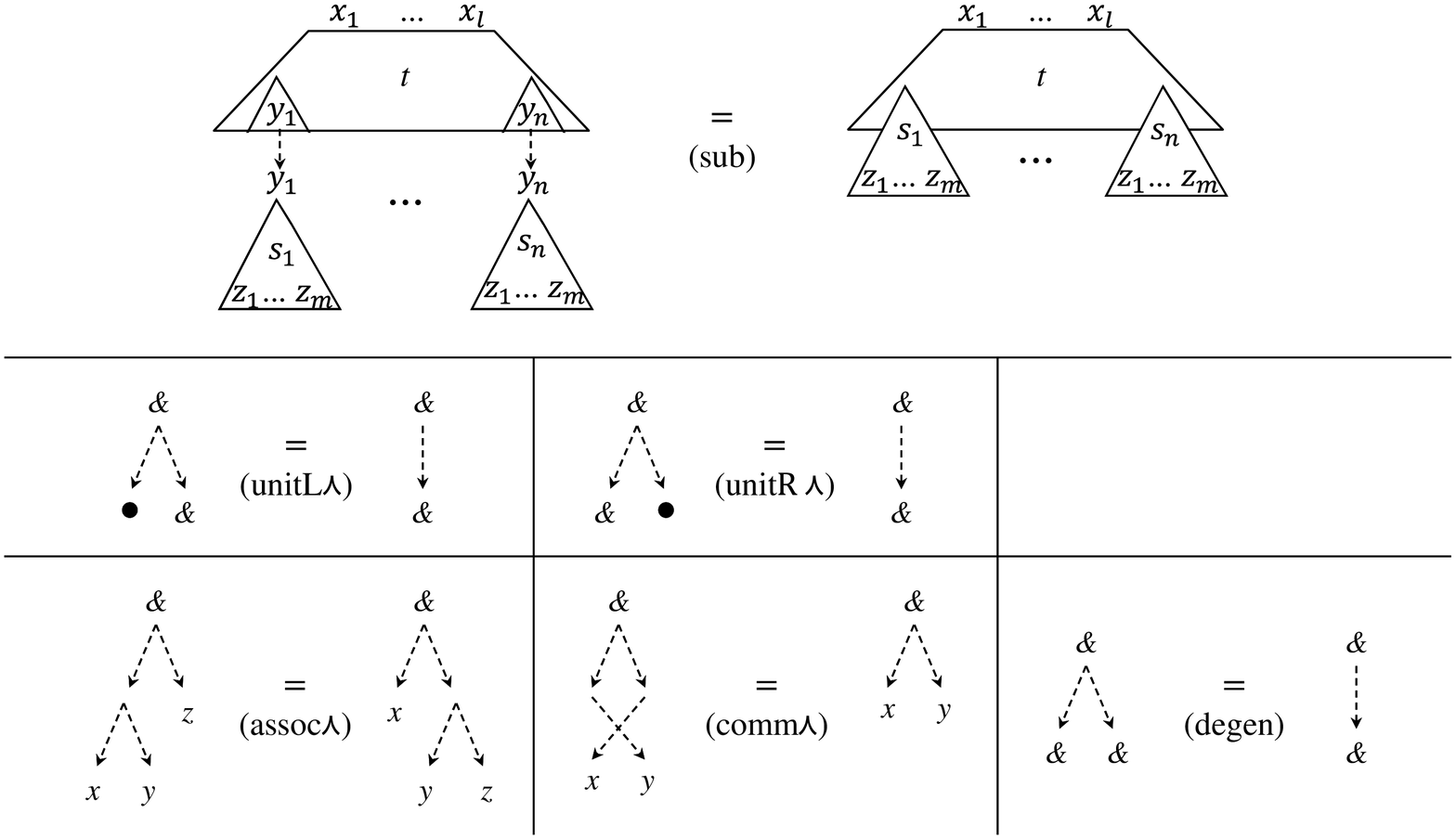}
\caption{Graphical representation of the axioms \AxG (continued)}
\label{fig:graphAxCont}
\smallskip
\hrule height 0.5pt
\bigskip
\[
\renewcommand{\arraystretch}{1.2}
\begin{array}[h]{lcllllllll}
\text{(tmnl)}&%
t &\;=\;& \EMP_Y\quad\text{for all }\ju{t}{Y}{\seq{}} \\
\text{(fst)}& \pi_{X,Y}^X \at \pa{s \opl t} &\;=&\; s \\
\text{(snd)}& \pi_{X,Y}^Y \at \pa{s \opl t} &\;=&\; t \\
\text{(dpair)} & \cpair{t_1}{t_2}\at s &\;=&\; \ccpair{t_1\at s}{t_2\at s}\\
\text{(fsi)} & \ccpair{\pi_{X,Y}^X}{\pi_{X,Y}^Y} &\;=&\; 
\Id_{X,Y}\\
\text{(bmul)}  &\EMP_{\dmark}\X\EMP_{\dmark} &=&\EMP_{\dmark}\at \mult \\
\text{(bcomul)} &\dup \at \nil_0  &\;=&\; \nil_0\X\nil_0\\
\text{(unR$\at$)}& t\at \Id &=& t \\
\text{(unL$\at$)}& \Id \at t  &\;=&\; t \\
\text{(assoc$\at$)}&(s \at t) \at u &\;=&\; s \at (t \at u) \\
\text{(bunit)} &\EMP_{\dmark} \at \nil_0 &=&\Id_0\\
\text{(compa)} &\dup\at\mult &=& (\mult\X\mult) \at (\Id\X\fn{c}\X\Id) \at
(\dup\X\dup) \\
\text{(comm$\uniSym$)}& \Uni{s}{t} &=&  \Uni{t}{s} &&\\
\text{(unit$\uniSym$)}& \Uni{}{t} &=& t \;=\; \Uni{t}{} \\
\text{(assoc$\uniSym$)}& \Uni{\Uni{s}{t}}{u} &=& \Uni{s}{\Uni{t}{u}} \\
\text{(degen')}& \Uni{t}{t} &=&  t &&\\
\end{array}
\]
\caption{Derived theory} 
\label{fig:der-th}
\end{rulefigh}

\subsec{Meaning of axioms}

The Fig.~\ref{fig:graphAx} and Fig.~\ref{fig:graphAxCont} show
how each axiom in \AxG can understood graphically and intuitively by
regarding UnCAL terms as graphs.
Not only giving us intuition, this graphical interpretation
can be stated formally.
In Appendix \ref{sec:graphmodel},
we will show that these graphs are obtained as the interpretions
of terms using a categorical model of \AxG 
in a particalar category for UnCAL,
i.e., the category $\gc$ of UnCAL graphs.

We explain the meaning of the axioms in \AxG in further detail.
The axiom (sub) 
is similar to the \beta-reduction in the \lmd-calculus.
With the axiom (SP), it induces the axioms for cartesian product \cite{Lambek-Scott}
(cf. the \textbf{derived theory} in \Sec \ref{sec:derived}).
It induces also a characteristic property of
the UnCAL graphs, namely,
shared graphs are copyable.
For example, the following two graphs are bisimilar (thus identified)
in the original formulation.
\begin{center}
\includegraphics[scale=.34]{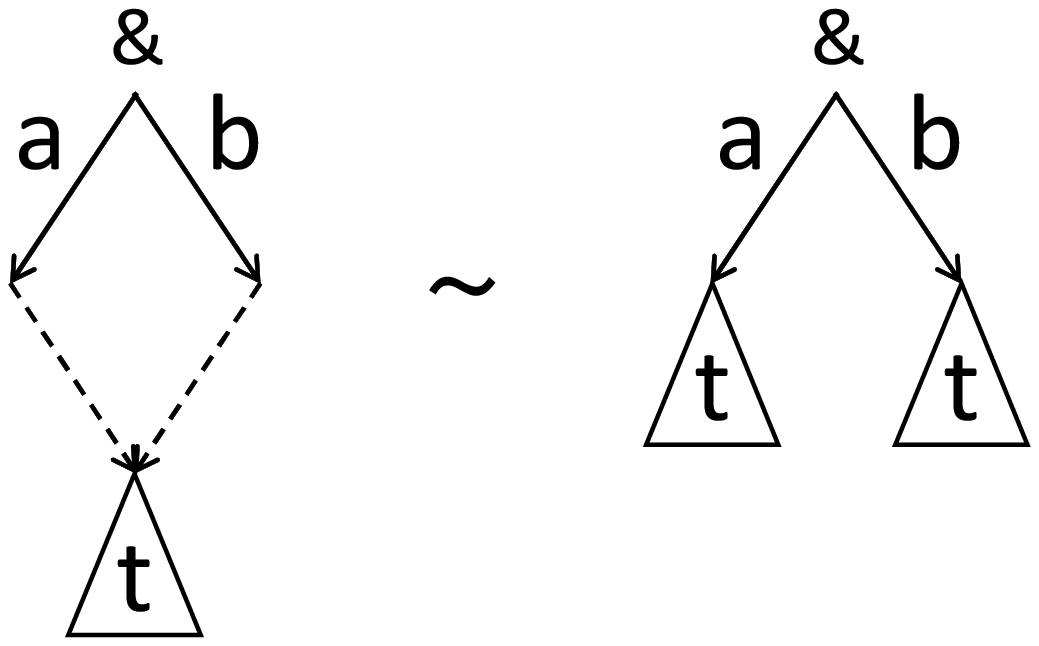}    
\end{center}
In our equational theory,
it is a theorem by (sub):
$$\mult\at (\pa{\code a\co \dmark \opl\code b\co \dmark}\at t)
\;= \;\mult\at\pa{\code a\co{t}\opl\code b\co{t}}.$$
Thus UnCAL graphs should not be interpreted in 
a mere monoidal category,
unlike other monoidal categorical formulations of graphs or DAGs,
such as \cite{Gibbons,Hassei, GS-monoidal,FiorePROP},
where shared graphs are not copyable.
The cartesian structure also provides a canonical commutative comonoid
with comultiplication \Delta.

Two terms are paired with a common root by
$\Uni{s}{t}=\mult\at \pa{s \opl t}$. 
The commutative monoid axioms states that this pairing $\Uni{-}{-}$ can be 
parenthesis-free in the nested case.
The degenerate bialgebra axioms state the 
compatibility between the commutative monoid and comonoid structures.
The degenerated bialgebra is suitable to model directed acyclic graphs
(cf.\ \cite[\Sec 4.5]{FiorePROP}), where 
it is stated for a strict monoidal category (known as a PROP) \cite{MacLanePROP}.
The monoid multiplication $\mult$ expresses a branch in a tree,
while the comultiplication $\Delta$ expresses a sharing.
Commutativity expresses that there is no order between the branches of a node,
cf.\ (commu$\uniSym$) in the \textbf{derived theory}.

The axiom (c1) and (c2) delete trivial cycles 
as graphs (see Fig. \ref{fig:graphAx} in Appendix).
Similar axioms can be found in other
axiomatisations of bisimulation, e.g.\ \cite{MilnerRegular,BET,Esik00}.

Parameterised fixed-point axioms axiomatise 
\cyc as a fixed point operator.
They (minus (CI)) are equivalent to the axioms for
Conway operators of 
\citet{BE}.
Beki\u{c} law is well-known in denotational semantics 
(cf. \cite[\Sec 10.1]{Winskel}),
which says that the fixed point of a pair 
can be obtained by computing the fixed point of each of its components
independently and composing them suitably
(see also Fig. \ref{fig:graphAx} in Appendix).
Conway operators have also arisen in work independently 
of Hyland and 
\citet{Hassei}, who established a connection
with the notion of traced cartesian categories \cite{JSV}.

\subsec{On commutative identities}\label{sec:CI}
There are equalities that holds in the cpo semantics but Conway operators do not satisfy;
e.g.\ $\cy{t} = \cy{t \at t}$ does not hold by the Conway axioms.
The axiom (CI) fills this gap.
It corresponds to
the commutative identities of 
\citet{BE},
which ensures that 
all equalities that hold in the cpo semantics
do hold.
See also \cite[Section 2]{alex-plot}  for a useful overview around 
this.
This form is taken from it and adopted to the UnCAL setting.

\subsec{Derived theory} \label{sec:derived}
The law (sub) is useful to simplify terms. 
The law has not been formulated in UnCAL literature, 
because no simultaneous
substitution has been formulated.
It will also be used to
compute normal forms of UnCAL terms to prove completeness for bisimulation
in \Sec \ref{sec:alg}.
The Fig. \ref{fig:der-th} shows some UnCAL theorems, which are 
formally derivable from the axioms,
or are proved using induction on typing derivations.
We write $\Uni{}{t}$ for $\Uni{\vnil}{t}$.

The meaning of these theorems are as follows.
\begin{enumerate}[itemsep=.5em]
\item The first three lines with (SP) in \AxG indicate 
that UnCAL has the cartesian product (cf.\ \cite{Lambek-Scott,domains-lambda}).
(fst), (snd), (fsi) and (dpair) are proved by expanding the abbreviations 
and (sub),(SP) in \ELu.
(tmnl) is proved by induction on typing derivations and \ELu.

\item (bmul), (bcomul), (bunit) and (compa) are a part of the axioms of commutative bialgebra. In our setting these are theorems
because it has the cartesian product,
see Remark \ref{th:bialgebra}.
\item (unR$\at$), (unL$\at$) and (assoc$\at$) state that $\at$ is the composition
of morphisms. These are proved by (sub).
\item \label{item:U}
The rest of theorems (comm$\uniSym$)-(degen') states that
$\uniSym$ has the unit $\nil$, and is associative, commutative, and idempotent.
These are derivable from the axioms of
\textbf{Degenerated commutative bialgebra}
through the abbreviation
$$\Uni{s}{t} = \mult \at \pa{s \opl t}.$$

These theorems for $\uniSym$ are very important in UnCAL.
They validate that the notation $\uni{s}{t}$ can be understood
as the set union operation. 
As a graph, the constructor $\uniSym$ represents a commutative and idempotent branch.

Since our modelling is categorical,
we did not directly choose these (comm$\uniSym$)-(degen') as axioms.
Rather, we chose the categorical monoid laws \cite{MacLane}
in a cartesian category as the axioms
(\textbf{Degenerated commutative monoid}) with
the unit $\nil$ and multiplication $\mult$.
See the proof of Prop.\ \ref{th:cl-cat} for details of the categorical
modelling.
\end{enumerate}

The axioms \AxG include a few redundant axioms, but they are
useful for understanding.
Actually, the axioms (c1) and (unitR$\mult$) are derivable.
The axiom (unitR$\mult$) is obtained from (unitL$\mult$) and (comm$\mult$).
For (c1), the proof is
\[
\cy{\dmark} =^{\text{(unitL}\mult)} \cy{\mult\at (\nil_0\X\Id)} =^{\text{(nat})}
 \cyc^\dmark(\mult_{\seq{y,\dmark}})\at\nil_0 =^{\text{(c2)}} y\at\nil_0
= \nil_0.
\]

\section{Categorical Semantics} 
\label{sec:interp}

In this section, 
we give categorical semantics of UnCAL graphs, and show its categorical 
completeness.

We interpret edges of an UnCAL graph as \Hi{morphisms} (of the opposite directions), the vertical composition $\at$
as the \Hi{composition  of morphisms}, and \cyc as 
a \Hi{fixed point operator} in a suitable category.
Thus the target categorical structure should have a notion of 
fixed point,
which has been studied in iteration theories of 
\citet{BE}.
In particular,
iteration categories \cite{IterCat} 
are suitable, which are traced cartesian categories \cite{JSV}
additionally satisfying the commutative identities axiom \cite{BE}.

We write $\terminal$ for the terminal object, $\X$ for the cartesian product,
$\pi_1,\pi_2$ for the first and second projections,
$<-,->$ for pairing, and $\dup=<\id,\id>$ for diagonal in a cartesian category.

\DefTitledN[def:iter-cat]{\cite{IterCat,BE}}
A \W{Conway operator} in a cartesian category $\CC$ 
is a family of functions
$
\dagg{-}:\CC(A\X X,X) \to \CC(A,X)
$
satisfying:
\begin{meqa}
\dagg{f \o (g \X \id_X)} &= \da f \o g\\
\dagg{\da f} &= \dagg{f \o (\id_A \X \dup)}\\
f \o <\id_A,  \dagg{g\o <\pi_1, f>} > &= \dagg{f \o <\pi_1, g>}.
\end{meqa}
An \W{iteration category} is a cartesian category having a Conway operator
additionally satisfying the ``commutative identities''
\[
\da{ \pa{f \o (\id_X \X \rho_1),\ooo,f \o (\id_X \X \rho_m)} }
=
\Delta_m \o \dagg{f \o (\id_X \X \Delta_m)} : X \to A^m
\]
where
\begin{itemize}
\item ${f}:{X \X A^m}\to{A}$
\item diagonal $\Delta_m\deq <\id_A, \ccc, \id_A>:{A}\to {A^m}$
\item ${\rho_i}:{A^m}\to{A^m}$
such that $\rho_i=<q_{i1},\ooo, q_{im}>$ 
where each $q_{ij}$ is one of projections $\pi_1,\ooo,\pi_m:{X^m}\to{X}$.
\end{itemize}
An \W{iteration functor} between iteration categories is a cartesian functor
that preserves Conway operators.
\oDefTitled

We may denote by $(\CC,\dagg{-})$  an iteration category, and
take it as the basic structure for the interpretation of an UnCAL theory.
A typical example of iteration category is the category of cpos and
continuous functions \cite{BE,Hassei}, where the least fixed point operator is a
Conway operator, see \Sec \ref{sec:cpo-model}.

\Def[def:L-monoid]
Given a set $L$, an $L$-\W{monoid} 
$(A, \den{-}^A_L)$ in 
an iteration category $(\CC,\dagg{-})$
consists of
\begin{enumerate}
\item a commutative monoid object $(A,\, \unit^A \!:\! \terminal\to A,\;
\mu^A \!:\! A\X A \to A)$ in \CC
satisfying 
$$\;\da{(\mu^A)} = \id_A,$$
\item a function $\den{-}^A_L: L \to \CC(A,A)$
 that assigns to each $\ell \in L$, a morphism 
$
\den{\ell}_L^A : A \to A
$
in \CC.
\end{enumerate}

Let $A$ in $\CC$ and $B$ in $\calD$ be $L$-monoids.
We say that 
an iteration functor 
$F:\CC\to\calD$ \Hi{preserves $L$-monoid structures} if 
$F(A)=B$, $F$ preserves monoid structures, and 
$F(\den{\ell}^A_L) = \den{\ell}^B_L$ for every $\ell\in L$.
\oDef

\Remark[th:bialgebra]
An $L$-monoid $A$ 
is not merely a monoid, but also
automatically
a \Hi{degenerated commutative bialgebra} (cf.\ \cite{FiorePROP})
in a cartesian category \CC.
Namely $A$ is also a comonoid $(A,!,\Delta)$
that satisfies the compatibility
\[
\Delta \o \unit^A = \unit^A \X \unit^A,\quad
\dup\o\mu^A \;=\; (\mu^A\X\mu^A) \o (\Id\X {c}\X\id) \at(\dup\X\dup), \quad
\mu^A\o\dup \;=\; \id
\]
where $c = <\pi_2,\pi_1>$.
The last equation is by
$\mu^A \o \Delta = \mu^A \o <\id, \id> = \mu^A \o <\id, \dagg{\mu^A}>
=^{\text{(fix)}}
\dagg{\mu^A} = \id$.
Thus, $L$-monoid structure is suitable to model branch and sharing 
in UnCAL.
\oRemark

\yy{-2em}
\begin{rulefigt}\normalsize
\y{-2em}
\arraycolsep = 2mm
\[
\begin{array}[h]{|l|@{~}lll|lll}\hline
\rtit{Mark} 
& Y = \seq{\var{y_1},\ooo,\var{y_n}}
&&\\ \cline{2-2}
 &  \ju{ {\var y_i}_Y }{Y}{\dmark} &\mapsto& \pi_i : \A{Y}\to A
\\ \hline
\rtit{Emp}
&&&\\ 
\cline{2-2}
&  \ju{ \EMP_Y }{Y}{\seq{}} &\mapsto& !_{\A{Y}}  : \A{Y}\to \terminal
\\ \hline
\rtit{Nil}
&&&\\ 
\cline{2-2}
&  \ju{ \nil_Y }{Y}{\dmark} &\mapsto& \unit^A\,\o\; !_{\A{Y}} : \A{Y}\to\terminal\to A 
\\ \hline
\rtit{Man}
&&&\\ 
\cline{2-2}
&  \ju{ \mult_{\seq{\var{y_1},\var{y_2}}} }{{\var{y_1},\var{y_2}}}{\dmark} &\mapsto& \mu^A : A\X A\to A
\\[.5em] \hline
\rtit{Com}& 
 \ju s{Y}{Z} &\mapsto& g : \A Y\to\A Z\\
&\ju t{X}{Y} &\mapsto& h : \A X\to\A Y\\ 
\cline{2-2}
&\ju{s \at t}{X}{Z} &\mapsto& g \o h : \A{X}\to \A{Z}
\\[.5em] \hline
\rtit{Label}& 
 \ell \in L & \mapsto& \den{\ell}^A_L : A \to A\\
& \ju{ t }{Y}{\dmark} &\mapsto& g : \A{Y} \to A  \\ 
\cline{2-2}
&  \ju{ \ell \co t }{Y}{\dmark}& \mapsto
  & \den{\ell}^A_L \o g : \A{Y}\to A
\\[.5em] \hline
\rtit{Pair}& 
  \ju{ s }{Y}{X_1} &\mapsto& g : \A{Y}\to\A{X_1}\\
& \ju{ t }{Y}{X_2} &\mapsto& h : \A{Y}\to\A{X_2}\\ 
\cline{2-2}
  &  \ju{ \pa{s \opl t} }{Y}{X_1\conc X_2} 
&\mapsto& <g , h> : \A{Y} \to \A{X_1}\X\A{X_2}
\\[.5em] \hline
\rtit{Cyc}& 
  \ju{ \qquad\quad t\;\, }{Y\!\conc\! X}{ X} &\mapsto& g : \A{Y}\X\A{X}\to\A{X}\\ 
\cline{2-2}
  &  \ju{\cyX{t} }{Y\phantom{Y\conc\,}}{X} 
&\mapsto& \dagg{g} : \A{Y} \to \A{X}
\\[.5em] \hline
\rtit{Def}&
\quad\ju{ t }{Y}{ \dmark}   &\mapsto& g : \A{Y}\to A\\
\cline{2-2}
 & \ju{ \xsub{x}{t} }{Y}{\var x} &\mapsto& g : \A{Y}\to A
\\ \hline
\end{array}
\]
\caption{Categorical interpretation $\den{-}_\CC^A$}
\label{fig:interp}
\end{rulefigt}


\subsection{Interpretation}\label{sec:interp}
Let $A$ be an arbitrary $L$-monoid in an iteration category \CC.
We give the interpretation of a term as a morphism on $A$.
We first notice that UnCAL is 
a single sorted system.
Hence we interpret the singleton context $\seq{\dmark}$ as $A$,
and
a context $\seq{\var{x}_1,\ooo,\var{x}_n}$ as $A^n$.
We interpret
a term $(\ju{ \!t\! }{Y}{X})$ as
a morphism $$\den{t}^A_\CC: \A{Y} \to \A{X}$$ in \CC.
(The super and subscripts of $\den{-}$ may be omitted hereafter.)
The table in Fig.\ \ref{fig:interp} shows 
the interpretation
according to the typing rules. 
This table is read as, for example in (Com),
$
 \ju{ s }{Y}{Z}\; \mapsto g : \A{Y}\to\A{Z}
$
means that when a term $s$ is interpreted as $g$, i.e., $\den{s}=g$,
and also $\den{t}=h$, 
the interpretation $\den{s \at t}$ is defined to be $g\o h$.
Now, it may be clear why we chose this judgment notation.
The source and target contexts of a judgment corresponds to the source
and target of a morphism in \CC in semantics. 
This is also a usual principle in categorical type theory.
All the abbreviations we introduced in \Sec\ref{sec:syntax} 
are justified by the semantics.
For example, why we took the abbreviation
${s \X t} = {\pa{s\at \pi_1 \opl t\at \pi_2}}$ is due to
\[
\den{\pa{s\at \pi_{X,Y}^X \opl t\at \pi_{X,Y}^Y}}
= <\den{s\at \pi_{X,Y}^X}, \den{t\at \pi_{X,Y}^Y}>
= <\den{s} \o {\pi_{X,Y}^X}, \den{t} \o {\pi_{X,Y}^Y}>
= \den{s} \X \den{t}
\]

An \W{(UnCAL) model} $A$ of a theory $E$ is an $L$-monoid in \CC
such that for every additional axiom $\ju{s=t}Y X$ of $E$,
$\den{s}^A_\CC=\den{t}^A_\CC$ holds.
Thus if a theory is pure, the notions of models and $L$-monoids
coincide. 
We simply say ``a model'' to mean ``a model of $E$''
if the theory or axioms $E$ we are assuming
is clear from the context.

We show another important principle of categorical type theory.

\LemmaTitled[th:subst-compo]{Substitution as composition}
Let $A$ be a model in \CC.
Under the assumption of Def. \ref{def:subst},
\[
\den{t\qesub} = \den{t}\o <\vec{\den{ s}}> \; .
\]
\oLemma
\Proof
By induction on the typing derivation of $t$.
\QED

As a corollary, renamed terms have the same meaning
in \CC. Thus, the names of markers are unimportant in the categorical
semantics.

\Cor
Let $\ju{ t }{Y}{X}$. Consider a renamed term
$ \ju{ t\, [Y \mapsto Y'] }{Y'}{X}$, 
where $|Y|=|Y'|$.
Then $\den{t} = \den{\, t\, [Y \mapsto Y']\,}$ in \CC.
\oCor

\ThTitled[th:sound]{Soundness}
Let $(A,\den{-}_L^A)$ be a model in an iteration category \CC.
For any theorem $\ju{s = t} Y X$, we have
$\den{s}^A_\CC = \den{t}^A_\CC$. 
\oTh
\Proof
By induction on the derivation of proof of $Y \pr s = t \;: X$.
We check every axiom is sound.
The axiom for (sub) is sound by Lemma \ref{th:subst-compo}.
Since \CC is an iteration category,
the axioms for Parameterised fixed point are sound. 
Since $A$ is an $L$-monoid, 
the axioms for commutative monoid, degenerated bialgebra 
and deleting trivial cycle are sound.
The induction step is routine.
\QED


\subsection{Categorical completeness}\label{sec:comp}
We first define an operation that renames markers in the target context,
which is need to identify judgments in a categorical semantics.

\Def[def:ren]
Suppose contexts $X , X'$ with $|X| =|X'|$, and
$\ju t Y {X}$.
Then a renaming $\;\ju {t \rename{X}{X'} } {Y} {X'}\;$
of target context
is inductively defined as follows.
\[
\arraycolsep = 1mm
\begin{array}[h]{rclllllllllllllllllll}
t \rename{\dmark}{x'} &\deq& \xsub{x'}t
\quad \text{for } t= y,\nil_Y,\mult,(\ell\co t)
\\
\EMP_Y \rename{\seq{}}{\seq{}} &\deq& \EMP_Y
\\
(t_1\at t_2)  \qeren &\deq& (t_1 \qeren) \at t_2 \\
\pa{t_1\opl t_2} \rename{X_1,X_2}{X_1',X_2'}
 &\deq& \pa{t_1\rename{X_1}{X_1'} \opl t_2\rename{X_2}{X_2'}}\\
 &&\text{for } \ju {t_1} Y {X_1},\; \ju{t_2} Y {X_2}\\
\cyX{t}      \qeren &\deq& \fn{cycle}^{X'}({t\; \{X\maps X' \}}) \\
\xsub{x}{t} \rename{x}{x'} &\deq& \xsub{x'}{t}
\end{array}
\]
\oDef

We establish the completeness of categorical model.
Let $E$ be arbitrary UnCAL axioms
(regardless of pure or having additional axioms).
For a theory generated by $E$,
we construct the classifying category 
\CE (cf.\ \cite{Crole,Hassei}).
As in ordinary categorical type theory,
we regard a term as an morphism, namely, we regard $\ju{t}{Y}{X}$ as
a morphism
$
t : |Y| \rTo |X|.
$
More precisely,
we identify several terms as the same morphism
by using the equality generated by $E$. We formulate it as follows.

An UnCAL theory $E$ defines an equivalence relation on well-typed terms,
which we denote by $=_E$.
Suppose contexts
$
Y=\seq{\vec y},
Y'=\seq{\vec{y'}},
X,X'
$ with 
$|Y|=|Y'|, |X|=|X'|$.
We identify a term $\ju{t}{Y}{X}$  with 
\[
\ju{t\sub{\vec {y_Y}}{\vec {y'_{Y'}}}\;\rename{X}{X'} }  {Y'}{X'}
\]
which is a term obtained by 
 bijectively renaming markers of $t$ in the source and target context suitably.
Define $\approx_E$ to be 
the symmetric transitive closure of the union of
the equivalence relation $=_E$ 
and this renaming identification, and
write an equivalence class of terms by $\approx_E$
as
$
\eclas{\ju t Y X}.
$
To establish categorical completeness,
we define the classifying category \CE by taking
\begin{itemize}
\item objects: natural numbers $n\in\Nat$
\item morphisms: $\eclas{\ju t Y X} : |Y| \to |X|$,\qquad
      identity: $\id_{|X|} \deq \eclas{\Id_X} : |X| \to |X|$
\item composition: 
$\eclas{\ju s Y Z} \o \eclas{\ju t X Y} \deq \eclas{\ju {s\at t} X Z}.$
\end{itemize}

\Prop[th:cl-cat]
\CE is
an iteration category
having an $L$-monoid $\GE$. 
\oProp
\Proof
We take

\noindent
\begin{tabular}[h]{lllll}
\DDOT terminal object: $0=|\seq{}|\in\Nat$
 \ADOT pair: $<\eclas{s},\eclas{t}> \deq \eclas{\ju{\pa{s\opl t}}{Y}{X_1+X_2}}$
\DOT product: the addition $+$ on \Nat
 \ADOT Conway: $\dagg{\eclas{\ju t {Y+X} X}}= 
\eclas{\ju {\cyX{t}} {Y} X}$
\DOT projections: $\eclas{\pi_{X,Y}^X},\eclas{\pi_{X,Y}^Y}$
 \ADOT monoid object: $\GE\deq 1=|\seq \dmark|\in\Nat$
\DOT unit: $\unit \deq \eclas{\nil_{\seq{\;}}}  : 0 \to 1$ 
 \ADOT multiplication: 
$\mu \deq \eclas{ \ju{\mult_{\seq{\var y_1,\var y_2}}}{\var y_1,\var y_2}{\dmark} }  : 1+1 \to 1$
\DOT counit: $!\deq \eclas{ \EMP_\dmark } : 1\to 0$
 \ADOT comonoid: $(\GE,\dup,!)$
\end{tabular}
and $\den{\ell}^\GE_L \deq \eclas{\ju {\ell\c\dmark} \dmark \dmark}
 \;\;\text{for every }\ell\in L$.
Then these data satisfy that $\CE$ is an iteration category
and $\GE$ is an $L$-monoid because of the axioms \AxG.
\QED

\Remark
As the definitions of \CE, pair of arrows, and the multiplication
shown in the above proof, our choice of the modified notation of UnCAL 
actually came
from this categorical structure.
Note that \CE is now the opposite category of 
an \Hi{iteration theory} by 
\citet{BE}
(N.B.\ a ``theory'' means here a Lawvere theory).
\oRemark

\Lemma[th:den-eclas]
$ \den{t}^\GE = \eclas{t}. $
\oLemma
\Proof
By induction on the typing derivation of $t$.
For the case of (Nil), Lemma \ref{th:subst-compo} is used.
For the case of (Label), (sub) axiom is is used.
Other cases are routine.
\QED

Now \GE is a particular model called a generic model in \CE.

\PropTitled{Generic model}
The $L$-monoid \GE in \CE is a model of $E$.
\oProp
\Proof
For every $(\ju {s = t} X Y)$ in $E$,
$
\den{s}^\GE=\eclas{s} = \eclas{t} = \den{t}^\GE
$
by Lemma \ref{th:den-eclas}.
\QED

\ThTitled[th:cat-comp]{Categorical soundness and completeness}
$\ju {s=t} Y X$ is derivable from $E$ (including \AxG) in \ELu iff\;\;
$\den{s}_\CC^A = \den{t}_\CC^A$ holds 
for all iteration categories \CC and all models $A$ of $E$ in \CC.
\oTh
\Proof
Soundness has been established in Thm.\ \ref{th:sound}.
Suppose the assumption of the theorem. Then in particular
for the classifying category \CE  and the generic model \GE, 
$\den{s}_\CE^\GE = \den{t}_\CE^\GE$ holds. 
Hence $\eclas{s}=\eclas{t}$ meaning that $s=t$ is derivable from $E$.
\QED

\smallskip

\Th[th:universality]%
For any model $A$ of $E$ in an iteration category \CC,
\phantom{MMMMMMMM}
\begin{wrapfigure}[5]{r}{.28\linewidth}\y{-3em}
\begin{diagram}[2em]
&  \Term &\rTo^{\den{-}^\GE}& \CE \\
\quad\qquad&\dTo<{\den{-}^A}&\ldTo>{\Psi^A}\\
&\CC
\end{diagram}
\end{wrapfigure}
there exists 
a unique iteration functor $\Psi^A : \CE \rTo \CC$
such that $\Psi^A( \GE ) = A$, and it 
preserves $L$-monoid structures.
Pictorially, 
it is expressed as
the right picture, where $\Term$ denotes the set of all well-typed terms.
\oTh
\Proof
We write simply \Psi for $\Psi^A$.
Since $\Psi$ preserves $L$-monoids, we have
$\den{t}^A = \Psi(\den{t}^\GE) = \Psi(\eclas{t})$ for any term $t$,
hence the mapping $\Psi$ is required to satisfy 
\[
\arraycolsep = 1mm
\renewcommand{\arraystretch}{1}
\begin{array}[h]{lllllll}
  \Psi(& \eclas{{\var y_i}} &) &=& \pi_i \\
  \Psi(& \eclas{ \EMP } &) &=&   !_{\A{Y}}  \\
  \Psi(& \eclas{ \nil } &) &=&   \unit^A\,\o\; !_{\A{Y}} \\
  \Psi(& \eclas{ \mult } &) &=& \mu^{A}   \\
  \Psi(& \eclas{ \nxsub{x}{t} } &) &=& \Psi(\eclas{t})
\end{array}
\qquad
\begin{array}[h]{llllllll}
\Psi(& \eclas{ s \at t } &) &=&   \Psi(\eclas{s}) \o \Psi(\eclas{t})   \\
  \Psi(& \eclas{  \ell \co{t} } &) &=&   \den{\ell}^A_L \o \Psi(\eclas{t})  \\
  \Psi(& \eclas{ \pa{ s \opl t }} &) &=& <\Psi(\eclas{s}) , \Psi(\eclas{t})>   \\
  \Psi(& \eclas{ \cy{t} } &) &=&   \dagg{\Psi(\eclas{t})}
\end{array}
\]
This means that $\Psi$ is an iteration functor 
that sends the $L$-monoid \GE  to the $L$-monoid $A$. 
Such $\Psi$ is uniquely determined by these equations because 
$(A,\den{-}^A)$ is 
a model.
\QED

So, the interpretation $\den{-}^A$ determines
an iteration functor $\Psi^A$.
What is $\Psi^A$ from the viewpoint of UnCAL's graph transformation?
The composite $\Psi^A(\den{-}^\GE)$
sends a directed edge in an input term to a morphism in \CC
whose source and target nodes are correspondingly mapped.
This idea is more precisely pursued in the next section.


\section{Characterisation of Structural and Primitive Recursion}
\label{sec:str-rec}

In this section, we tackle to model 
the computation mechanism of UnCAL ``structural recursion on graphs''
using our categorical semantics.

\subsection{Overview}
A structural recursive function is defined by the syntax
\begin{equation}
\arraycolsep = 1mm
\begin{array}[h]{rcll}
  \texttt{sfun }f(\ell_1\co t) &=& e_{\ell_1}\\
                &\ccc&\\
                f(\ell_n\co t) &=& e_{\ell_n}
\end{array}
  \label{eq:diamond}
\end{equation}
where $\ell_1,\ccc,\ell_n$ are disjoint and cover all labels in $L$. 
We assume that the number $n$ of clauses is finite.
Some $\ell_i$ may be meta-variables and 
the case analysis will be given meta-theoretically
as in Example \ref{ex:aa} below.
Each $e_{\ell_i}$ is
an UnCAL term extended with the form $g(s)$ where $g$ is a label.  
To concentrate on the essential
part of modelling structural recursion, in this paper, we assume that
every $e_{\ell_i}$ does not involve other function calls $h(t)$ where
$h$ is defined by other structural recursive function defined by 
\code{sfun}-definition.
But we can use other functions defined by additional axioms, 
as in Example \ref{ex:aa}.

In the original formulation, given an input graph,
the graph algorithm computes a result graph
using the definition (\ref{eq:diamond}).
This relation between inputs and results becomes a
function $f$, which satisfies the equations 
\cite[Prop.\ 3]{Buneman}:
\begin{equation*}
\arraycolsep = .4mm
\renewcommand{\arraystretch}{1}
\x{3em}
\begin{array}[h]{lcllllll}
  f(& {{\var y_i}} &) &=& {\var y_i}\\
  f(& { \EMP } &) &=&   \EMP  \\
  f(& { \nil } &) &=&   \nil \\
\end{array}
\qquad
\begin{array}[h]{lcllllll}
  f(& { \nxsub{x}{t} } &) &=& \xsub{x}{f({t})}\\
  f(& { s \;\uniSym\; t } &) &=& f(s) \;\uniSym\; f(t)\\
  f(& \pa{  s \opl t } &) &=& \pa{f({s}) \opl f({t}) }
\end{array}
\qquad
\begin{array}[h]{lcllllllll}
f(& \quad\; \ell \co{t} \quad &)  &=&  e_\ell &\ccc(\star)
\\
  f(& { s \at t } &) &=&   f({s}) \at f({t}) &\ccc(\neqmark)  &\x{2.5em}
\stepcounter{equation}
(\theequation)\\
  f(& { \cy{t} } &) &=&   \cy{f({t})}           \quad&\ccc(\neqmark) 
\end{array}
\label{eq:my-sfun}
\end{equation*}
when $e_\ell$ does not depend on $t$. 
We intend that the part $(\star)$ represents a family of equations 
for all labels $\ell$
as (\ref{eq:diamond}).
Here ``$e_\ell$ depends on $t$'' means that
$e_\ell$ contains $t$ other than the form $f(t)$.
This is understandable naturally, as the example \code{f2} in Introduction
recurses 
the term $\mathbf{t}_G$ structurally. 
Combining the categorical viewpoint we have investigated, 
$f$ can be understood as a functor because it preserves $\at$ in 
$(\neqmark)$. Moreover,
$f$ preserves \cyc 
and cartesian products, hence it can be regarded as a traced cartesian functor.
Importantly, Buneman et al.~observed that
the above nine equations hold only when $e_\ell$ {does not} depend
on $t$, such as \code{f2}.

The case where $e_\ell$ depends on $t$ requires more
careful thought.
Two equations marked ($\neqmark$) {do not hold}
in this case (and other seven equations do hold). 
Crucially, \code{f1} in Introduction is already in this case, where \code{T} appears 
as not of the form $\code{f1}(t)$.
The following example shows why ($\neqmark$) do not hold.

\Example[ex:aa]
We suppose the labels \code{true} and \code{false}.
We abbreviate
the terms $\code{true}\co\vnil$ and $\code{false}\co\vnil$ 
as simply \code{true} and \code{false}.
We first define the additional axioms
\[
 \bigcur{
\code{head-a?}(\vnil)  &=& \vnil\\
\code{head-a?}(\code{a}\co t)  &=& \code{true}\\
\code{head-a?}(\ell\co t)  &=& \vnil  &\text{for }\ell\not= \code{a}
}
\]
for a predicate \code{head-a?} that checks whether the head is \code a.
Since this is not a recursive definition, we set it as axioms.
The UnCAL axioms $E$ we now assume is 
the union of \AxG and the additional axioms.
We define
the recursive function \fnaa that tests whether the argument contains 
``\code{a:a:}''. 
\[
\arraycolsep = 1mm
\begin{array}[h]{llcll}
  \code{sfun} &\code{aa?}(\code a\co t) &=& \code{head-a?}(t)\\
              &\code{aa?}(\ell\co t) &=& \code{aa?}(t) &\qquad\text{for
                }\ell\not= \code a
\end{array}
\]
The right-hand side $\code{head-a?}(t)$ of the first clause depends on $t$.
Then we have the inequalities:
\begin{Verbatim}[commandchars=\\\{\},codes=\mathcom]
aa?( (a:&)$\at$(a:\{\})) = aa?( a:a:\{\} ) = true 
                                  $\,\not=\;$ \vnil = \vnil$\at$\vnil = aa?(a:&) $\at$ aa?(a:\{\})
aa?( cycle(a:&) ) = aa?( a:a:cycle(a:&) ) = true 
                                  $\,\not=\;$ \{\} = cycle(\{\}) = cycle(aa?(a:&)) 
\end{Verbatim}
\oExample
This means that $f$ does not preserve \cyc in general, and even {is not functorial}. 
Thus the categorical view seems not helpful to understand this pattern of recursion.

This pattern of recursion is similar to 
the case of primitive recursion on an algebraic data type.
Since structural recursion is a generalisation of primitive recursion, 
primitive recursion can be encoded as a structural recursion.
The technique of paramorphism \cite{para} in functional programming 
is a way to represent primitive recursion in terms of ``fold''.
Using a similar idea, we will show that
the case $e_\ell$ depends on $t$ can also be derivable from 
the universality.

\subsec{Aims}
In this section, we will show the following.
\begin{enumerate}
\item We give a way to generate mathematically a function \Psi
(resp. \Upsilon)
on UnCAL terms from the syntactic definition (\ref{eq:diamond}) of 
a function $f$
for structural recursion in \Sec \ref{sec:char} (resp. primitive
recursion in \Sec \ref{sec:char-prim}).

\item We clarify what equations are satisfied for \Psi (resp. \Upsilon).
We call these equations \Hi{the characterisation} of 
structural recursive $f$ (\ref{eq:my-sfun}) 
(resp. primitive recursive (\ref{eq:my-pfun})).
\label{itm:my-sfun-k}

\item We show that the characterisation \Hi{uniquely determines}
a function. 
\label{itm:complete}
\end{enumerate}
Throughout this section, we suppose UnCAL axioms $E$,
(i.e. $E$ is the union of \AxG and additional axioms)
as stated in
\Sec \ref{sec:comp}. 
\FORreviewer{See also \Sec \ref{sec:el-ax}.}
We may use additional axioms 
to define built-in and utility functions as Example \ref{ex:aa}.

\subsection{Structural recursion on graphs}
\label{sec:char}

We first characterise structural recursion on graphs 
for the case
that $e_\ell$ may involve the form $f(t)$ in (\ref{eq:diamond}), 
but does not involve $t$ solely,
such as the function \code{f2} in Introduction.
To formally formulate the form $f(t)$, we extend our syntax to 
have the construct of function terms given by the following typing rule.
\[
\ninfrule{Fun}
 {f \in L \infspc
 \ju{ t }{Y}{\dmark} }
 {\ju{ f(t) }{Y}{\dmark}}
\]
We assume that $L$ contains also function symbols. 
The categorical semantics of this is given
by $\den{f(t)} = \den{f}_L \o \den{t}$.
The idea of this semantics is to represent the form $f(t)$ at the level
of categorical arrows.
Note that this does not gives the semantics of recursive calls.
It is no problem because the term of recursive call $f(t)$ is vanished
in modelling the structural recursion $f$
(cf. the construction of $e_\ell'$ below).

\subsec{Assumptions for structural recursion}\label{sec:ass-str}
Suppose the definition of structural recursive function $f$ is of the form
(\ref{eq:diamond}).
Let $e_\ell$ be the right-hand side of a clause in (\ref{eq:diamond}).
We assume that it is well-typed as 
\[
\ju{e_\ell}{Y}{W}
\]
where $W=\seq{\var x_1,\ooo,\var x_k}$ 
\FORreviewer{(i.e. there exist suitable contexts $W,Y$ for $e_\ell$)}
and
$$k \deq |W|.$$
\begin{enumerate}
\item We assume that a term $f(t)$ possibly appears in $e_\ell$
and is of the type
\[
 \ju{f(t)}{Y}{W}
\]
\item We assume that all the right-hand sides $e_{\ell_i}$ in (\ref{eq:diamond})
have the same $k$ and source and target contexts $Y,W$.
\FORreviewer{{\bf We use this number $k$ throughout this section.}}
\end{enumerate}
We define $e'_\ell$ to be a term obtained from $e_\ell$, by
replacing every $f(t)$ with 
$\pa{\nnxsub{x_1}{\var{x_1}}\opl\ccc\opl \nnxsub{x_k}{\var{x_k}}}$ of type $W$
and assume
$$\;\ju{e'_\ell}{W}{W}\;.$$
By construction, 
$\ju{e_\ell \;\,=\;\, e'_\ell \at f(t)}{Y}{W}  
$ holds.

\subsec{The case $k=1$}
We first examine the simplest case $k=|W|=1$
to get intuition of 
our categorical characterisation of structural recursion.
The idea of constructing $e'_\ell$ above
is that since $e_\ell$ involves several $f(t)$'s, we first separate
these from $e_\ell$. Now $e'_\ell$ is a ``skeleton'' of $e_\ell$ where 
every occurrence
of the recursive call $f(t)$ is replaced with the marker \dmark, 
considered as a hole.
Then we model 
the execution of the structural recursion (\ref{eq:diamond})
as the operation that replaces all labelled edges ``$\ell \co$'' in a term
with $e'_\ell$ with a suitable composition.
This is achieved by taking 
the $L$-monoid $\Mdel \deq (\GE,\den{-}_L^{\Mdel})$ in \CE by
$$
\den{\ell}_L^{\Mdel}
 \deq \eclas{\ju{e'_\ell }{\dmark}{\dmark}} : \GE \to \GE,
$$
which is moreover a model of $E$.

\Remark 
For example, in case of the example \code{f2} in Introduction defined by
\[
          \code{sfun f2}(\ell\co t) = \code{a}\co\code{f2}(t)
\]
we take
$
e_\ell =               \code{a}\co\code{f2}(t),\quad     
  e'_\ell =           {\code{a}\co\dmark},        \quad
 \den{\ell}_L^\Mdel = [ \dmark \pr{\code{a}\co\dmark} :\dmark ]_E.
$
\oRemark

Notice that $\Mdel$ is different from the $L$-monoid $\GE=(\GE,\den{-}_L^\GE)$
of the generic model,
where $\den{\ell}_L^{\GE} = \eclas{\ell\co \dmark}$.
Now we consider the mapping from the generic model to the model $\Mdel$.
In the situation of Theorem \ref{th:universality},
the unique iteration functor $\Psi^{\GE} : \CE \to\CE$ exists 
and it maps the $L$-monoid \GE to the $L$-monoid $\Mdel$.
Thus, what the functor $\Psi^\GE$ does is 
that every morphism for $\ell$ (coming from a labelled edge $\ell$ in a term) 
is replaced with
$e'_\ell$  as
\begin{diagram}[height=1.5em]
  \GE     &&  \GE\\
  \uTo<\ell  &\mapsto& \uTo>{e'_\ell}\\
  \GE     &&  \GE
\end{diagram}
which is actually we wanted to model.
Namely, we model the meaning of the syntactically given function $f$ as 
a functor $\Psi^\GE$.

More precisely, it is modelled as follows.
We may omit the superscript of $\Psi$, hereafter.
Since $\GE$ is a model, the choice of a representative in 
an equivalence class applied to \Psi is irrelevant,
i.e.,
if $s \approx_E t$, then $\Psi(\eclas{s}) =\Psi(\eclas{t})$.
This means that it is safe to write simply
$\Psi(t)$ for a suitably renamed term $t$ avoiding name clash, and then
it gives a {syntactic} translation on UnCAL terms
by the unique (arrow part) function 
\begin{equation*}\label{eq:psi}
\Psi : \CE(\GE^{|Y|},\GE^{|X|})\to\CE(\GE^{|Y|},\, \GE^{|X|})  
\end{equation*}
on terms of source context $Y$ and target context $X$, characterised as
\begin{equation}
\arraycolsep = .4mm
\renewcommand{\arraystretch}{1}
\begin{array}[h]{lcllllll}
  \Psi(& {{\var y_i}} &) &=& \var y_i\\
  \Psi(& { \EMP } &) &=&   \EMP  \\
  \Psi(& { \nil } &) &=&   \nil \\
\end{array}
\quad
\begin{array}[h]{lcllllll}
  \Psi(& { \mult } &) &=& \mult\\
  \Psi(& { \nxsub{x}{t} } &) &=& {\Psi({t})}\\
  \Psi(& {  \pa{s \opl t} } &) &=& \pa{\Psi({s}) {\opl} \Psi({t})}  
\end{array}
\quad
\begin{array}[h]{lcllllllll}
  \Psi(& \ell \co{t} &) &=&   e'_\ell \at {\Psi({t})}  \\
  \Psi(& { s \at t } &) &=&   \Psi({s}) \at \Psi({t})   \\
  \Psi(& { \cy{t} } &) &=&   \cy{\Psi({t})}
\end{array}
\label{eq:my-sfun}
\end{equation}
Then one can clearly see that \Psi traverses a term 
in a structural recursive manner.
Note that the case for $\xsub{x}{t}$, it actually translates
equivalence classes as
$$ \Psi( \eclas{ \ju{\xsub{x}{t}}Y x } ) = \Psi(\eclas{\ju{ t }Y \dmark})
 =\Psi(\eclas{t}).$$

Here comes an important point.
{This pattern \Psi derived from the universality in Thm.\ \ref{th:universality}
is exactly the same as 
what {Buneman et al.}
called the
\W{``structural recursion on graphs''}}
for the case that $e_\ell$ does not depend on $t$
\cite[Proposition 3]{Buneman}.
Actually, we could make the characterisation
{more precise} than the original 
and the subsequent developments \cite{Buneman,ICFP10,LOPSTR,ICFP13,PPDP13}.

\begin{enumerate}[itemsep=.5em]
\item Buneman et al.~stated only that 
(\ref{eq:my-sfun})
is a \Hi{property} \cite[Prop.\ 3]{Buneman} of 
a ``structural recursive function on graphs'' defined by the algorithms.
This means that there could be many functions that satisfy the {property}.
Our result of the universality stated in Theorem \ref{th:universality}
provides more precise information.
A function satisfying 
the characterisation (\ref{eq:my-sfun}) 
\Hi{uniquely} determines a function \Psi, and there are no functions that satisfy the characterisation
other than \Psi.
\label{itm:sound-complete}

\item The characterisation mathematically clarifies what are the 
\Hi{structures} of
``structural recursion on graphs''.
The structures preserved by it 
are \Hi{iteration category} and \Hi{$L$-monoid}  structures.
\end{enumerate}

\ExampleTitled{Replace all labels with \code{a}}
\[
          \code{sfun f3}(\ell\co t) = \code{a}\co\code{f3}(t)
\]
Example of execution is:
\begin{Verbatim}[commandchars=\\\[\],codes=\mathcom,fontsize=\small]
f3( people:{ population:425017\nnil, ethnicGroup:"Celtic"\nnil,
            ethnicGroup:"Portuguese"\nnil, ethnicGroup:"Italian"\nnil })
\narone a:{ a:a\nnil, a:a\nnil, a:a\nnil }
\end{Verbatim}
Define the particular $L$-monoid $(\GE, \den{-}_L)$
in \CE by
$
\den{\ell}_L = \eclas{\ju {\code{a}\c\dmark} \dmark \dmark}.
$
Then the unique iteration functor $\Psi:\CE \to \CE$ gives
the structural recursive function defined by \code{f3}.
Actually, it computes
\begin{Verbatim}[commandchars=\\\[\],codes=\mathcom,fontsize=\small]
\Psi\!\!\!( people:{ population:425017\nnil, ethnicGroup:"Celtic"\nnil,
            ethnicGroup:"Portuguese"\nnil, ethnicGroup:"Italian"\nnil })
= a:{ a:a\nnil, a:a\nnil, a:a\nnil }
\end{Verbatim}
\oExample

\subsec{The case $k \ge 1$}
Again consider the setting of \Sec \ref{sec:ass-str}.
The method we have taken can be generalised to the case for
any $k \ge 1$, i.e. $e_\ell$ is of the type
\[
\ju{e_\ell}{Y}{\var x_1,\ooo,\var x_k}
\]
This is merely by taking
$\GE^k$ (instead of $\GE$) as a model.
Every object $\GE^k$ forms a monoid
with unit and multiplication given by
\[
\begin{array}[h]{lll}
\nu^{\GE^k} &:& \terminal \to \GE^k  \\
\nu^{\GE^k} &\deq& \eclas{\pa{\nil_0\opl\ooo\opl\nil_0}}  \\
\mu^{\GE^k} &:& \GE^k\X \GE^k \to \GE^k\\
\mu^{\GE^k} &\deq&  
 \eclas{ \ju{\pa{\uni{x_1}{y_1},\ooo, \uni{x_k}{y_k}}}
       {x_1,\ooo,x_k,y_1,\ooo,y_k}{ \dmark_1,\ooo,\dmark_k }}
\end{array}
\]
It forms the $L$-monoid $\Mdel \deq ({\GE^k},\den{-}_L^{\GE^k})$ by
$$
\den{\ell}_L^{\Mdel}
 \deq \eclas{\ju{e'_\ell }{W}{W}} : \GE^k \to \GE^k$$
which is moreover a model.
Thus, the unique iteration functor $\Psi^{\GE^k} : \CE \to\CE$ exists 
such
that $\Psi^{\GE^k}(\GE)=\GE^k$ and preserves $L$-monoid structures.
Similarly to the case $k=1$,
it gives the unique function 
on UnCAL terms
\begin{equation}
\label{eq:psik}
\Psi : \CE(\GE^{|Y|},\GE^{|X|})\to\CE(\GE^{k\cdot |Y|},\, \GE^{k\cdot
  |X|})
\end{equation}
which is characterised as
\begin{equation}
\arraycolsep = .4mm
\renewcommand{\arraystretch}{1}
\begin{array}[h]{lcllllll}
  \Psi(& {{\var y_i}} &) &=& \pa{y_i^1\opl\ccc\opl y_i^k}\\
  \Psi(& { \EMP } &) &=&   \EMP  \\
  \Psi(& { \nil } &) &=&   \pa{\nil\opl\ooo\opl\nil} \\
\end{array}
\;
\begin{array}[h]{lcllllll}
  \Psi(& { \mult } &) &=& \mu^{\GE^k}\\
  \Psi(& { \nxsub{x}{t} } &) &=& {\Psi({t})}\\
  \Psi(& \pa{  s \opl t } &) &=& \pa{\Psi({s}) {\opl} \Psi({t})  }
\end{array}
\;
\begin{array}[h]{lcllllllll}
  \Psi(& \ell \co{t} &) &=&   e'_\ell \at {\Psi({t})}  \\
  \Psi(& { s \at t } &) &=&   {\Psi({s}) \at \Psi({t})}   \\
  \Psi(& { \cy{t} } &) &=&   \cy{\Psi({t})}
\end{array}
\label{eq:my-sfun-k}
\end{equation}
Note that \Psi returns a term in a context $Y^1,\ccc,Y^k$,
where $Y^1 = \seq{y_1^1,\ooo,y_m^1},\; Y^2 = \seq{y_1^2,\ooo,y_m^2}, \ooo$,
and the superscripts indicate $i$-th copy of $Y\; (1\le i \le k)$.
As in the case of $k=1$, 
the equations (\ref{eq:my-sfun-k})
 \W{uniquely determine} the structural recursive function \Psi.
For this unique function $\Psi$,
we will write $$\msrec{ (e'_\ell)_{\ell\in L} }
\qquad\text{or simply}\quad \msrec{ e' }
$$
for $e' = (e'_\ell)_{\ell\in L}$.

This is our semantical explanation of ``structural recursion on graphs''.
As a bonus, 
the characterisation reveals 
minor {errors} in the previous 
definitions of structural recursion in the literature 
\cite{Buneman,ICFP10,LOPSTR}.
The case for $\nil$ must be 
$\Psi( { \nil } ) =  \pa{\nil\opl\ooo\opl\nil}$.
But
in the literature mentioned above, this was 
$\Psi( { \nil } ) =   \nil$ (N.B.\
$\pa{\nil\opl\ooo\opl\nil} \not= \nil$).
This shows the importance of
identification of the mathematical structure (i.e. a monoid $\GE^k$)
to provide a right definition.
The papers \cite{ICFP13,PPDP13} correctly handle this case by considering
a suitable monoid as our formulation.

\ExampleTitled[ex:abab]{\cite{ICFP10} abab}
We demonstrate structural recursion on graphs of the case $k=2$.
The function \code{abab}
changes all edges of even distance from the root
to \code{a}, odd distance edges to \code{b}. It is defined by
\[
\code{sfun abab}(\ell\co t) = 
 \pa{ \xsub{z1}{\code a\co{{z2}}} \opl \xsub{z2}{\code b\co{{z1}}}} \at \code{abab}(t) 
\]
Example of execution is:
\begin{Verbatim}[commandchars=\\\{\},codes=\mathcom]
            $z1$ @ abab(p:q:r:\{\}) $\narone$ a:b:a:\{\}.
\end{Verbatim}
The original execution process 
was hard to understand,
because the structural recursion is officially defined 
by the graph algorithms \cite{Buneman}.
\cite{ICFP10} gave 
an informal explanation how it is executed
using the following figure
\[
\mathtt{abab}(
\xymatrix @C=8pt {
*{\o} \ar[r]^{\texttt{p}} & *{\bullet} \ar[r]^{\texttt{q}} & *{\bullet} \ar[r]^{\texttt{r}} & *{\bullet}
}
)
\ =\ 
\xymatrix @C=20pt {
*{\o} \ar[rd]^(.2){\texttt{a}} & *{\bullet} \ar@{-->}[rd]^(.2){\texttt{a}} & *{\bullet} \ar[rd]^(.2){\texttt{a}} & *{\bullet}
\\
*{\bullet} \ar@{-->}[ru]_(.2){\texttt{b}} & *{\bullet} \ar[ru]_(.2){\texttt{b}} & *{\bullet} \ar@{-->}[ru]_(.2){\texttt{b}} & *{\bullet}
}
\ =\ 
\xymatrix @C=8pt {
*{\o} \ar[r]^{\texttt{a}} & *{\bullet} \ar[r]^{\texttt{b}} & *{\bullet} \ar[r]^{\texttt{a}} & *{\bullet}
}
\]
but it is still mysterious 
for outsiders.
What happens exactly in the computation process?
Now we take $e'_\ell$ to be the term
\[
\ju {\pa{\xsub{z_1}{\code{a}\co{\var{z_2}}} \opl
\xsub{z_2}{\code{b}\co{\var{z_1}}}}}
{z_1,z_2}
{z_1,z_2}
\]
This is the case $k=2$.
Here we give
a formal \Hi{equational proof}
using our characterisation of structural recursion and the theorems
(sub) and (asso$\at$):
\[
\arraycolsep = 0mm
\begin{array}[h]{lllllllll}
&\var{z_1}\at \srec{e'_\ell}( \code p\co {\code q\co {\code r\co {\nil}}} )
\\
=\quad &\var{z_1}\at (e'_\ell\at e'_\ell \at e'_\ell\at \pa{\nil\opl\nil}) 
= \var{z_1}\at (
e'_\ell &\at &<\xsub{z_1}{\code{a}\co{\var{z_2}}} &\opl \xsub{z_2}{\code{b}\co{\var{z_1}}}>\\
&&\at &<\xsub{z_1}{\code{a}\co{\nil}} &\opl \xsub{z_2}{\code{b}\co{\nil}}>)
\\
=\quad & \multicolumn{5}{l}{(\var{z_1} \at e'_\ell) \at 
<\xsub{z_1}{\code{a}\co{\code{b}\co{\nil}}} \opl  \xsub{z_2}{\code{b}\co{\code{a}\co{\nil}}}>
\;=\; \code{a}\co {\code{b}\co {\code{a}\co {\nil}}}}.
\end{array}
\]
This \emph{equational proof} has not been possible so far \cite{Buneman,ICFP10,LOPSTR,ICFP13,PPDP13}.
\oExample

\subsection{Primitive recursion on graphs}
\label{sec:char-prim}

\subsec{Assumptions for primitive recursion}\label{sec:ass-prim}
Suppose the definition of structural recursive function $f$ is of the form
(\ref{eq:diamond}).
Let $e_\ell$ be the right-hand side of a clause in (\ref{eq:diamond}),
which is well-typed as 
\[
\ju{e_\ell}{ }{W}
\]
where $W=\seq{\var x_1,\ooo,\var x_k}$ 
and
$$k \deq |W|.$$
\begin{enumerate}
\item We assume that terms $f(t),t$ possibly appear in $e_\ell$
and are of the types
\[
 \ju{f(t)}{ }{W} \qquad \;\ju{t}{ }{\dmark}
\]
\item We assume that all the right-hand sides $e_{\ell_i}$ in (\ref{eq:diamond})
have the same $k$, the empty source context and the target context $W$.
\end{enumerate}
We define $e'_\ell$ to be a term obtained from $e_\ell$, by
replacing every $f(t)$ with 
$\pa{\nnxsub{x_1}{\var{x_1}}\opl\ccc\opl \nnxsub{x_k}{\var{x_k}}}$ of type $W$,
and every $t$ (not in the form $f(t)$) with $\dmark$,
and assume 
$$\;\ju{e'_\ell}{W,\dmark}{W}
.\;$$
By construction, 
$\ju{e_\ell \;\,=\;\, e'_\ell \at <f(t),t>}{ }{W}   
\label{eq:eabs}
$ 
holds.

The marker $\dmark$ in $e'_\ell$ 
is regarded as a hole which will be filled with
$t$, and 
$\pa{\nnxsub{x_1}{\var{x_1}}\opl\ccc\opl \nnxsub{x_k}{\var{x_k}}}$ 
is regarded as other holes filled with $f(t)$.
We call this type of recursion \Hi{primitive recursion on graphs},
because it is similar to 
the case $f(S(n))=e(f(n),n)$
of primitive recursion on natural numbers, which can
use both $n$ and $f(n)$ at the right-hand side.
Now we take another monoid $\GE^k\X \GE$ as an UnCAL model.
$\GE^k\X\GE$ forms a monoid as in the case of $\GE^k$,
moreover a $L$-monoid by 
$$\den{\ell}_L^{\GE^k\X \GE} : \GE^k\X \GE \to \GE^k\X \GE,\quad
\den{\ell}_L^{\GE^k\X \GE}
 \deq \eclas{\ju{ \pa{e'_\ell \opl\; (\ell \co \dmark)}}{W,\dmark}{W,\dmark}}.$$
Since $\GE^k\X\GE$ is a model, there exists 
the unique (arrow part) function
$$
\Psi : \CE(\GE^{|Y|},\GE^{|X|})\to\CE(\GE^{(k+1)\cdot|Y|}, \GE^{(k+1)\cdot|X|})
$$
by Thm. \ref{th:universality}.
Finally, we define the function $$
\Phi : \CE(\terminal,\GE^{|X|})\to\CE(\terminal,\, \GE^{k\cdot|X|})\,;
\quad
\Phi(t) \deq \pi \at \Psi(t)
$$
where we take $Y=\seq{}$ for \Psi and
\pi
is a projection that projects $\GE^{k\cdot|X|}$-components from
$\GE^{(k+1)\cdot|X|}$,
given by
$$\ju {\pi \deq <X^1,\ccc,X^k>}{X^1,\ccc,X^k,X}{X^1,\ccc,X^k},
$$
and
$X^i=\seq{x_1^i,\ooo,x_m^i}$, where 
the superscripts indicate that it is the $i$-th copy 
taken from $X\; (1\le i \le k)$.
Then $\Psi(t) = \pa{\Phi(t),t}$ holds for any term 
$\ju t {}\dmark$.
Hence using (\ref{eq:my-sfun}),
$\Phi$ is the \Hi{unique function} satisfying
\begin{equation}
\arraycolsep = .4mm
\renewcommand{\arraystretch}{1}
\begin{array}[h]{lcllllll}
\phantom{X}\\
  \Phi(& { \EMP_0 } &) &=&   \EMP_0  \\
  \Phi(& { \nil_0 } &) &=&   \pa{\nil_0\opl\ooo\opl\nil_0} \\
\end{array}
\;
\begin{array}[h]{lcllllll}
  \Phi(& \uni s t &) &=& \uni{\Phi(s)}{\Phi(t)}\\
  \Phi(& { \nxsub{x}{t} } &) &=& {\Phi({t})}\\
  \Phi(& \pa{  s \opl t } &) &=& \pa{\Phi({s}) {\opl} \Phi({t})}  
\end{array}
\;
\begin{array}[h]{lcllllllll}
  \Phi(& \ell \co{t} &) &=&   e'_\ell  \at \pa{\Phi({t})\opl t} \\
  \Phi(& { s \at t } &) &=&  \pi \at (\Psi({s}) \at \Psi({t}))   \\
  \Phi(& { \cy{t} } &) &=&   \pi \at \cy{\Psi({t})}
\end{array}
\label{eq:my-pfun}
\end{equation}
This $\Phi$ completely characterises \Hi{what 
{Buneman et al.}
called the structural recursion on graphs for the case
that $e_\ell$ depends on $t$}.
This result provides more precise information than the existing results.

\begin{enumerate}[itemsep=.5em]
\item 
The equations (\ref{eq:my-pfun}) are 
a unique characterisation of
{primitive recursion} on UnCAL graphs.

\item 
Buneman et al.~showed by a counterexample 
that $f(s\at t) = f(s)\at f(t)$ and $f(\cy{t}) =
\cy{f(t)}$ do not hold in general for a primitive recursive $f$.
They did not analyse the precise reason of it and 
what equations \emph{do} hold for them.
Now we could give the laws
for $\at$ and $\cyc$ in (\ref{eq:my-pfun}).
\label{item:law1}

\item These laws 
can be read as how to compute primitive recursion \Phi for the cases 
for ${ s \at t } $  and
$\cy{t}$ using the structural recursion \Psi (N.B.\ not \Phi).
But this way is not mandatory.
Since \Phi respects the equality $\approx_E$,
taking any term $u$ which is equivalent (w.r.t.\ $\approx_E$)
to ${ s \at t } $  (resp.\ $\cy{t}$),
and next computing $\Phi(u)$ is also possible.
The following example illustrates this situation.
In any case,
the laws of \Phi for $\at$ and $\cyc$ in (\ref{eq:my-pfun})
hold.
\label{item:law2}
\end{enumerate}
Note that $\Phi$ is a set-theoretical total function,
hence it is completely determined (in other words, terminating).
For this unique function $\Phi$,
we will write $$\mprec{ (e'_\ell)_{\ell\in L} }
\qquad\text{or simply}\quad \mprec{ e' }.
$$

\Example[ex:prim-diff]
We model the primitive recursive function \code{aa?} given in 
Example \ref{ex:aa}.
Take an $L$-monoid
\[
\arraycolsep = .4mm
\begin{array}{llrllllll}
\den{\code{a}}_L &= 
[{x,\dmark} &\pr&  <{\code{head-a?}(\dmark)} &\opl  \code{a}\co \dmark >
&: {x,\dmark}]_E
\\
 \den{\ell}_L &= 
[ {x,\dmark} &\pr&  <x &\opl  \ell\co \dmark >
 &:{x,\dmark}]_E
\text{ for }\ell\not = \code{a}.
\end{array}
\]

We can compute $\code{aa?}(\cy{\aa\co\dmark})$ as follows
by using the law for \cyc in (\ref{eq:my-sfun}).
\begin{meqa}
 \Phi(\cy{\aa\co\dmark}) &=\; 
\pi\at \cy{\Psi(\aa\co\dmark)}\\
&=\; \pi\at  \cy{\pa{\code{head-a?}(\dmark)\opl \aa\co \dmark}} \\
&=\; \pi\at \pa{\code{head-a?}(\dmark)\opl \aa\co \dmark}\at 
\pa{\code{head-a?}(\dmark)\opl \aa\co \dmark}\at \cy{\aa\co\dmark}\\
&=\; \pi\at \pa{\code{head-a?}(\aa\co\dmark)\opl \aa\co\aa\co \dmark}
\!\at \cy{\aa\co\dmark}
\!=\! \code{true}.
\end{meqa}
There is another way to compute it by firstly unfolding
the argument using the fixed point law (fix) without relying 
the structural recursive function \Psi.
Now $e'= \code{head-a?}(\dmark)$.
\begin{meqa}
 \Phi(\cy{\aa\co\dmark}) &= 
\Phi(\aa\co\aa\co\cy{\aa\co\dmark}) \\
&= e' \at \pa{ e'\at \pa{\Phi(\cy{\aa\co\dmark}),\cy{\aa\co\dmark}}
, \aa\co\cy{\aa\co\dmark}}\\
&= \code{head-a?}(\aa\co\cy{\aa\co\dmark})
= \code{true}
\end{meqa}
\oExample

In summary, taking a suitable $L$-monoid, we could model
various types of recursion principle uniformly by the universality
of the generic model in the categorical semantics of UnCAL.


\subsection{Application to the fusion law}\label{sec:fusion}

We show an application of our characterisation to 
an optimisation for structural recursion, known as \Hi{fusion law}.
The following theorem is called a fusion law,
which states that the composition of two structural recursive functions
can be expressed as a single one.
It means that recursing an 
input term \Hi{only once} is enough (at the right-hand side), rather than
twice (i.e.\ at the left-hand side, firstly \ssrec~recurses an input
and secondly $h$ recurses the result), hence it provides an optimisation.

\Th[th:fusion]
Suppose that
$\ju{e_\ell}{\dmark}{\dmark}$ and $\ju{d_\ell}{X}{X}$ are terms
for every $\ell\in L$.
If a structural recursive function $h$ satisfies
$h(e_\ell) = d_\ell$ for all $\ell \in L$, then
\begin{equation}
h \o \srec{ e } = \srec{d }.  
\label{eq:fusion-str}
\end{equation}
where $e = (e_\ell)_{\ell\in L},\; d = (d_\ell)_{\ell\in L}$.

\oTh
This has been proved in \cite[Theorem 4]{Buneman}.
The original proof was quite involved, which spent 3.5 pages long.
It was proved by induction on terms and using a technical
lemma lifting an equation on finite trees to that on infinite graphs.
This involved proof method was unavoidable
because in the original formulation,
\begin{itemize}[itemsep=.5em]
\item graphs are the basic data of the formulation, 
\item terms denote only a subset of the set of all graphs, and
\item infinite graphs are
allowed as possible graphs (but no infinite terms are allowed),
\item thus terms are not enough to capture
all graphs (cf.\ discussion \Sec \ref{sec:problems}).
\item 
The statement (\ref{eq:fusion-str}) was actually the statement
on finite and infinite graphs and it was not only on finite terms, 
\item so the technical lifting lemma from finite trees to infinite trees
was necessary.
\end{itemize}
Similar fusion laws were stated in \cite{ICFP13} 
for an ordered variant of UnCAL
but no proof was given. 
Because the formulation in \citet{ICFP13} follows the original
formulation \cite{Buneman}, the formal proofs for the fusion laws, if exist, 
would have the same problems.

Our algebraic formulation in this paper cleanly avoids these problems.
We can now give a much simpler and conceptual proof, which
does not even rely on induction.
Because we do not use graphs at the level of syntactic formulation,
now the statement (\ref{eq:fusion-str}) is actually a statement on
UnCAL terms.
The proof proceeds as follows.

\Proof
Let $k \deq |W|$. 
The proof is due to 
the following diagram, which commutes by Thm. \ref{th:universality}.
Here $\Term(Y,X)$ is the set of all well-typed terms
having the source context $X$ and the target context $Y$,
$\Psi^A, \Psi^B$ are the arrow part functions of functors,
and $\CE(\GE^{|Y|},\GE^{|X|})$ is a hom-set,
cf. (\ref{eq:psik}).
\y{-.5em}
\begin{diagram}[height=2.3em]
   \Term(Y,X) &\rTo^{\den{-}^\GE}& \CE(\GE^{|Y|},\GE^{|X|}) \\
 \dTo<{\den{-}^A}&
  \ldTo<{\Psi^A}\ldTo(2,4)>{\Psi^B}\\
 \CE(\GE^{|Y|},\GE^{|X|})\\
 \dTo<h\\
 \CE(\GE^{k\cdot |Y|},\GE^{k\cdot |X|})
\end{diagram}
It expresses that the composition of two 
iteration functors $\Psi^A$ and $h$,
which both preserves $L$-monoids,
is again an iteration functor $\Psi^B$ that preserves $L$-monoids, 
and such $\Psi^B$
exists uniquely by the universality.

More precisely,
Let $A=(\GE,\den{-}^A_L), B=(\GE^{k},\den{-}^B_L)$ 
be models of $E$ in \CE.
Suppose that $h$ is an iteration functor such that $h(\GE)=\GE^{k}$ and
preserves $L$-monoid structures.
Then by 
Thm.\ \ref{th:universality}, two iteration functors
$$\Psi^A =\srec{e},\quad \Psi^B =\srec{d}$$
such that $\Psi^A(\GE)=\GE, \Psi^B(\GE)=\GE^{k}$, 
$\Psi^A$ maps the $L$-monoid \GE to the $L$-monoid $A$,
and $\Psi^B$ maps the $L$-monoid \GE to the $L$-monoid $B$,
exist uniquely and the above diagram commutes.
Observe that to give an iteration functor $h$ sending the $L$-monoid $A$ 
to $B$ is equivalent to giving a
structural recursive function $h$, satisfying
$$
h(\den{\ell}^A_L) = \den{\ell}^B_L \;\text{ for all }\ell\in L.
$$
This is exactly the condition stated in Thm.\ \ref{th:fusion},
hence 
we have proved the fusion law.
\QED

The fusion law for the case of primitive recursion \cite{Buneman}
is
\[
 \prec{d} \o \prec{ e } 
   = \prec{\; (\prec{d} (\prec{e}(\ell\co \dmark))_{\ell\in L}\; }
\]
So we prove:
\[
 (\prec{d} \o \prec{ e }) (s)
   = \prec{\; (\prec{d} (\prec{e}(\ell\co \dmark))_{\ell\in L}\; }(s)
\]
This is proved straightforwardly
by induction on the typing derivation of $s$ and 
the characterisation (\ref{eq:my-pfun}).
The difference from the original proof is that now 
we do not rely on a technical and involved
lemma lifting an equation on finite trees to that on infinite graphs.
The use of induction on $s$ is validated by the result
that \AxG completely characterises bisimulation of UnCAL graphs,
which will be proved in the next section.


\section{Completeness for Bisimulation}\label{sec:alg}

In this section, we prove that our axioms \AxG completely
characterise the extended bisimulation used in the original formulation
of UnCAL.
Namely, \AxG is a {complete} syntactic axiomatisation 
of the original bisimulation equality on UnCAL graphs.
We prove it 
by connecting it with the algebraic axiomatisations of bisimulation
by Bloom and \Esik \cite{BE,Esik00,Esik02}.
We first review their characterisation result 
using the notion of \mu-terms by following the formulation
in \cite[\Sec 2.5]{Esik02}.

\subsection{Equational logic for \mu-terms}\label{sec:ELmu}
Let \Sig be a signature, i.e.\
a set of function symbols equipped with arities.
We define \mu-terms by
$$
t \;::=\; x \;\|\; f(t_1,\ooo,t_n) \;\|\; \mux t\,,
$$
where $x$ is a variable, and $f$ is an $n$-ary function symbol in \Sig.
For a 
set $V$ of variables, we denote by $\Tm(V)$ the set of all \mu-terms
generated by $V$.
We also define the \mu-notation on vectors
$\mu (x_1,\ooo,x_n).(t_1,\ooo,t_n)$
by induction on $n$: if $n=1$, it is 
$\mu x_1.t_1$. If $n \gt 1$, we define
\begin{meqa}
\mu (x_1,\ooo,x_n).(t_1,\ooo,t_{n}) \deq (\mu (x_1,\ooo,x_{n-1}).s,\;
       \mu x_n.t_n[\mu (x_1,\ooo,x_{n-1}).s/(x_1,\ooo,x_{n-1})])
\end{meqa}
where $s = (t_1,\ooo,t_{n})[\mu x_n.t_n/x_n]$. The notation
$[t/(x_1,\ooo,x_{n})]$ means $$[(\pi_1 t,\ooo, \pi_n t)/(x_1,\ooo,x_{n})]$$
where $\pi_i (x_1,\ooo,x_{n}) = x_i$.
We now regard each label $\ell\in L$ as an unary function symbol
and consider the signature $\Sig=L\union\set{0^{(0)},+^{(2)}}$
(the superscripts denote the arities).

\subsec{Axioms for bisimulation}
We assume the following equational axioms on 
\mu-terms to capture bisimulation.

\begin{framed}
\begin{meq}
\renewcommand{\arraystretch}{1.3}
\arraycolsep = 1mm
\begin{array}[h]{rllllllllll}
\axtit{Conway equations} \x{25em}\ \\
\mux t[ s / x] &=& t[\, \mux s[t / x]\, /x \,]\\
\mux \muy t &=& \mux t[ x / y]  \\
\axtit{Group equations} \\
\qquad (\mux t[ 1\cdot x / x],\ooo,t[ n\cdot x / x])_1 &=& \muy (t[ y / x_1,\ooo, y / x_n])\\
\axtit{Axioms for branches and bisimulation}\\
\end{array}
\\
\arraycolsep = .5mm
\begin{array}[h]{cclcclclllllllllllllll}
s+(t+u) &=& (s+t)+u \qquad & s+t&=&t+s  & t+0=t &\\
\mux x &=& 0 & \mux (x + t) &=&
t &\text{for }t\text{ not containing }x
\end{array}
\end{meq}
\end{framed}

The axiom of group equations \cite{Esik02,Esik-group} is a schema read
as follows. The notation $(-)_1$ at the left-hand side
denotes the first component of a vector.
Let $n$ be a natural number and
$G=(\set{1,\ooo,n},\cdot)$ be a finite group of order $n$.
Given a vector
$x=(x_1,\ooo,x_n)$ of distinct variables, define
$i\cdot x=(x_{i\cdot 1},\ooo,x_{i\cdot n}).$
Thus, $i\cdot x$ is obtained by permuting the components of $x$ according 
to the $i$-th row of the multiplication table of $G$.
Varying all possible $n\in \Nat$ and finite groups $G$ of order $n$, we generate
concrete axioms of "group equations" 
by using the multiplication $\cdot$ of $G$.
This form is taken from \cite[p.280]{Esik02}.
Group equations are shown to be an alternative 
form of the commutative identities. For further details, 
see \cite{Esik02}, or \cite[\Sec 18 A simple \mu-language]{Esik-group}.

\W{The fixed point law}
\[
\mux t \;=\; t[\mux t /x]
\]
is an instance the first axiom of Conway equations by taking $s = x$.

\subsec{Equational logic}
We call \ELmu the standard equational logic of \mu-terms with
the above axioms.
We write
$\;\prMu s = t\;$ if an equation $s=t$
is derivable from the above axioms in
\ELmu. 
For example, idempotency is derivable:
\[
\prMu\;\; t + t \,=\, t
\]
The proof is $t = \mu x.(x+t) = (\mu x.(x+t)) +t = t+t$,
which uses the axiom $\mux (x + t) =t$
and the fixed point law.
Since \mu-terms can be regarded as a representation of process terms 
of regular behavior (or synchronization trees \cite{BE})
as \citet{MilnerRegular} showed,
the standard notion of bisimulation between two \mu-terms can be
defined (e.g. \cite[page 9]{Sew95}). We write $s \bisim t$ if they are bisimilar.

\ThTitledN[th:Esik]{\cite{BET,BE,Esik00}(\cite[Thm.\ 2.]{Sew95})}
The equational logic \ELmu  completely axiomatises 
the bisimulation, i.e.
for \mu-terms $s$ and $t$,
$$\prMu s = t \;\;\Longleftrightarrow\;\; s \bisim t.$$
\oTh

\subsection{Characterising UnCAL normal forms}\label{sec:alg-sem}

Our strategy to show completeness of our axioms \AxG for bisimulation
is that we reduce our problem to Thm.\ \ref{th:Esik} of the case of 
\mu-terms.
Hence, we need to relate UnCAL terms and \mu-terms.
Clearly, UnCAL terms have richer term constructs than \mu-terms.
But taking suitable sets of axioms as rewrite rules,
we can obtain certain \Hi{normal forms}, which are simpler 
and relate directly to \mu-terms.
Of course, we should carefully choose axioms from \AxG,
because some axioms, such as (fix), cause infinite rewrites.
We choose the axioms (sub) and (\Bekic) from \AxG and
orient each axiom from left to right as a rewrite rule.
Then we have the following rewrite system \RR.
\[
\begin{array}[h]{lrlllllllllllll}
\urule{(sub)} &
t \at \pa{s_1\opl\ccc\opl s_n} &\to& t\qesub
\\
\urule{(Beki\u{c})\;} & \cyc^{X+Y}(\ccpair{t}{s})
&\to&
\renewcommand{\arraystretch}{.8}
\begin{array}[h]{lll}
  \cpair{\pi_{Z,X}^X}{\cyY{s}} \,\at\, \\
  \cpair{\Id_Z} {\cyX{t \at \cpair{\Id_{Z\!+\! X}}{\cyY{s}}}}
\end{array}
\end{array}
\]

\subsec{Termination} The rewrite system \RR is terminating, i.e.
strongly normalising (SN)  \cite{Baader}.
We prove it by translating \RR to another rewrite system on 
terms of \lmdG-calculus, which is a simply-typed \lmd-calculus extended with fixed point
operator given in \Sec \ref{sec:lambdaG}.
(The \lmdG-calculus does not depend on the result of completeness
for bisimulation we establish in this section, hence it is not a circular argument.)
We use the translation $\tra{-}$ from UnCAL terms to 
\lmdG-terms given in Fig. \ref{fig:trans}
in Appendix.
Applying this translation to \RR,
we obtain the rewrite rules $\RRg$ on \lmdG-terms:
\[
\begin{array}[h]{lrlllllllllllll}
\urule{(sub')} &
(\lmd \vec y.t)\; ({s_1\opl\ccc\opl s_n}) &\to& t\qesub
\\
\urule{(\Bekic')} &
\fix_{m+n}(\lmd(\vec x,\!\vec y).\,(\,\btt,\; \bss\,)) &\to&
(\fixm{\lmd\vec x.\; (\lmd \vec y.\btt) \;\; \fixn{\lmd \vec y. \bss}     },\\
&&&\;\,
\fixn{\lmd \vec y.\; (\lmd \vec x.\bss) \;\;  \fixm{\lmd \vec x.\, (\lmd
    \vec y.\btt) \; \fixn{\lmd \vec y. \bss}}})
\end{array}
\]
where $\btt$ and $\bss$ are short for $t\; (\vec x,\vec y)$ and $s\; (\vec x,\vec y)$,
respectively.

The rewrite system $\RRg$ is SN.
Basic reasons are as follows. The rule (sub') is essentially the \beta-reduction rule of the simply-typed
\lmd-calculus, hence terminating.
Applying (\Bekic') rule,
the number of the tuples at the argument of $\fix$ is reduced.
Namely, the subscript of $\fix$ (i.e. $m$ or $n$) in the right-hand side of (\Bekic')
is always smaller than that in the left-hand side (i.e $m+n$),
hence the rule (\Bekic') is terminating.

More formally, we can prove SN of $\RRg$
using a general established method called {\GS}~\cite{IDTS,IDTS00},
which is based on Tait's computability method to show SN.
\TGS has succeeded to prove termination of various recursors such as 
the recursor in G\"{o}del's System T.
The basic idea of \GS is to check whether the arguments of 
recursive calls in the the right-hand side of a rewrite rule 
are ``smaller'' than the left-hand sides' ones. It is similar 
to Coquand's notion of 
``structurally smaller'' \cite{CoqPat}, but more relaxed and extended.
The rewrite system $\RRg$ 
can be seen as a rewrite system of the format of inductive datatype system
given in \cite{IDTS00}. Note that $s$'s and $t$ are metavariables
in an inductive datatype system,
and the abstraction $\lmd x.t$ is written as $[x]t$ in \cite{IDTS00}.
Since $\RRg$ fits into the 
\GS using
the well-founded order 
$
\cyc^m \gt \cyc^n 
$
where natural numbers $m \gt n$, \RRg is SN.
A larger rewrite system involving these rules
has been proved to be SN by using \GS in \cite{cydat}.

It is straightforward to show that 
if we have $s \to_\RRd t$ using \RRd, then
we have $\tra{s} \to_{\RRg} \tra{t}$ using \RRg.
Suppose that there is an infinite rewrite sequence:
\[
s_0 \to_\RRd s_1 \to_\RRd s_2 \to_{\RRd} \ccc 
\]
It is translated to an infinite rewrite sequence in the \lmdG-calculus:
\[
\tra{s_0}\to_{\RRg} \tra{s_1}\to_{\RRg}
\tra{s_2}\to_{\RRg} \ccc 
\]
But this is impossible because $\RRg$ is SN. So \RRd is SN.

\subsec{Confluence}
The rewrite system $\RR$ is confluent  because it is left-linear
(i.e. there is no repetition of the same metavariable at every left-hand side)
and there is no overlapping between the rules \cite{Klop,KOR}.

\subsec{Unique normal forms}
Since \RR is confluent and terminating, any term has the unique normal form
by rewriting using \RR \cite{Baader}.
Given an UnCAL term,
we write $$\fn{nf}(t)$$ for the unique normal form of $t$.

\Lemma[th:cy-tup]
If $t$ is of type $X+Y$ where $n=|X+Y|\gt 1$, then $\nf t$ 
is of the form $<t_1,\ooo,t_n>$.
\oLemma
\Proof
A possible term $t$ of such a type is either $<t_1,\ooo,t_n>$ or 
$\cyc^{X+Y}{(<s,t>)}$. For the latter case, applying (\Bekic), 
$\cyc^{X+Y}{(<s,t>)}$ becomes 
a tuple involving $\cyX s, \cyY t$ and $\at$. 
The $\at$ is consumed by (sub),
and applying (\Bekic) repeatedly
the subterms $\cyX s, \cyY t$ finally become $\cyc^1$-terms.
Hence the normal form of $t$ is a tuple.
\QED

We call a raw term $t$ \W{value} if it follows the following grammar
\[
\begin{array}[h]{lclllllllllllllllllll}
  s,t &::=& {\var{y}}
  &|&   \ell\co{t} 
  &|&   \cyc^x{\xsub{x}{t}} 
  &|&   \nil
  &|&   \mult
  &|&   \mult\at\cpair{s}{t} 
  &|&   \xsub{{x}}{t} \\
  &|&   <s,t>
  &|&   \EMP
\end{array}
\]
Note that variables $x,y$ are arbitrary, which
may be the same markers or may be different.

\Lemma
We define 
$
\MM \deq \set{ t \| \ju t Y X \text{ and $t$ is a value}}.
$
For any
$
\ju t Y X,
$
we have $\nf t \in \MM$.
\oLemma
\Proof
By induction on typing derivations of $t$.
\begin{itemize}
\item Case (Emp)(Pair),(Nil),(Man),(Mark). It is in \MM.
\item Case (Def) $t=\xsub{x}s$. Let $s'=\nf s$. By I.H., $s' \in \MM$. Hence
  $\xsub{x}{s'}\in \MM$.
\item Case (Label) Similar to (Def).
\item Case (Com) $t = u \at s$. 
  \begin{itemize}
\item Case $t = \mult\at s$. Then $\nf s = <s_1,s_2>$ and
$s_1,s_2$ are normal forms. By I.H., 
$s_1,s_2\in\MM$. Then
$\nf t = \mult\at<s_1,s_2> \in \MM$.

  \item Case $u\not=\mult$.
Let $u',s'$ be the normal forms of $u,s$.
By I.H., these are in \MM. 
By Lemma \ref{th:cy-tup}, $s' = <s'_1,\ooo,s'_n>$ 
($n$ is possibly $0,1$). So
$\nf{u' \at <s'_1,\ooo,s'_n>} = u'\; [\vec x\mapsto s'_1,\ooo,s'_n]\in\MM$.

  \end{itemize}

\item Case (Cyc) $t=\cyX s$ with $n=|X|$.
  \begin{itemize}
  \item  $n \gt 1$. Then $\nf s =<s_1,\ooo,s_n>$ and every $s_i \in \MM$.
By Lemma \ref{th:cy-tup},
$\nf t = \nf{\cyX{<s_1,\ooo,s_n>}}$ is a tuple consisting of $s_i$, hence in \MM.

\item $n =1$. Similar to (Def).
  \end{itemize}

\end{itemize}\y{-2em}
\QED

\Prop
If $t\in\MM$ is of type \dmark, then $t$ follows the grammar
\[
\begin{array}[h]{lclllllllllllllllllll}
\MM_\dmark \ni s,t&::=& {\var{y}}
  &|&   \ell\co{t} 
  &|&   \cyc^x{\xsub{x}{t}} 
  &|&   \nil
  &|&   \mult
  &|&   \mult\at\cpair{s}{t} 
\end{array}
\]
\oProp
\Proof
Clear from the type.
\QED

We regard $\MM_\dmark$ as a term set.
As the final normalisation process, we generate another 
term set \Nm from $\MM_\dmark$
by replacing every sole $\mult_\seq{y_1,y_2}$
(which is not of the form $\mult_\seq{y_1,y_2} \at <s,t>$) in terms of $\MM_\dmark$, with 
the ``\eta-expansion'' $\mult_\seq{y_1,y_2}\at<y_1,y_2>$.
This process uses the axiom (\eta\!\mult) in \AxG.
Thus, we have
\[
\begin{array}[h]{llclllllllllllllllllll}
\Nm \;\ni\;\; &s,t \;::=\;& {\var{y}}
  &\;\;|\;\;&   \ell\co{t} 
  &\;\;|\;\;&   \cyc^x \xsub{x}t
  &\;\;|\;\;&   \nil
  &\;\;|\;\;&   \mult\at<{s},{t} >
\\
\Tm(V)\;\ni\;\;  &s,t \;::=\;&
  {y}
  &\;\;|\;\;&  \ell(t)
  &\;\;|\;\;&  \mu x.{t} 
  &\;\;|\;\;&  0
  &\;\;|\;\;&  {s}+{t}
\end{array}
\]

We call an element of \Nm \W{UnCAL normal form}, which is always of type \dmark.
Every UnCAL normal form 
bijectively corresponds to a \mu-term in $\Tm(V)$, i.e. 
$
\Nm \iso \Tm(V),
$
because each the above construct corresponds to the lower one.
Hereafter, we may identify normal forms in \Nm and \mu-terms as above.

\Def[def:bisim-UnCAL]
We say that well-typed UnCAL terms \(s\) and \(t\)
are \W{bisimilar}, denoted by $s \bisim t$, if the UnCAL graphs
\(\den{s}_\gc^{\set{\dmark}}\) and \(\den{t}_\gc^{\set{\dmark}}\) are
extended bisimilar described in Appendix \Sec \ref{sec:UnCALgraphs}.
\oDef

For a \mu-term  \(t\),
its \Hi{chart} \cite[page 9]{Sew95}
is extended bisimilar to the UnCAL graph \(\den{t}_\gc^{\set{\dmark}}\).
This is shown by induction on the term syntax.
Then the bisimilarity for \mu-terms
agrees with the bisimilarity for UnCAL normal forms in \Nm.

\subsection{Completeness of the axioms for bisimulation}
\label{sec:bisim}

By Thm.\ \ref{th:Esik}, we have known that
\ELmu is complete for bisimulation. 
We show the completeness of \AxG in \ELu for bisimulation,
using the following Lemma \ref{th:nf-corr} that relates 
the problem of
\ELmu with that \ELu of through UnCAL normal forms.

\Lemma[th:nf-corr]
For UnCAL normal forms $n,m\in\Nm$, if
$\prMu n = m$ then there exists a marker $x$ such that
$$\;\ju{n=m}{Y}{\seq x}$$ is derivable from \AxG in \ELu.
Note that $x$ is mostly $\dmark$, but consider the case $n=\cyc^x(t)$.
\oLemma
\Proof
By induction on proofs of \ELmu.
For every axiom in \ELmu, there exists a corresponding axiom
in \AxG or an \ELu theorem, hence it can be emulated in \ELu.
\QED

\ThTitled[th:complete]{Completeness} 
Assume a pure UnCAL theory.
For any terms $s$ and $t$ of type $x$,
\begin{center}
$\ju{s=t}{Y}{\seq x}\;$ is derivable from \AxG in \ELu\quad iff\quad
$\; s \bisim t  $.
\end{center}
\oTh
\Proof
$[\Rightarrow]:$
Because for every axiom $s=t$ in \AxG, we have $s\sim t$,
and
the bisimulation is closed under contexts and substitutions \cite{Buneman}.

\noindent
$[\Leftarrow]:$
Suppose $s \bisim t$. Since for every rewrite rule in \RR,
both sides of the rule is bisimilar, $\fn{nf}$ preserves the bisimilarity. 
So we have
$\fn{nf}(s) \bisim s \bisim t \bisim \fn{nf}(t)$.
Since normal forms correspond to \mu-terms, using Thm.\ \ref{th:Esik},
we have $\prMu \fn{nf}(s) =  \fn{nf}(t)$.
By Lemma \ref{th:nf-corr}, we have a theorem
$\ju{\fn{nf}(s)=\fn{nf}(t)}{Y}{\seq x}$.
Thus $s=t$ is derivable.
\QED


\section{Instances of UnCAL Models}\label{sec:models}

The categorical semantics provides
a \Hi{generic framework} of semantics of UnCAL.
It is not only providing a single semantics of UnCAL, but also
captures various \Hi{concrete models} of UnCAL 
by the general form of categorical semantics,
because it is \Hi{parameterised}
by an \Hi{arbitrary} iteration category \CC.
To give a {concrete} model, we 
\Hi{choose} a concrete iteration category \CC 
with a {concrete} $L$-monoid in it. 
We have given the syntactic model \GE in \CE in \Sec \ref{sec:comp}.
In this section, we will give several other
concrete models.

\subsection{The bisimulation model}
\label{sec:bisim-model}

Buneman et al.~formulated that
UnCAL graphs were identified by a sort of bisimulation called
extended bisimulation \cite[Definition 3]{Buneman}.
We give this bisimulation model as 
an example of {UnCAL model}.
We define an equivalence relation
$\sim_E$ on well-typed terms
by the symmetric transitive closure of the union of
the renaming identification,
the bisimulation $\bisim$ on UnCAL terms given in Def. \ref{def:bisim-UnCAL}
and 
the congruence generated by the additional axioms.
Crucially, our equational axiomatisation in \ELu is complete for 
bisimulation.

\ThTitled[th:completeB]{Completeness for bisimulation} 
\label{thm:compaxbisim}
Let $E$ be UnCAL axioms (including \AxG).
Then $\ju{s=t}{Y}{X}\;$ is derivable in \ELu 
iff\;
$\; s \bisim_E t  $.
\oTh
\Proof
Immediate corollary to Thm.\ \ref{th:complete}.
\QED

Thus, the equivalence relation generated by $E$ 
in the equational logic \ELu is 
exactly the same as bisimulation with additional axioms. Hence
the generic model \GE in the classifying category \CE
considered in \Sec \ref{sec:comp}
is also the terms modulo bisimulation model.

The original graph theoretic definition of 
UnCAL graphs, extended bisimilarity, and constructors
form also an UnCAL model, which is described
in detail in Appendix \ref{sec:graphmodel}.

\subsection{A CPO model}
\label{sec:cpo-model}

We next give a new model of UnCAL.
One of the most natural examples of iteration category 
is the category \CPO of cpos (complete partial orders) having the least element $\bot$
and continuous functions.
It is well-known that the category \CPO is 
traced cartesian \cite{Hassei}, moreover
an iteration category \cite{BE}, where 
the least fixed point operator
(calculated by
$\lub_{n\in\Nat} f^n(\bot)$ for a continuous function $f$)
is a Conway operator $\dagg{-}$.
In the category \CPO,
a cycle is interpreted as a suitable infinite unfolding.

Let $L^\omega$ denote the set of all finite and infinite strings of labels $L$.
We define
$\DD $ to be the set of
all non-empty subsets of $L^\omega$
modulo the equivalence relation generated by the preorder $\sLE$, where 
$A \sLE B$ iff
for all $u \in A$, there exists $w \in B$ such that $u$ is a prefix of $w$.
This $\DD$ with the order $\sLE$ 
(extended to the quotient sets) forms a cpo, actually a Hoare powerdomain
$\PP(L^\omega)$
\cite{Abramsky-Jung}.
The least element $\bot$  is the singleton 
$\set{\epsilon}$ of the empty string.
We now define an $L$-monoid structure on \DD by taking 
$\eta^D\deq \bot=\set{\epsilon}$, $\mu^D$ as the binary join $\lub$, and
$\ell^D (A) \deq \set{\ell\, w \| w \in A}$ for $\ell \in L$.

Then $\DD$ satisfies all the axioms in \AxG (see the examples below).
The categorical interpretation $\den{-}^{\DD}_\CPO$,
induced from Fig.\ \ref{fig:interp},
translates
an UnCAL term $\ju{t}{Y}{X}$ to a continuous function
$$\den{t}^{\DD}_\CPO:\DD^{|Y|}\to \DD^{|X|}.$$

\Example[ex:cpo-interp]
Since \DD is a model, any UnCAL theorem $s=t$ is valid using \DD, i.e.
$\den{s}^{\DD}_\CPO = \den{t}^{\DD}_\CPO$ holds.
We demonstrate the interpretation of theorems
(except for \ref{item:as}).
\begin{enumerate}[label=(\arabic*)]
\item 
$\den{(\code a\co{\var x}) \at \xsub{x}{\code b\co\vnil}}
 = (\lmd x.\code{a}^\DD(x))\, \set{\code{b}}
 = \set{{\code a \code b }}
 = \den{\code a \co \code b \co \vnil}
$
 \label{item:betaex}
\item $\den{\uni{\code a \co \vnil}{\empty} \;\uniSym\; 
     \uni{\code b \co \code c\co \vnil}{\code a \co \vnil}}
\;=\; \set{\code a , \code b \code c }
\;=\; \den{\uni{\code a \co \vnil}{\code b \co \code c \co \vnil}}
$
\item $\den{\cyc(\dmark)}
= \dagg{ \lmd x. x } = \bot = \set{\epsilon}
= \den{\,\vnil\,},\;$
 \label{item:c1}
\item $\den{ \ju{\cyc^{\dmark}(\mult_{\seq{\dmark,\var y}})}{\var{y}}{\dmark} }
= {\lmd y.\, \dagg{ \lmd x. \mu^\DD(x , y)} }
= {\lmd y.\, \lub_{n\in\Nat} ( (\lmd x. \mu^\DD(x , y))^n ) (\bot)}
= {\lmd y.\, y}
= \den{\, \ju{\var{y}}{\var{y}}{ \dmark} \,}$
 \label{item:c2}
\item $\den{\cyc(\code{a}\co \dmark)}
= \dagg{ \lmd x. \code{a}^\DD(x) }
= {\lub_{n\in\Nat} (\lmd x. \code{a}^\DD(x))^n}(\bot)
= \lub \set{ \set{\code{a}^n} \| n \in \Nat}
= \set{\code{a}^\omega}
$
 \label{item:as}
\end{enumerate}
\oExample
There are several benefits of this model.
\begin{enumerate}[itemsep=.5em]
\item The cpo model explains in a mathematically rigorous way 
how the infinite expansion of a cyclic graph occurs,
i.e. it is the least fixed point calculated by
the least upper bound of an \omega-chain
(see~\ref{item:c2} and~\ref{item:as}).

\item The cpo model gives \Hi{intuition}
why the axioms hold. For example, ``$\vnil$'' can be understood
as $\bot$, and a reason why the axioms (c1) and (c2) hold 
can be understood by the fixed points (see~\ref{item:c1} and~\ref{item:c2})
and the structure of cpos.

\item This \Hi{motivates} the following idea:
a \lmd-calculus with fixed point operator
may be used to calculate values and
to check equality (e.g.\ \ref{item:betaex}) of UnCAL terms.
We will realise this idea in the following subsection.
\end{enumerate}

\subsection{The \lmdG-calculus model}
\label{sec:lambdaG}

UnCAL is very similar to an ordinary functional programming language,
in the sense that UnCAL programs are given by structural recursive definitions.
Also, in the examples (1)--(5) and the item (iii) in \Sec \ref{sec:cpo-model},
we observed some similarities between UnCAL  and a \lmd-calculus.

But UnCAL's target data was graphs 
(rather than data of algebraic data types) and
the computation of UnCAL was realised by graph algorithms.
Thus it has been considered that UnCAL is
far from an ordinary functional programming language.
But there must be some connection to functional programming.
What is the exact connection?

Using our categorical semantics, we can now formally 
answer to the question.
In this subsection,
we show a connection between UnCAL and functional programming
by giving an interesting translation from UnCAL terms to \lmd-terms.
Crucially, this translation is 
\W{automatically obtained as an instance of our categorical
semantics}.
That is, we first define an applied simply-typed \lmd-calculus called
the \lmdG-calculus, and next show that it forms an UnCAL model.
Then it automatically induces a categorical interpretation, which gives
a sound and complete translation from
UnCAL terms and programs 
to \lmd-terms in the \lmdG-calculus.

\subsec{The \lmdG-calculus}

Suppose a label set $L$ is given.
The \lmdG-calculus is
a simply-typed lambda calculus extended with pairs, constants,
fixed point operator with the axioms
in Fig. \ref{fig:axioms-FL} and possibly additional axioms \Ax.
We assume three base types, $\fn{B}$ (for Booleans), 
$\fn{G}$ (for graphs), and $\fn{L}$ (for labels).
We identify $\textbf{1}\X \tau = \tau\X \textbf{1} = \tau$.
We also identify $\tau_1\X(\tau_2\X\tau_3) = (\tau_1\X\tau_2)\X\tau_3$,
simply write $\tau_1\X \tau_2\X\tau_3$ and corresponding elements as 
$(t_1,t_2,t_3)$, and 
assume that $\pi_i$ is suitably defined for $n$-case.
We use the notation $\lmd(x_1,\ooo,x_n).t$ (or $\lmd\vec x.t$) to denote
$\lmd x^{\sig_1\X\ccc\X\sig_n}. t\usub{x_1,\ooo,x_n}{\pi_1 x_1,\ooo,\pi_n x}$.
We also identify a curried term with its uncurried one, i.e.,
by $\lmd(x_1,\ooo,x_n).\lmd(y_1,\ooo,y_m). t$, we mean
$\lmd(x_1,\ooo,x_n,y_1,\ooo,y_m). t$.
This is because in our use of \lmdG-calculus, we are 
mainly interested in up to first-order types.

The typing rules are exactly the standard rules for the simply-typed 
\lmd-calculus (e.g.\ \cite{Crole})
using the typed constants given below.
Note that the constant \oo corresponds to $\nil$ in UnCAL, 
semantically corresponds to $\bot$ in the cpo semantics.
It is a formalisation of the ``black hole''
(``apparent undefinedness'') used in a call-by-need calculus by \citet{NaHa09}.

As usual, every axiom (including \Ax)
consists of a pair of well-typed terms of the same type
and under the same context.
Most of the axioms for \lmdG are obtained by
translating the axioms \AxG in \ELu using the 
the translation $\tra{-}$ in Fig. \ref{fig:trans} in Appendix.
The axioms of parameterised fixed point are slightly different, but
they are a version of the axioms of Conway parameterised fixed point
 \mu-terms \cite{Esik02}.
(amalg) is the axiom schema called \Hi{amalgamation} \cite[Def. 3.2]{alex-plot}
or \Hi{functorial implication} \cite[Def. 8]{iter-alg},
which is very close to 
the commutative identities (CI) in \AxG.
Note that (c1) is derivable as in \AxG
(but we include it for readability).
The \Hi{\lmdG-theory} is an equational theory obtained by ordinary equational
logic of \lmd-calculus using the axioms in Fig. \ref{fig:axioms-FL}
and additional axioms \Ax. 

\begin{rulefigw}
\[
\y{-1em}
\begin{array}{llllll}
\multicolumn{2}{l}{\textbf{The \lmdG-calculus}}\\
\textit{Types}&
\tau :: = \textbf{1} \| \GG \| \fn{L} \| \fn{B} \| \tau \X \tau' \| \tau \to \tau' 
\\
\textit{Terms}&
  e \;::=\;\; x \;\|\; \lmd x.e \;\|\; e_1\, e_2 
\;\|\; (e_1,e_2) 
\;\|\; c \qquad\text{(constant)}
\\
\textit{Constants}\quad&
() : \textbf{1}\qquad \pi_i : \tau_1\X\tau_2\to \tau_i
\quad
\text{for $i=1,2$}
\qquad 
\oo:\GG\qquad
\union:\GG\X\GG\to\GG
\\
&- \co - : \fn{L}\X \GG \to \GG\qquad
\fix_n :(\GG^n\to\GG^n)\to\GG^n
\qquad
\ell : \fn{L}\quad
\text{for $\ell \in L$}
\\
&\TRUE:\fn{B}\quad\FALSE:\fn{B}\quad
\textsf{if - then - else -} : \fn{B}\X\GG\X\GG\to\GG\quad
-\syeq- : \fn{L}\X\fn{L}\to\fn{L}
\end{array}
\]
\y{-.5em}
\[
\arraycolsep = .2mm
\normalsize
\begin{array}[h]{lrllllllll}
\textit{Axioms}\\
\axtit{Calculus}\\
\urule{(beta)}&    (\lmd x.t)\; s &\;=&\; t \usub{x}{s}\\
\urule{(eta)}&    (\lmd x.t\, x)  &\;=&\; t \\
\axtit{Cartesian}
\\
\urule{(fst)}& \pi_1\; (s , t) &\;=&\; s \\
\urule{(snd)}& \pi_2\; (s , t) &\;=&\; t \\
\urule{(fsi)} & \opair{\pi_1\; t}{\pi_2\; t} &\;=&\; t\\
\axtit{Deleting trivial cycle}
\\
\urule{(c1)} & \fix(\lmd x. x) &\;=&\; \oo \\
\orule{(c2)} & \fix(\lmd x.(x \union t)) &\;=&\; t \\[11em]
\end{array}
\begin{array}[h]{lrllllllll}
\axtit{Parameterised fixed point} 
\\
\urule{(Cw1)}&
\fix(\lmd x. t\usub{x}{s} ) &\;=&\; t\usub{x}{\fix(\lmd x. s\usub{x}{t})}\\
\urule{(Cwn2)}&
\fix(\lmd x. \fix(\lmd y. t)) &\;=&\; \fix(\lmd x. t\usub{y}{x} )  \\
\urule{(amalg)}
&\multicolumn{3}{l}{\fix (\lmd (x_1,\ooo,x_n).(t_1,\ooo,t_n))}\\
& &\hspace{-2em}=&\hspace{-2em} (\fix(\lmd y.s),\ooo,\fix(\lmd y.s))\\
&\multicolumn{3}{c}{\text{if }  t_i\usub{x_1,\ooo,x_n}{y,\ooo,y} = s \text{ for all } t_i}
\\
\axtit{Commutative monoid with idempotency}
\\
\urule{(unitL$\union$)} &\oo \union t &\;=&\; t\\
\urule{(unitR$\union$)} &t \union \oo &\;=&\; t\\
\urule{(assoc$\union$)} &s \union (t \union u) &\;=&\; (s \union t) \union u \\
\orule{(com$\union$)} &s \union t &\;=&\; t \union s
\\
\orule{(degen)} &t \union t &\;=&\; t\\
\axtit{Conditionals}\\
\urule{(ift)} &\ifthen\TRUE t e &\;=\;& t \\
\urule{(ife)} &\ifthen\FALSE t e &\;=\;& e \\
\urule{(eqt)} & \ell \syeq \ell \;&\;=\;& \TRUE \text{ for }\ell\in L\\
\urule{(eqe)} & \ell \not\syeq \ell' &\;=\;& \FALSE \text{ for }\ell\not=\ell'\in L
\end{array}
\]
\y{-1em}
\begin{minipage}[t]{\textwidth}\small
\textit{Note.}
We use the abbreviation $s \union t \deq \union\;(s,t)$.
In (eta) and (c2), $x$ does not appear in $t$. 
The subscript $n\in\Nat$ of $\fix$
may be ommited for simplicity.
\end{minipage}
\caption{The \lmdG-calculus}
\label{fig:axioms-FL}
\end{rulefigw}


\subsec{Categorical semantics}
We define the category \catLam of first-order \lmdG-terms
by taking
\begin{itemize}[itemsep=.5em]
\item \textit{objects:} types $\GG^n$ \;$(n\in\Nat)$
\item \textit{arrows:} 
arrows from $\GG^{m}$ to $\GG^n$
are equivalence classes
$[\ju{t}{}{\GG^m\to\GG^n}]$ of closed terms
quotiented by the congruence generated
by renaming, the axioms of \lmdG and additional axioms \Ax
\item \textit{composition:}
$[\pr \lmd y.s: \GG^k\to\GG^n]\o [\pr \lmd x.t : \GG^m\to\GG^k] 
= [\pr \lmd x. \; s[t/y] : \GG^m\to\GG^n]$
\item \textit{identity:}
$[\pr \lmd x.x: \GG^n\to\GG^n]$
\end{itemize}
Then, because of the axioms,
it is immediate that \catLam forms
an iteration category, 
where the Conway operator $\dagg{-}$ is given by \fix.
Moreover, $\GG$ forms an $L$-monoid by taking
\begin{meqa}
  [\oo]: \textbf{1} \to \GG, \qquad
  [\union ] : \GG\X\GG \to \GG, \qquad
  \den{\ell}_L^\GG = [\lmd x . \ell \co x].
\end{meqa}
The categorical interpretation in Fig.\ \ref{fig:interp}
induces the interpretation $\den{-}_\catLam^\GG$ 
that
interprets an UnCAL term $\ju{t}{Y}{X}$
as a closed \lmd-term of first-order type:
$$ \den{t}_\catLam^\GG \;:\; \GG^{|Y|} \to \GG^{|X|}. $$
The interpretation $\den{-}_\catLam^\GG$
is concretely described as the interpretation
function $\tra{-}$ in Fig. \ref{fig:trans} in Appendix.
The examples considered in Example \ref{ex:cpo-interp} are now
interpreted as follows.
\begin{enumerate}[label=(\arabic*)]
\item 
$\den{(\code a\co{\var x}) \at \xsub{x}{\code b\co\vnil}}
 = [(\lmd x.\code{a}\co x)\; (\code{b}\co \oo)]
 = [{\code a \co \code b \co \oo }]
$
 \label{item:betaex}
\item $\den{\uni{\code a \co \vnil}{\empty} \;\uniSym\; 
     \uni{\code b \co \code c\co \vnil}{\code a \co \vnil}}
= [{\code a \co \oo} \union {\oo}  \union
     \unio{\code b \co \code c\co \oo}{\code a \co \oo}]
= [{\unio{\code a \co \oo}{\code b \co \code c \co \oo}}]
$
\item $\den{\cyc(\dmark)}
= [\fix( \lmd x. x )] = [\oo]$
 \label{item:c1}
\item $\den{ \ju{\cyc^{\dmark}(\mult_{\seq{\dmark,\var y}})}{\var{y}}{\dmark} }
= [\lmd y.\, \fix({ \lmd x. {x}\union{y}})]
= [\lmd y.\, y$]
 \label{item:c2}
\item $\den{\cyc(\code{a}\co \dmark)}
= [\fix(\lmd x.\code{a}\co x)] 
= [\code{a}\co \fix(\lmd x.\code{a}\co x)] 
= [\code{a}\co \code{a}\co \fix(\lmd x.\code{a}\co x)] $
 \label{item:as}
\item $\den{\ju{\mathbf{t}_G}{y_1,y_2}{\dmark} } = 
\den{\code a \cc{\Uni{\code b\co{\var x}}{\code c\co{\var x}}} \;\at\;
 {\cy{\nxsub{x}{\code  d\cc{ \UniT{\code p\co{\var{y_1}}} {\code q\co{\var{y_2}}} {\code r\co{\var{x}}}}
   \;} }}}\\
\phantom{M} = [\lmd(y_1,y_2).\; (\lmd x. \code a\co \unio{\code b\co x}{\code c\co x})
  \; \fix (\lmd x. \code d \co ( (\code p\co y_1)\union(\code q\co y_2)\union(\code r\co x)))]$
\end{enumerate}
The last two examples correspond to the examples considered in
Introduction, \Sec \ref{sec:ireview}.

\Prop[th:corrspond-ELu]
Given UnCAL's additional axioms \EE, 
we define \lmdG's additional axioms $\EE'$ by
$
\EE' \deq \set{ \tra{s} = \tra{t} \|  (s = t) \in \EE}.
$
Suppose $\ju{s}{Y}{X}$ and $\ju{t}{Y}{X}$. Then
\begin{center}
$\ju{s = t}{Y}{X}$ is derivable in \ELu
\;iff\; $\den{s}_{\catLam}^\GG=\den{t}_{\catLam}^\GG$\; holds in the \lmdG-theory.
\end{center}
\oProp
\Proof
\protect{[$\Rightarrow$]}: By soundness of the categorical semantics.

[$\Leftarrow$]: We define the inverse translation $(-)^\circ$
of $\den{-}_{\catLam}$,
i.e. a mapping from 
the image of $\den{-}_{\catLam}^\GG$ to $\CE(m,n)$ by
\[
\arraycolsep = 1mm
\begin{array}[h]{llllllllll}
&\inv{ \lmd  \vec y. y_i } &=& \eclas{i} 
&\inv{ \lmd y. s\; t } 
&=& \inv{\lmd y. s} \o <\id_{\GE^m},\inv{\lmd y.t}>  \\
&\inv{ \lmd  y. (s, t) } &=& \pa{\inv{\lmd  y.s} \opl \inv{\lmd  y.t}}
&\inv{ \lmd  y. \fix(t) } &=& \dagg{\inv{\lmd  y.t}} \\
&\inv{ \lmd  y. \ell\co t } &=& \den{\ell}_L \o {\inv{\lmd  y.t}}
&\inv{ \lmd  y. () } &=& \eclas{\EMP}  \\
&\inv{ \pi_i } &=& \eclas{\pi_i} 
&\inv{ \lmd  y.\oo } &=& \eclas{\vnil} 
\qquad\inv{ \union } = \eclas{\mult}
\end{array}
\]
Note that 
for $\inv{t}$,
any subterm of $t$, other than \fix, is up to first-order type.
It defines a partial function 
$$(-)^\circ : \cat{Lam}(\GG^{m},\GG^{n}) \to \CE(m,n).$$
which is well-defined, i.e., for closed terms $t, t' :\GG^m\to\GG^n$,
where any subterm of $t,t'$, other than \fix, is up to first-order type,
the following holds:
\begin{center}
if $t=t'$ holds in the \lmdG-theory,
then $t^\circ = {t'}^\circ$ is an UnCAL theorem.
\end{center}
This is proved by induction on the proof of $t=t'$.
We need to check that for every axiom $s=s'$ of the \lmdG-calculus,
$s^\circ = {s'}^\circ$ is an UnCAL theorem.
By applying $(-)^\circ$ to an axiom, 
we can recover the corresponding axiom in \AxG or obtain an UnCAL theorem.
Note that the axioms for \lmdG were obtained by
translating each of axiom \AxG to that of \lmdG
using $\den{-}_{\catLam}^\GG$.
Then we have $\invv{\den{t}_\catLam^\GG}= [t]_E$ for all UnCAL terms $t$,
by induction on typing derivations. 
Suppose $\den{s}_{\catLam}^\GG=\den{t}_{\catLam}^\GG$.
Applying $(-)^\circ$, we obtain
$[s]_E = (\den{s}_{\catLam}^\GG)^\circ = (\den{t}_{\catLam}^\GG)^\circ = [t]_E$.
\QED

Hence any provable equation $s=t$ over UnCAL terms in \ELu
is also provable using \lmdG-calculus.
If we have suitable operational semantics for \lmdG-calculus,
we may prove it by simplifying terms by the operational semantics in \lmdG.
This methodology is especially needed
when one wants to prove the form $s = v$
where $v$ is a suitable ``value'' in UnCAL.
It is actually possible via a functional programming language
called FUnCAL, which we will see in \Sec \ref{sec:FUnCAL}.

\subsubsec{Axiomatisation of the recursion operators}

To translate UnCAL programs defined by structural recursion to \lmdG,
next we axiomatise the recursion operators formally in the \lmdG-calculus.
We assume additionally new constants $\fn{srec}$ for structural recursion,
$\fn{prec}$ for primitive recursion,
and the following axioms in \lmdG.

\arraycolsep = .4mm
\renewcommand{\arraystretch}{1}
\[
\begin{array}[h]{llllllll}
\axtitt{Axioms for recursion operators} \\
\multicolumn{5}{l}{
\lssrec:
(\fn{L}\to\GG^k\to\GG^k)\to(\GG^n\to\GG^m)\to(\GG^{k\cdot n}\to\GG^{k\cdot m})
}\\
  \lsrec{e} (&\lmd \vec y.{ y_i}&)  &=& 
    \pi'_i \\
  \lsrec{e}(&\lmd y.s \; t  &) &=&  
   \lmd y'.(\lsrec{e}\,\lmd y. s)
            \; (y', (\lsrec{e}\,{\lmd y.t}) \; y')\\
  \lsrec{e} (&   \lmd y.(s, t)   &) &=& \lsrec{e}\,{(\lmd y.s)} \;\times'\; \lsrec{e}\,{(\lmd y.t)}\\
  \lsrec{e}(&\lmd y. { \fix({t}) } &) &=& 
     \lmd y'.\;\fix(\, (\lsrec{e}{\,\lmd y.t})\;{y'} \, )
\\
  \lsrec{e}(&\lmd y. \ell \co{t} &) &=&
     \lmd y'.\,   e\; \ell\;  (\,(\lsrec{e}\, \lmd y. t)\; y'\,)  \\
  \lsrec{e} (&\lmd y. \EMP &)  &=& \lmd y'.  \EMP  \\
  \lsrec{e} (&\pi_i &)  &=& \pi'_i  \\
  \lsrec{e} (&\lmd y. \oo &)  &=&  \lmd y'.({\oo},\dots,{\oo}) \\
  \lsrec{e} (& \union  &) &=& \union'
\\[1em]
\multicolumn{5}{l}{
\lsprec:
(\fn{L}\to\GG^{k+1}\to\GG^{k})\to \GG^m \to \GG^{k\cdot m}
}\\
\multicolumn{5}{l}{
\lprec{e}t = \pi\; \bigl(\lsrec{( \lmd \ell \lmd x.\; (e\;\ell\; x,\;
  \ell\co \pi' x ))} t \bigr)
}
\end{array}
\]
The types of the recursion operators $\lssrec$ and $\lsprec$ 
correspond to those of functions \Psi in \Sec \ref{sec:char} and \Upsilon in \Sec
\ref{sec:char-prim}, respectively,
which are actually parameterised by $k,m,n\in\Nat$
(but we omit to attach subscripts).

The operator $\pi'$ selects the final coordinate of a tuple $x$,\;
$\pi$ selects the rest of it, and
$\pi_i'$ is the $i$-th projection of a tuple of tuples.
Moreover, $\X'$ is the concatenating 
operator of two tuples under \lmd-binder,
$\union'$ is the ``zip'' 
operator of two tuples by $\union$ under \lmd-binder.
Here, by an operator, we mean a suitable \lmd-term.

We call this extension 
with additional axioms translated from UnCAL's additional axioms
\emph{the extended \lmdG-calculus} 
and 
its theory \emph{the extended \lmdG-theory}.
These axioms are a formalisation of the characterisation we have obtained
in (\ref{eq:my-sfun}).

\subsec{Relating UnCAL and \lmdG}

Suppose the structural recursion $\msrec{e}$ for 
the $L$-indexed expressions $e = (e_\ell)_{\ell\in L}$ of UnCAL, 
where $\ju {e_\ell}W W$ is given.
We assume that $e$ for $\msrec{e}$ 
must satisfy the condition that there exists a \lmdG-term 
$\traG e : \fn{L} \to \GGG{W} \to \GGG{W}$
such that
$$\traG{e}\; \ell \;=\; \den{e_\ell}_\catLam^\GG$$ 
holds in the extended \lmdG-theory.
This assumption means that a structural recursive definition
is given by a suitable case analysis. For example,
the function \code{aa?} in Example \ref{ex:aa} is
represented as $\msrec e$ where 
$e_\code{a}= \code{head-a?}(t)$
and $e_\ell= \code{aa?}(t)$ for $\ell\not= \code a$.
In this case, we can take
\[
\traG e \deq \lmd \ell\, t.\, \ifthen{\ell\syeq\code a}{\code{head-a?}(t)}{\code{aa?}(t)}
\]
Similary for $\mprec{e}$, we assume that for $\jud {W,\dmark} e W$,
there exists a \lmdG-term 
$\traG e : \fn{L} \to \GG^{|W|+1} \to \GGG{W}$
such that
$\traG{e}\; \ell \;=\; \den{e_\ell}_\catLam^\GG$
holds in the extended \lmdG-theory.

The axiomatisation is sound and complete as follows.

\Prop
For any term $\ju{t}{Y}{X}$,\;
\begin{meq}
\msrec{e}(t) = u \quad \text{iff} \quad
\lsrec{(\lmd\ell.\traG{e}\;\ell)}(\den{t}_\catLam) = \den{u}_\catLam
\;\text{
in the extended \lmdG-theory},
\\
  \mprec{e}(t) = u \quad \text{iff} \quad
\lprec{(\lmd\ell.\traG{e}\; \ell)}(\den{t}_\catLam) = \den{u}_\catLam
\;\text{
in the extended \lmdG-theory}.
\end{meq}
\oProp
\Proof
\protect{[$\Rightarrow$]}:
By induction on the typing derivation of $t$.

[$\Leftarrow$]: As in Prop.\ \ref{th:corrspond-ELu},
we define the inverse translation also for 
\lssrec~and \lsprec.
\begin{meqa}
  [\lsrec{(\lmd \ell.\traG{e}\; \ell)}{t}]^\circ &\deq \msrec{ e }(\inv{t})\\
  [\lprec{(\lmd \ell.\traG{e}\; \ell)}{t}]^\circ &\deq \mprec{ e }(\inv{t})
\end{meqa}
The rest is similar to Prop.\ \ref{th:corrspond-ELu}.
\QED

These theorems open possibilities of importing
the techniques and theory developed for the \lmd-calculus and
functional programming to UnCAL.
It enhances reasoning, computations and optimisations of
UnCAL programs.
One can translate an equational problem of UnCAL into the extended \lmdG-calculus,
then one can use reasoning and computation in \lmdG and retrieve 
the result by translating back it to UnCAL by $\inv{-}$ given in 
the proof of Prop. \ref{th:corrspond-ELu}.

\subsubsec{The language FUnCAL
as an efficient operational semantics of \lmdG}
\label{sec:FUnCAL}

Finally, we mention that 
our clarification of the connection between 
UnCAL and the extended \lmdG-calculus has 
been theoretically and concretely more precisely implemented.

The functional language FUnCAL \cite{MA}
gives an efficient operational semantics of \lmdG.
Rather than equational theories, 
FUnCAL focuses on evaluating and manipulating
UnCAL programs as usual functional programs. 
The syntax of FUnCAL is essentially the same as \lmdG and
an abstract machine for FUnCAL, which respects the axioms of 
the extended \lmdG-calculus
(hence it is a correct implementation),
has been designed and 
implemented.
The implementation is available at
\begin{center}
\texttt{https://bitbucket.org/kztk/funcal}
\end{center}
It is based on
\citet{NaHa09}'s semantics of call-by-need computations,
which extends \citet{Launch}'s natural semantics of lazy evaluation
with the black hole
\cite{cyclic-lmd,ETGRS}.
With an embedded implementation of the lazy semantics inspired
by~\citet{FiKS11} who used a monad to describe deferred
computation in a certain lazy semantics, it is shown to be efficient about 4
to 6 times faster in concrete examples than an ordinary implementation
of UnCAL \cite[Table 1]{MA}.  It can be regarded as effectiveness of
our semantic and axiomatic framework.


\section{Discussion: Variations of UnCAL and Related Graph Calculi}
\label{sec:concl}

The categorical structure we have clarified
leads us naturally to other variations and 
to relate the UnCAL with other graph calculi,
by modifying the current axioms \AxG.
In this section, we discuss variations of UnCAL and 
related graph calculi.

\begin{itemize}[leftmargin=0pt,itemsep=.9em]
\item \textit{Ordered branching UnCAL}

In the original UnCAL considered in this paper, 
the branch of a tree has no order, i.e. 
$\uniSym$ is commutative and idempotent,
which is due to the bisimulation equivalence.
The modification to ordered branches is 
pursued in \cite{ICFP13,PPDP13}, where
elaborated adaptations of the bisimulation and
graph-theoretic semantics of structural recursion for ordered branches
are developed.
In our case setting, the modification is simpler.
Ordered branch UnCAL can be axiomatised just by 
dropping the axioms marked ``\delmark'' in \AxG of Fig. \ref{fig:axioms}.
Correspondingly, the \lmdG-calculus for ordered branches
is obtained by dropping the axioms marked ``\delmark'' 
in the axioms of \lmdG.
The ordered variant of Thm.\ \ref{th:Esik}
can be proved straightforwardly by accordingly modifying 
bisimulation as in \cite{ICFP13,PPDP13},
the axioms and
the proof of \cite{BE,BET} or \cite{Salomaa,MilnerRegular}
to deal with ordered branches.

\item \textit{Why not monoidal, rather than cartesian? ---
forbidding copy of graphs}

To choose the axioms of monoidal product, instead of cartesian product,
means to forbid copying of graphs
in substitution.
This setting is more common in modelling graphs,
such as \cite{Gibbons,Hassei, GS-monoidal}.
Degenerated bialgebras in a strict monoidal category 
(known as a PROP \cite{MacLanePROP}) are used by \citet{FiorePROP} to 
model directed acyclic graphs.
That work was originated by Milner's question \cite[Introduction]{FiorePROP}
concerning extensions of his \Hi{bigraphs} \cite{ax-bigraph}.
A similar approach can also be found in
Plotkin's algebraic modelling of bigraphs \cite{Plot-bigraph}.
Exploring a link to bigraphs would be interesting.

\item \textit{Why not traced, rather than iteration category? 
--- cyclic sharing theories} 

Dropping the axiom (CI), it no longer forms an iteration category, but
still is a \Hi{traced} cartesian category \cite{JSV}.
Hasegawa clarified that to model graphs in calculi using 
a letrec-style notation, both linear and non-linear treatments 
of constructs are necessary. This clarification led him to
\Hi{cyclic sharing theories} and its semantics:
cartesian-center traced monoidal categories \cite{HasseiTLCA,Hassei}.
It was also used in \cite{Hassei} to model Milner's \Hi{action calculi} \cite{ActionCalc}.
These calculi and models may be useful for a variation of UnCAL.

\item \textit{How about premonoidal trace? --- ``arrows'' and loops in Haskell}

The notion of trace operator can also be adapted
to a further-weakened setting called premonoidal categories \cite{premonoidal}.
It has been used to model recursive computation with side-effects \cite{Benton},
and has been implemented as a feature called arrows with loops \cite{Paterson}
in Haskell.
Modifying our axioms to referring 
the axioms of premonoidal trace (or arrows with
loops) 
may be practically interesting.
Here the idea of cyclic terms and arrow interpretation \cite{FLOPS}
may be incorporated.

\item \textit{Why not closed? --- higher-order types}

Introducing \Hi{cartesian closed} (or monoidal closed) axioms 
\cite{Lambek-Scott}, 
we obtain a higher-order version of UnCAL.
It will be related to other higher-order
graph calculi, such as the cyclic \lmd-calculus \cite{cyclic-lmd}
and higher-order cyclic sharing theories \cite{Hassei}.
Moreover, as a framework of equational logic for graphs and 
structural/primitive recursion,
introducing higher-order types is a natural choice, as
we have done in the \lmdG-calculus.
But our consideration of the \lmdG-calculus is restrictive than 
the cyclic \lmd-calculus, etc.,
because the \lmdG-calculus has only first-order 
fixed point operator.
\end{itemize}

Making these connections explicit and
such various ideas of variations and extensions 
are a powerful benefit of category theoretic characterisation,
because category theory is a general language to bridge
different worlds of mathematics and computational structures.

\bigskip

\textit{Acknowledgments} 
We are grateful to Zhenjiang Hu and the members of 
the Bidirectional Graph Transformation Project
at NII for various discussions on topics related to UnQL/UnCAL.
We also acknowledge the stimulated discussion in 
the meeting 
of the Cooperative Research Project ``Logical Approach to Metaprogramming''
(2013-2015)
at
RIEC, Tohoku University.
We are especially grateful to Atsusi Ohori for his comments on relevance to 
object-oriented database.
This work was supported in part by JSPS KAKENHI Grant Number 24300001 and  
25540002 (Hamana), 15K15966 (Matsuda), 23220001 and 15H05706 (Asada).


\bibliographystyle{apalike}
\bibliography{bib}

\appendix
\section{The Graph Model of UnCAL}\label{sec:graphmodel}

In this appendix,
we reorganise the original graph theoretic definition of 
UnCAL graphs, bisimilarity, and constructors \cite{Buneman}
as an instance of our categorical model.

\subsection{UnCAL graphs}
\label{sec:UnCALgraphs}

We define the UnCAL graphs formally. 
Let 
$\Lab$ be a set of 
labels. 
An \W{UnCAL graph} $G$ 
is a quadruple $$(V,E,I,O)$$ where $V$ is a set of \Hi{vertices},
$E \subseteq V \times (\Lab \cup\set{\eps}) \times V$
is a set of \Hi{edges}, and
a function  $I : X \rightarrow V$ and a relation $O \subseteq V \times Y$ 
are the connection of roots $X$ to vertices, 
and that of vertices to leaves $Y$, respectively,
where $X$ and $Y$ are sets of markers.
A \Hi{finite} UnCAL graph is an UnCAL graph whose set of vertices is finite.
We write the set of UnCAL graphs as $\gs{Y}{X}$.

Two UnCAL graphs are
bisimilar if the infinite trees obtained by unfolding sharings and
cycles are identical after short-cutting all the $\eps$-edges.
Formally, it is defined as follows.
We write $v
\to^{l} u$ if there is an edge $(v,l,u) \in E$ between vertices $v, u \in
V$ in an UnCAL graph $G = (V,E,I,O)$.  An \W{extended simulation} $\mathcal{X}$
between an UnCAL graph $G_1=(V_1,E_1,I_1,O_1)$ and $G_2 = (V_2,E_2,I_2,O_2)$
is a relation satisfying the following conditions: 
\begin{enumerate}
\item 
if $(v,u) \in \mathcal{X}$, for any path $v=v_0 \to^\eps
\dots \to^\eps v_n \to^{l} v_{n+1}$ where $n \ge 0$, 
there is a path $u=u_0
\to^\eps \dots \to^\eps u_m \to^{l} u_{m+1}$ where $m \ge 0$ and $(v_{n+1},u_{m+1}) \in
\mathcal{X}$, 
\item 
if $(v,u) \in \mathcal{X}$,
for any path $v=v_0 \to^\eps \dots \to^{\eps} v_n$ where $n \ge 0$ such that $(v_n,\amarker{x}) \in O_1$, 
there is a path $u=u_0 \to^\eps \dots \to^{\eps} u_m$  where $m \ge 0$ such that $(u_m,\amarker{x}) \in O_2$.
\end{enumerate}
Moreover, $\mathcal{X}$ is called \W{extended bisimulation} if its
opposite relation is also an extended simulation relation 
between $G_2$ and $G_1$.
Two UnCAL graphs $G_1, G_2 \in \gs{Y}{X}$ are \W{extended bisimilar}, 
denoted by $G_1 \sim G_2$, if there is an extended bisimulation $\mathcal{X}$ between $G_1$ and $G_2$ such that
$(I_1(\amarker{x}),I_2(\amarker{x})) \in \mathcal{X}$ for any $\amarker{x} \in X$.
We define $\gsb{Y}{X}$ to be the quotient of $\gs{Y}{X}$ by the extended bisimilarity.

\subsection{The category of UnCAL graphs and a model}

UnCAL graphs form a category $\gc$ 
whose objects are sets of markers, and
whose hom-sets are $\gsb{Y}{X}$.
Namely, a morphism $G \in \gsb{Y}{X}$ is an UnCAL graph
(up to extended bisimilarity)
with roots $X$ and leaves $Y$.
It is clear that the following definitions are all well-defined with
respect to extended bisimilarity.
For an object $X$, the identity morphism is
   \[
 (X, \emptyset, \id_X, \Delta_X)
   \]   
where $\Delta_X$ is the diagonal relation on $X$.
For $G_1 \in \gsb{Y}{X}$ and $G_2 \in \gsb{Z}{Y}$,
their composition $G_1 \com G_2 \in \gsb{Z}{X}$ is
    \[
    \bb
(V, E, I_1,O_2)
\mbox{ where }~
      \bbt
      \bbt
        (V_1,E_1,I_1,O_1) &=& G_1 \\
        (V_2,E_2,I_2,O_2) &=& G_2 \\
      \ee\\
      \bbt
        V &=& V_1 \cup V_2\\
        E &=& E_1 \cup E_2 \cup \set{ (v,\eps,I_2(\amarker{x})) \mid (v,\amarker{x}) \in O_1 }
      \ee
      \ee     
    \ee    
    \]

We show that $\gc$ is an iteration category.
$\gc$ is cartesian: for $X_1$ and $X_2$, their product is the disjoint union $X_1+X_2$ (with some canonical renaming of
markers).
The $i$-th projection $\pi_i \in \gsb{X_1+X_2}{X_i}$ is
   \[
 (X_i, \emptyset, \id_{X_i}, \set{(x,x) \in X_i \times (X_1+X_2)\mid x \in X_i}).
   \]   
For $G_i \in \gsb{Y}{X_i}$, their pairing morphism $\pf{G_1}{G_2} \in \gsb{Y}{X_1+X_2}$ is
    \[
    \bb
 (V_1 \cup V_2, E_1 \cup E_2, I_1 \cup I_2, O_1 \cup O_2 ) \\
      \qquad \mbox{ where }~
      \bbt
        (V_1,E_1,I_1,O_1) &=& G_1 \\
        (V_2,E_2,I_2,O_2) &=& G_2 \\
      \ee
    \ee
    \]
where we rename elements in $V_1$ and $V_2$ so that they are disjoint.
The terminal object is the empty set $\emptyset$, and
the unique morphism $\langle\rangle_X \in \gsb{X}{\emptyset}$ is
    \[
       (\emptyset,\emptyset,\emptyset,\emptyset).
    \]

The category $\gc$ has a Conway operator satisfying the commutative identity:
For $G \in \gsb{Y+X}{X}$, $\dagg{G} \in \gsb{Y}{X}$ is
    \[
    \bb
(V, E', I, \set{ (v,\amarker{y}) \in O \mid \amarker{y} \in Y })
     \\
      \qquad \mbox{ where }
      \bbt
        (V,E,I,O) = G \\
        E' = E \cup \set{ (v,\eps,I(\amarker{x})) \mid (v,\amarker{x}) \in O }\\
        \ee
    \ee    
    \]
It is clear that the above structures satisfy the
axioms of an iteration category.
The Fig.~\ref{fig:graphAx} and Fig.~\ref{fig:graphAxCont} actually show
how the axioms \AxG are interpreted in the category $\gc$ of UnCAL
graphs.

Next, we show that $\{\dmark\}$ has an $L$-monoid structure in $\gc$.
First, it has a monoid structure:
the unit $\eta \in \gsb{\emptyset}{\{\dmark\}}$ is
     \[
(\set{\dmark}, \emptyset, \set{ \defaultM \mapsto \dmark }, \emptyset)
     \]
and the multiplication 
$\mu \in \gsb{\{\dmark\}+\{\dmark\}}{\{\dmark\}}$ is
     \[
        (\set{\dmark}, \emptyset, \set{ \defaultM \mapsto \dmark }, \{(\dmark,\dmark_1), (\dmark,\dmark_2)\})
     \]
where we renamed $\{\dmark\}+\{\dmark\}$ to $\{\dmark_1, \dmark_2\}$.
It is clear that this satisfies the axioms of a commutative monoid and
the axiom
$
 \dagg{\mu} = \id.
$
Finally, for $\ell \in L$ 
   \[
\sem{\ell}^{\{\dmark\}}_L
=
   \bb
(\set{\mathrm{v},\mathrm{u}}, \set{(\mathrm{v},{\ell},\mathrm{u})}, \set{ \defaultM \mapsto \mathrm{v}}, \set{(\mathrm{u},\defaultM)}).
   \ee
   \]
Thus, we have an $L$-monoid $(\set{\dmark}, \sem{-}^{\{\dmark\}}_L)$,
 i.e., a model of pure theory in $\gc$.

The induced categorical interpretation $\den{-}_\gc^{\set{\dmark}}$ provides
the graph theoretic meaning of a term. 
For example,
\y{-1em}
$$
\den{
 \code a \cc{\Uni{\code b\co{\var x}}{\code c\co{\var x}}} \;\at\;
 {\cy{\nxsub{x}{\code  d\cc{ \UniT{\code p\co{\var{y_1}}} {\code q\co{\var{y_2}}} {\code r\co{\var{x}}}}
   \;} }}
}_\gc^{\set{\dmark}}
= \raisebox{-3.5em}{\includegraphics[scale=.2]{Fig/graG4.eps}}
$$
Each UnCAL graph in Fig. \ref{fig:graph-constr} is exactly the interpretation
of the corresponding term by $\den{-}_\gc^{\set{\dmark}}$.
The model
$(\set{\dmark}, \sem{-}^{\{\dmark\}}_L)$ in $\gc$ is a sound and complete model of \AxG
as shown in Theorem~\ref{th:complete}.

\bigskip

\begin{rulefigt}\normalsize
\arraycolsep = 2mm
\[
\begin{array}[h]{|l|@{~}lll|lll}\hline
\rtit{Mark} 
& Y = \seq{\var{y_1},\ooo,\var{y_n}}
&&\\ \cline{2-2}
 &  \ju{ {\var y_i}_Y }{Y}{\dmark} &\mapsto& \pr \pi_i : \GGG{Y}\to \GG
\\ \hline
\rtit{Emp}
&&&\\ 
\cline{2-2}
&  \ju{ \EMP_Y }{Y}{\seq{}} &\mapsto& \pr \lmd y.()  : \GGG{Y}\to \terminal
\\ \hline
\rtit{Nil}
&&&\\ 
\cline{2-2}
&  \ju{ \nil_Y }{Y}{\dmark} &\mapsto& \pr \lmd y.\nil : \GGG{Y}\to \GG 
\\ \hline
\rtit{Man}
&&&\\ 
\cline{2-2}
&  \ju{ \mult_{\seq{\var{y_1},\var{y_2}}} }{{\var{y_1},\var{y_2}}}{\dmark} &\mapsto& \pr \union : \GG\X \GG\to \GG
\\[.5em] \hline
\rtit{Com}& 
 \ju s{Y}{Z} &\mapsto& \pr g : \GGG Y\to\GGG Z\\
&\ju t{X}{Y} &\mapsto& \pr h : \GGG X\to\GGG Y\\ 
\cline{2-2}
&\ju{s \at t}{X}{Z} &\mapsto& \pr g \o h : \GGG{X}\to \GGG{Z}
\\[.5em] \hline
\rtit{Label}& 
 \ell \in L & \mapsto& \pr \den{\ell}^\GG_L : \GG \to \GG\\
& \ju{ t }{Y}{\dmark} &\mapsto& \pr g : \GGG{Y} \to \GG  \\ 
\cline{2-2}
&  \ju{ \ell \co t }{Y}{\dmark}& \mapsto
  & \pr \den{\ell}^\GG_L \o g : \GGG{Y}\to \GG
\\[.5em] \hline
\rtit{Pair}& 
  \ju{ s }{Y}{X_1} &\mapsto& \pr g : \GGG{Y}\to\GGG{X_1}\\
& \ju{ t }{Y}{X_2} &\mapsto& \pr h : \GGG{Y}\to\GGG{X_2}\\ 
\cline{2-2}
  &  \ju{ \pa{s \opl t} }{Y}{X_1\conc X_2} 
&\mapsto& \pr \lmd y^{\GGG{Y}}\!\!.\,(g\; y, h\; y) : \GGG{Y} \to \GGG{X_1}\X\GGG{X_2}
\\[.5em] \hline
\rtit{Cyc}& 
  \ju{ \qquad\quad t\;\, }{Y\!\conc\! X}{ X} &\mapsto& \pr g : \GGG{Y}\X\GGG{X}\to\GGG{X}\\ 
\cline{2-2}
&  \ju{\cyX{t} }{Y\phantom{Y\conc\,}}{X} 
&\mapsto& \pr \lmd y^{\GGG{Y}}\!\!.\,\fix(\lmd x^{\GGG{X}}\!\!.\,g\; (y,x)) : \GGG{Y} \to \GGG{X}
\\[.5em] \hline
\rtit{Def}&
\quad\ju{ t }{Y}{ \dmark}   &\mapsto& \pr g : \GGG{Y}\to \GG\\
\cline{2-2}
 & \ju{ \xsub{x}{t} }{Y}{\var x} &\mapsto& \pr g : \GGG{Y}\to \GG
\\ \hline
\end{array}
\]
In (Mark), if $n=1$ then the interpretation is $\lmd x.x:\GG\to\GG$.
\caption{Translation $\tra{-}$ from UnCAL terms to \lmdG terms}
\label{fig:trans}
\end{rulefigt}



\end{document}